\documentclass[aps, pra, superscriptaddress, 10pt, twocolumn, longbibliography,  titlepage=false]{revtex4-1}
\usepackage{graphicx,nicefrac}
\usepackage{amsmath}
\usepackage{amssymb}
\usepackage{subfigure}
\usepackage{natbib}
\usepackage{color}
\usepackage{blkarray}
\usepackage{dsfont}
\usepackage[breaklinks=true,colorlinks=true,linkcolor=blue,urlcolor=blue,citecolor=blue]{hyperref}
\usepackage[percent]{overpic}
\usepackage{mathrsfs}
\usepackage{enumitem}
\usepackage{xspace}
\usepackage{xcolor}
\usepackage{epigraph}
\usepackage{siunitx}
\usepackage{txfonts}

\def\ket#1{{\left| #1 \right\rangle}}
\def\G{\mathsf{G}}
\def\Gi{\mathsf{G}_{\mathsf{i}}}
\def\Gx{\mathsf{G}_{\mathsf{x}}}
\def\Gy{\mathsf{G}_{\mathsf{y}}}

\newcommand{\eg}{\emph{e.g.}}
\newcommand{\ie}{\emph{i.e.}}

\renewcommand\vec{\mathbf}

\begin{document}
\title{Detecting and tracking drift in quantum information processors}
\author{Timothy Proctor}
\thanks{tjproct@sandia.gov}
\affiliation{Quantum Performance Laboratory, Sandia National Laboratories, Albuquerque, NM 87185 and Livermore, CA 94550} 
\author{Melissa Revelle}
\affiliation{Sandia National Laboratories, Albuquerque, NM 87185}
\author{Erik Nielsen}
\affiliation{Quantum Performance Laboratory, Sandia National Laboratories, Albuquerque, NM 87185 and Livermore, CA 94550} 
\author{Kenneth Rudinger}
\affiliation{Quantum Performance Laboratory, Sandia National Laboratories, Albuquerque, NM 87185 and Livermore, CA 94550} 
\author{Daniel Lobser}
\affiliation{Sandia National Laboratories, Albuquerque, NM 87185}
\author{Peter Maunz}
\affiliation{Sandia National Laboratories, Albuquerque, NM 87185}
\author{Robin Blume-Kohout}
\affiliation{Quantum Performance Laboratory, Sandia National Laboratories, Albuquerque, NM 87185 and Livermore, CA 94550} 
\author{Kevin Young}
\affiliation{Quantum Performance Laboratory, Sandia National Laboratories, Albuquerque, NM 87185 and Livermore, CA 94550} 	
\date{\today}

\begin{abstract}
\noindent If quantum information processors are to fulfill their potential, the diverse errors that affect them must be understood and suppressed. But errors typically fluctuate over time, and the most widely used tools for characterizing them assume static error modes and rates. This mismatch can cause unheralded failures, misidentified error modes, and wasted experimental effort. Here, we demonstrate a spectral analysis technique for resolving time dependence in quantum processors. Our method is fast, simple, and statistically sound. It can be applied to time-series data from any quantum processor experiment. We use data from simulations and trapped-ion qubit experiments to show how our method can resolve time dependence when applied to popular characterization protocols, including randomized benchmarking, gate set tomography, and Ramsey spectroscopy. In the experiments, we detect instability and localize its source, implement drift control techniques to compensate for this instability, and then demonstrate that the instability has been suppressed.
\end{abstract}
\maketitle

\section*{\normalsize Introduction}
\noindent Recent years have seen rapid advances in quantum information processors (QIPs). Testbed processors containing 10s of qubits are becoming commonplace \cite{rol2017restless, otterbach2017unsupervised, friis2018observation, arute2019quantum} and error rates are being steadily suppressed \cite{rol2017restless, blume2016certifying}, fueling optimism that useful quantum computations will soon be performed. Improved theories and models of the types and causes of errors in QIPs have played a crucial role in these advances.  These new insights have been made possible by a range of powerful device characterization protocols  \cite{blume2016certifying, merkel2013self, knill2008randomized, magesan2011scalable, proctor2018direct, magesan2012efficient, cross2016scalable, barends2014rolling, carignan2015characterizing, gambetta2012characterization, kimmel2015robust} that allow scientists to probe and study QIP behavior.  But almost all of these techniques assume that the QIP is stable --- that data taken over a second or an hour reflect some constant property of the processor. These methods can malfunction badly if the actual error mechanisms are time-dependent \cite{dehollain2016optimization, epstein2014investigating, van2013quantum,  fong2017randomized, chow2009randomized, fogarty2015nonexponential, wan2019quantum}. 

Yet temporal instability in QIPs is ubiquitous \cite{wan2019quantum, harris2008probing, bylander2011noise,chan2018assessment, fogarty2015nonexponential, klimov2018fluctuations, megrant2012planar,muller2015interacting, meissner2018probing, klimov2018fluctuations, de2018suppression, merkel2018magnetic, burnett2019decoherence}. The control fields used to drive logic gates drift \cite{wan2019quantum}, $T_1$ times can change abruptly \cite{burnett2019decoherence}, low-frequency $1/f^\alpha$ noise is common \cite{bylander2011noise}, and laboratory equipment produces strongly oscillating noise (\eg, $\SI{50}{\hertz}$/$\SI{60}{\hertz}$ line noise and $\sim$$\SI{1}{\hertz}$ mechanical vibrations from refrigerator pumps). These intrinsically time-dependent error mechanisms are becoming more and more important as technological improvements suppress stable and better-understood errors.  As a result, techniques to characterize QIPs with time-dependent behavior are becoming increasingly necessary.

In this article we introduce and demonstrate a general, flexible~and powerful methodology for detecting and measuring time-dependent errors in QIPs. The core of our techniques can be applied to time-series data from any set of repeated quantum circuits --- so they can be applied to most QIP experiments with only superficial adaptations --- and they are sensitive to both periodic instabilities (\eg, $\SI{50}{\hertz}$/$\SI{60}{\hertz}$ line noise) and aperiodic instabilities (\eg, $1/f^\alpha$ noise). This means that they can be used for routine, consistent stability analyses across QIP platforms, and that they can be applied to data gathered primarily for other purposes, \eg, data from running an algorithm or error correction. Moreover, we show how to use our methods to upgrade standard characterization protocols --- including randomized benchmarking (RB) \cite{knill2008randomized, magesan2011scalable, proctor2018direct, magesan2012efficient, cross2016scalable, barends2014rolling, carignan2015characterizing, gambetta2012characterization} and gate set tomography (GST) \cite{blume2016certifying,merkel2013self} --- into time-resolved techniques. Our methods therefore induce a suite of general-purpose drift characterization techniques, complementing tools that focus on specific types of drift \cite{cortez2017rapid,bonato2017adaptive,wheatley2010adaptive, harris2008probing, klimov2018fluctuations, chan2018assessment, bylander2011noise, young2012qubits, gupta2018machine, granade2016practical, granade2017qinfer, huo2017learning, kelly2016scalable, huo2018temporally, rudinger2018probing}. We demonstrate our techniques using both simulations and experiments. In our experiments, we implemented high-precision, time-resolved Ramsey spectroscopy and GST on a $^{171}$Yb$^{+}$ ion qubit. We detected a small instability in the gates, isolated its source, and modified the experiment to compensate for the discovered instability. By then repeating the GST experiment on the stabilized qubit, we were able to show both improved error rates and that the drift had been suppressed.

\begin{figure*}[]
\includegraphics[width=18.cm]{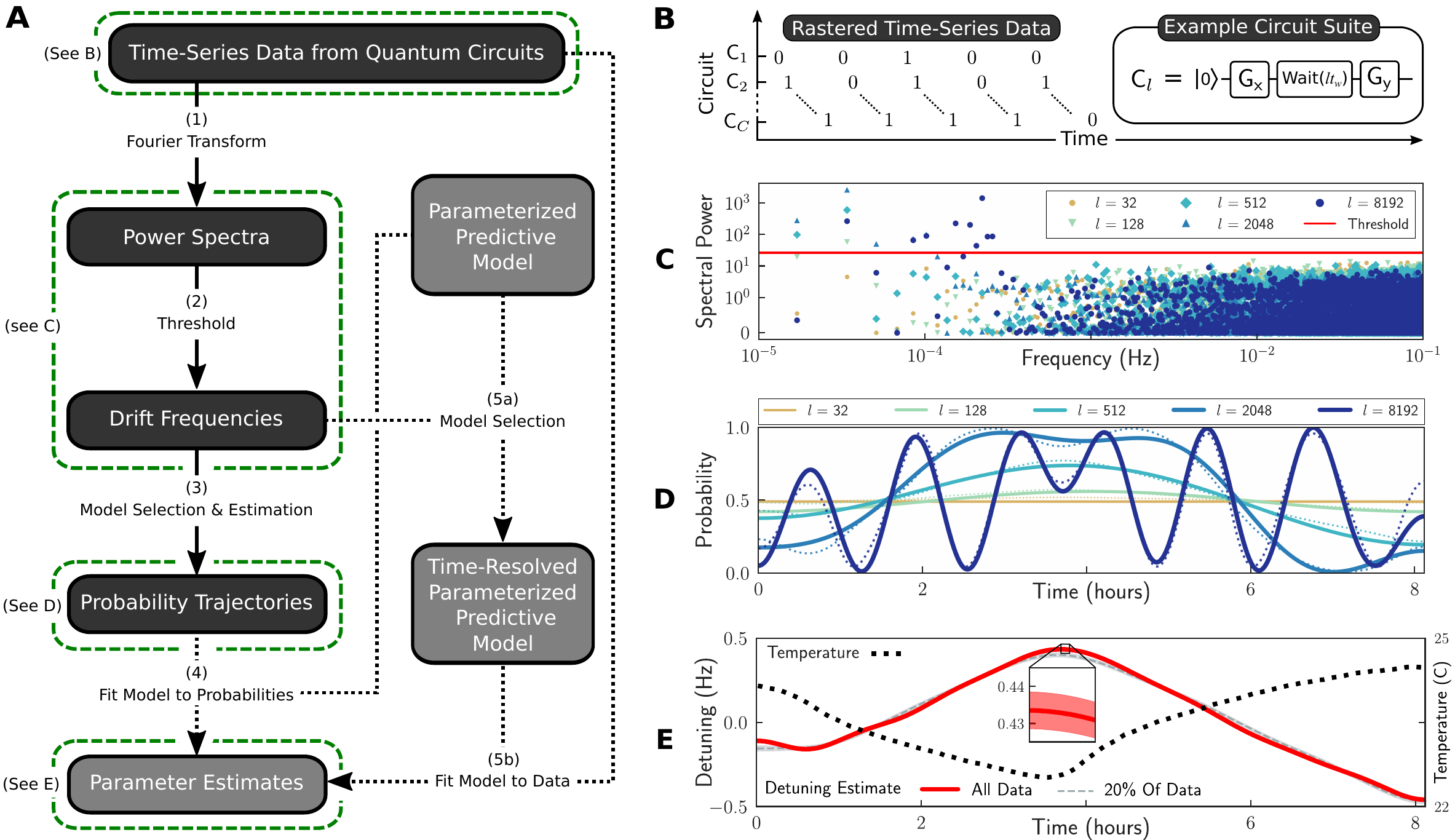}
\caption{{\bf Diagnosing time-dependent errors in a quantum information processor}. {\bf A.} A flowchart of our methodology for detecting and quantifying drift in a QIP, using time-series data from quantum circuits. The core steps (1--3) detect instability, identify the dominant frequencies in any drift, and estimate the circuit outcome probabilities over time. They can be applied to data from any set of quantum circuits, including data collected primarily for other purposes. Additional steps (4 and/or 5) estimate time-varying parameters (\eg, error rates) whenever a time-independent parameterized model is provided for predicting circuit outcomes. Such a model is readily available whenever the circuits are from an existing characterization technique, such as Ramsey spectroscopy or gate set tomography.~{\bf B.} An example of circuits on which this technique can be implemented --- Ramsey circuits with a variable wait time $lt_{w}$ --- as well as an illustration of data obtained by ``rastering'' (each circuit is performed once in sequence and this sequence is repeated $N$ times).~{\bf C-E.} Results from performing these Ramsey circuits on a $^{171}$Yb$^{+}$ ion qubit ($l=1,2,4,\dots,8192$, $t_w \approx  \SI{400}{\micro\second}$, $N=6000$).~{\bf C.} The power spectra observed in this experiment for selected values of $l$. Frequencies with power above the threshold almost certainly appear in the true time-dependent circuit probabilities, $p_l(t)$.~{\bf D.}~Estimates of the probability trajectories (unbroken lines are estimates from applying step 3 of the flowchart; dotted lines are the probabilities implied by the time-resolved detuning estimate shown in E). ~{\bf E.} The standard Ramsey model $p_l(t) = A + B\sin(2\pi l t_w\Omega)$, where $\Omega$ is the qubit detuning, is promoted to a time-resolved parameterized model (step 5a) and fit to the data (step 5b) using maximum likelihood estimation, resulting in a time-resolved detuning estimate (red unbroken line). The detuning is strongly correlated with ambient laboratory temperature (black dotted line), suggesting a causal relationship that is supported by further experiments (see the main text). The detuning can still be estimated to high precision using only 20\% of the data (gray dashed line), which demonstrates that our techniques could be used for high-precision, targeted drift tracking while also running application circuits. The shaded areas are $2\sigma$ ($\sim 95\%$) confidence regions}.
\label{fig:flowchart}
\end{figure*}

\vspace{-0.2cm}
\section*{\normalsize Results}
\noindent
{\bf Instability in quantum circuits.} 
Experiments on QIPs almost always involve choosing some quantum circuits and running them many times. The resulting data is usually recorded as counts \cite{blume2016certifying,merkel2013self, kimmel2015robust, knill2008randomized, magesan2011scalable, proctor2018direct, magesan2012efficient, cross2016scalable, barends2014rolling, carignan2015characterizing, gambetta2012characterization} for each circuit --- \ie, the total number of times each outcome was observed for each circuit. Dividing these counts by the total number of trials yields frequencies that serve as good estimates of the corresponding probabilities averaged over the duration of the experiment. But if the QIP's properties vary over that duration, then the counts do not capture all the information available in the data, and time-averaged probabilities do not faithfully describe the QIP's behavior.  The counts may then be irreconcilable with any model for the QIP that assumes that all operations (state preparations, gates, and measurements) are time independent. This discrepancy results in failed or unreliable tomography and benchmarking experiments \cite{dehollain2016optimization, epstein2014investigating, van2013quantum, fong2017randomized, fogarty2015nonexponential, chow2009randomized, wan2019quantum}.

Time-resolved analysis of the data from any set of circuits can be enabled by simply recording the observed outcomes (``clicks'') for each circuit in sequence, rather than aggregating this sequence into counts. We call the sequence of outcomes $\vec{x} = (x_1, x_2, \ldots, x_N)$ obtained at $N$ data collection times $t_1, t_2, \ldots, t_N$ a ``clickstream.'' There is one clickstream for each circuit. We focus on circuits with binary 0/1 outcomes (see Appendix~\ref{app:1} for discussion of the general case), and on data obtained by ``rastering'' through the circuits.  Rastering means running each circuit once in sequence, then repeating that process until we have accumulated $N$ clicks per circuit (see Fig.~\ref{fig:flowchart}B).  Under these conditions, the clickstream associated with each circuit is a string of bits, at successive times, each of which is sampled from a probability distribution over $\{0,1\}$ that may vary with time. If this probability distribution does vary over time, then we say that the circuit is temporally unstable. In this article we present methods for detecting and quantifying temporal instability, using clickstream data from any circuits, which are summarized in the flowchart of Fig.~\ref{fig:flowchart}A.

Our methodology is based on transforming the data to the frequency domain and then thresholding the resultant power spectra. From this foundation we generate a hierarchy of outputs: 1) yes/no instability detection; 2) a set of drift frequencies; 3) estimates of the circuit probability trajectories; and 4) estimates of time-resolved parameters in a device model. To motivate this strategy, we first highlight some unusual aspects of this data analysis problem.

Formally, a clickstream $\vec{x}$ is a single draw from a vector of independent Bernoulli (``coin'') random variables $\vec{X}=(X_1, X_2, \dots, X_N)$ with biases $\vec{p} = (p_1,p_2,\dots,p_N)$. Here $p_{i}=p(t_i)$ is the instantaneous probability to obtain 1 at the $i^{\rm th}$ repetition time of the circuit, and $p(\cdot)$ is the continuous-time probability trajectory. The naive strategy for quantifying instability is to estimate $\vec{p}$ from $\vec{x}$ assuming nothing about its form. However, $\vec{p}$ consists of $N$ independent probabilities and there are only $N$ bits from which to estimate them, so this strategy is flawed. The best fit is always $\vec{p} = \vec{x}$, which is a probability jumping between 0 and 1, even if the data seems typical of draws from a fixed coin. This is overfitting. 

To avoid overfitting, we must assume that $\vec{p}$ is within some relatively small subset of all possible probability traces. Common causes of time variation in QIPs are not restricted to any particular portion of the frequency spectrum, but they are typically sparse in the frequency domain, \ie, their power is concentrated into a small range of frequencies. For example, step changes and $1/f^{\alpha}$ noise have power concentrated at low frequencies, while $\SI{50}{\hertz}$/$\SI{60}{\hertz}$ line noise has an isolated peak, perhaps accompanied by harmonics. Broad-spectrum noise does appear in QIP systems, but because it has an approximately flat spectrum, it acts like white noise --- which produces uncorrelated stochastic errors that are accurately described by time-independent models. So, we model variations as sparse in the frequency domain, but otherwise arbitrary. Note that we do not make any other assumptions about $p(t)$. We do not assume that it is sampled from a stationary stochastic process, or that the underlying physical process is, \eg, strongly periodic, deterministic, or stochastic.

\vspace{0.2cm}
\noindent
{\bf Detecting instability.} The expected value of a clickstream is the probability trajectory, and this also holds in the frequency domain. That is, $\mathbb{E}[\tilde{\vec{X}}] = \tilde{\vec{p}}$ where $\mathbb{E}[\cdot]$ is the expectation value and $\tilde{\vec{v}}$ denotes the Fourier transform of the vector $\vec{v}$ (see the Methods for the particular transform that we use). In the time domain, each $x_i$ is a very low-precision estimate of $p_i$. In the frequency domain, each $\tilde{x}_{\omega}$ is the weighted sum of $N$ bits, so the strong, independent shot noise inherent in each bit is largely averaged out and any non-zero $\tilde{p}_{\omega}$ is highlighted. Of course, simply converting to the frequency domain cannot reduce the total amount of shot noise in the data. To actually suppress noise we need a principled method for deciding when a data mode $\tilde{x}_{\omega}$ is small enough to be consistent with $\tilde{p}_{\omega}=0$. One option is to use a regularized estimator inspired by compressed sensing \cite{donoho2006compressed}. But we take a different route, as this problem naturally fits within the flexible and transparent framework of statistical hypothesis testing \cite{lehmann2006testing,shaffer1995multiple}. 

We start from the null hypothesis that all the probabilities are constant, \ie, $\tilde{p}_{\omega} = 0$ for every $\omega >0$ and every circuit. Then, for each $\omega$ and each circuit we conclude that $|\tilde{p}_{\omega}| > 0$ only if $|\tilde{x}_{\omega}|$ is so large that it is inconsistent with the null hypothesis at a pre-specified significance level $\alpha$.  If we standardize $\vec{x}$, by subtracting its mean and dividing by its variance, then this procedure becomes particularly transparent: if the probability trace is constant, then the marginal distribution of each Fourier component $\tilde{X}_{\omega}$ for $\omega > 0$ is approximately normal, and so its power $|\tilde{X}_{\omega}|^2$ is $\chi_1^2$ distributed. So if $|\tilde{x}_{\omega}|^2$ is larger than the $(1 - \alpha)$-percentile of a $\chi^2_1$ distribution, then it is inconsistent with $\tilde{p}_{\omega}=0$.  To test at every frequency in every circuit requires many hypothesis tests.  Using standard techniques \cite{lehmann2006testing,shaffer1995multiple}, we set an $\alpha$-significance power threshold such that the probability of falsely concluding that $|\tilde{p}_{\omega}|>0$ at any frequency and for any circuit is at most $\alpha$ (\ie, we seek strong control of the family-wise error rate; see Appendix~\ref{app:1}).

We now demonstrate this drift detection method with data from a Ramsey experiment on a $^{171}$Yb$^{+}$ ion qubit suspended above a linear surface-electrode trap~\cite{blume2016certifying} and controlled using resonant microwaves. Shown in Fig.~\ref{fig:flowchart}B, these circuits consist of preparing the qubit on the $\hat{x}$ axis of the Bloch sphere, waiting for a time $lt_{w}$ ($l=1,2,4,\dots,8192$, $t_{w} \approx \SI{400}{\micro\second}$), and measuring along the $\hat{y}$ axis. We performed $6000$ rasters through these circuits, over approximately 8 hours. A representative subset of the power spectra for this data are shown in Fig.~\ref{fig:flowchart}C, as well as the $\alpha$-significance threshold for $\alpha=5\%$. The spectra for circuits containing long wait times exhibit power above the detection threshold, so instability was detected. These data are inconsistent with constant probabilities. Ramsey circuits are predominantly sensitive to phase accumulation, caused by detuning between the qubit and the control field frequencies, so it is reasonable to assume that it is this detuning that is drifting. The detected frequencies range from the lowest Fourier basis frequency for this experiment duration --- approximately $\SI{15}{\micro\hertz}$ --- up to approximately $\SI{250}{\micro\hertz}$.  The largest power is more than $1700$ standard deviations above the expected value under the null hypothesis, which is overwhelming evidence of temporal instability.

\begin{figure*}[t!]
\includegraphics[width=17.5cm]{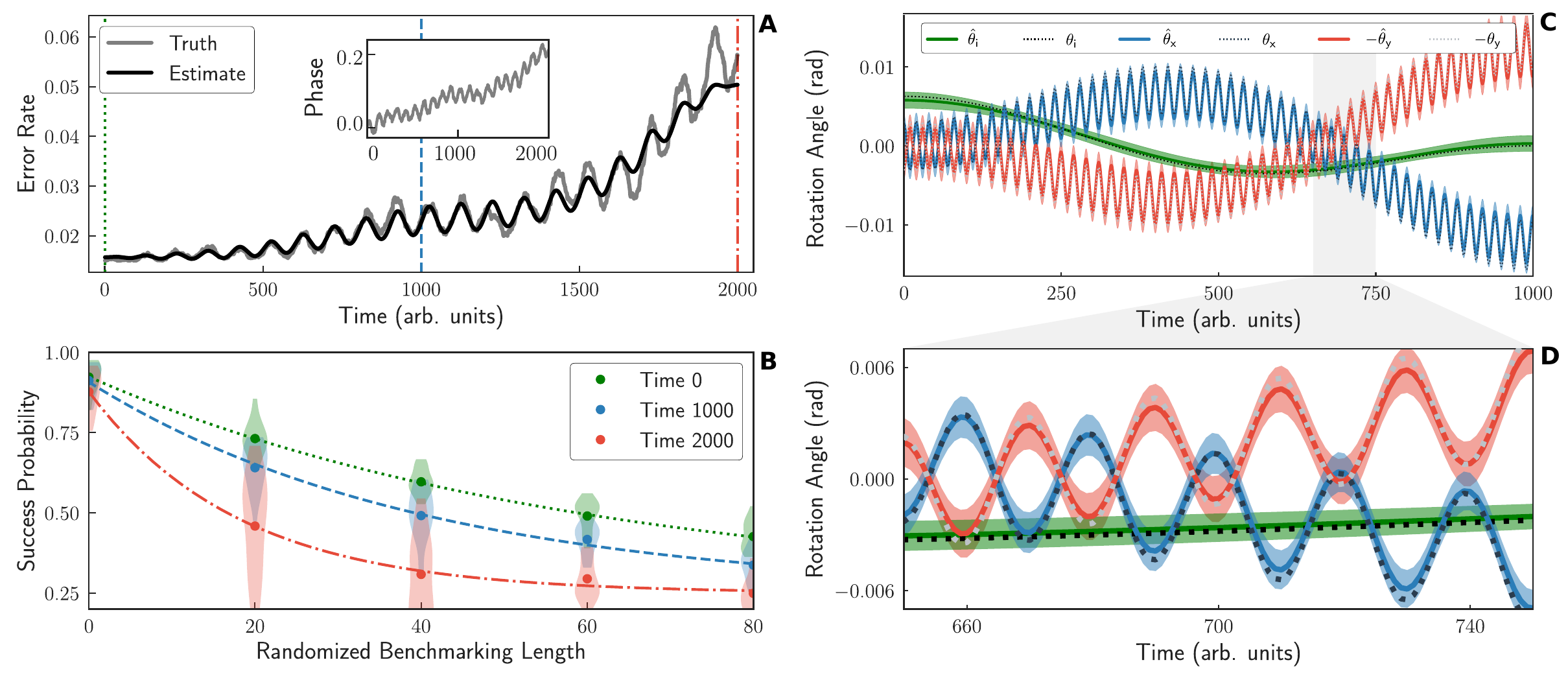}
\caption{{\bf Time-resolved benchmarking \& tomography on simulated data.}~{\bf A-B.} Time-resolved RB on simulated data for gates with time-dependent phase errors.~{\bf A.} Inset: the simulated phase error over time. Main plot: the true, time-dependent RB error rate ($r$) versus time (grey line) and a time-resolved estimate obtained by applying our techniques to simulated data (black line).~{\bf B.} Instantaneous average-over-circuits (points) and per-circuit (distributions) success probabilities at each circuit length, estimated by applying our spectral analysis techniques to the simulated time-series data, and fits to an exponential (curves), for the three times denoted by the vertical lines in A. Each instantaneous estimate of $r$, shown in A, is a rescaling of the decay rate of the exponential fit at that time.~{\bf C-D.} Time-resolved GST on simulated data, for three gates $\Gi$, $\Gx$ and $\Gy$ that are subject to time-dependent coherent errors around the $\hat{z}$, $\hat{x}$ and $\hat{y}$ axes, respectively, by angles $\theta_{\mathsf{i}}$, $\theta_{\mathsf{x}}$ and $\theta_{\mathsf{y}}$. The estimates of these rotation angles (denoted $\hat{\theta}_{\mathsf{i}}$, $\hat{\theta}_{\mathsf{x}}$ and $\hat{\theta}_{\mathsf{y}}$) track the true values closely. The shaded areas are $2\sigma$ ($\sim 95\%$) confidence regions.}
\label{fig:tr-rb}
\end{figure*}

\vspace{0.2cm}
\noindent
{\bf Quantifying instability.} Statistically significant evidence in data for time-varying probabilities does not directly imply anything about the scale of the detected instability. For instance, even the weakest periodic drift will be detected with enough data. We can quantify instability in any circuit by the size of the variations in its outcome probabilities. We can measure this size by estimating the probability trajectory $\vec{p}$ for each circuit (step 3, Fig.~\ref{fig:flowchart}A). As noted above, the unregularized best-fit estimate of $\vec{p}$ is the observed bit-string $\vec{x}$, which is overfitting. To regularize this estimate, we use model selection. Specifically, we select the time-resolved parameterized model $p(t) = \gamma_0 + \sum_{k}  \gamma_kf_k(t)$, where $f_k(t)$ is the $k^{\rm th}$ basis function of the Fourier transform, the summation is over those frequencies with power above the threshold in the power spectrum, and the $ \gamma_k$ are parameters constrained only so that each $p(t)$ is a valid probability. We can then fit this model to the clickstream for the corresponding circuit, using any standard data fitting routine, \eg, maximum likelihood estimation. 

Estimates of the time-resolved probabilities for the Ramsey experiment are shown in Fig.~\ref{fig:flowchart}D (unbroken lines). Probability traces are sufficient for heuristic reasoning about the type and size of the errors, and this is often adequate for practical debugging purposes. For example, these probability trajectories strongly suggest that the qubit detuning is slowly drifting. To draw more rigorous conclusions, we can implement time-resolved parameter estimation.

\vspace{0.2cm}
\noindent
{\bf Time-resolved benchmarking \& tomography.} The techniques presented so far provide a foundation for time-resolved parameter estimation, \eg, time-resolved estimation of gate error rates, rotation angles, or process matrices. We introduce two complementary approaches, which we refer to as ``non-intrusive'' and ``intrusive'', that can add time resolution to any benchmarking or tomography protocol. The non-intrusive approach is to replace counts data with instantaneous probability estimates in existing benchmarking/tomography analyses (step 4, Fig~\ref{fig:flowchart}A). It is non-intrusive because it doesn't require modifications to existing analysis codes. In constrast, the intrusive approach builds an explicitly time-resolved model and fits its parameters to the time-series data. We now detail and demonstrate these two techniques.

All standard characterization protocols, including all forms of tomography \cite{blume2016certifying,merkel2013self} and RB \cite{knill2008randomized, magesan2011scalable, proctor2018direct, magesan2012efficient, cross2016scalable, barends2014rolling, carignan2015characterizing, gambetta2012characterization}, are founded on some time-independent parameterized model that describes the outcome probabilities for the circuits in the experiment, or a coarse-graining of them (\eg, mean survival probabilities in RB). When analyzing data from these experiments, the counts data from these circuits are fed into an analysis tool that estimates the model parameters, which we denote $\{ \gamma_i\}$. To upgrade such a protocol using the non-intrusive method, we: (i) use the spectral analysis tools above to construct time-resolved estimates of the probabilities; (ii) for a given time, $t_j$, input the estimated probabilities directly into the analysis tool in place of frequencies; (iii) recover an estimate of the model parameters, $\{ \gamma_i(t_i)\}$ at that time; and (iv) repeat for all times of interest $\{t_j\}$. This non-intrusive approach is simple, but statistically \emph{ad hoc}. 

The intrusive approach permits statistical rigor at the cost of a more complex analysis. It consists of (i) selecting an appropriate time-resolved model for the protocol and (ii) fitting that model to the time-series data (steps 5a-5b, Fig.~\ref{fig:flowchart}A). In the model selection step, we expand each model parameter $ \gamma$ into a sum of Fourier components: $ \gamma \to   \gamma_0+ \sum_{\omega}  \gamma_{\omega} f_{\omega}(t)$, where the $ \gamma_{\omega}$ are real-valued amplitudes, and the summation is over some set  of non-zero frequencies. This set of frequencies can vary from one parameter to another, and may be empty if the parameter in question appears to be constant. To choose these expansions we need to understand how any drift frequencies in the model parameters would manifest in the circuit probability trajectories, and thus in the data. 

To demonstrate the intrusive approach, we return to the Ramsey experiment. In the absence of drift the probability of ``1'' in a Ramsey circuit with a wait time of $lt_{w}$ is $p_{l}= A + B \exp(- l/l_{0}) \sin(2\pi l t_w \Omega)$, where $\Omega$ is the detuning between the qubit and the control field, $\nicefrac{1}{l_0}$ is the rate of decoherence per idle, and $A, B \approx \nicefrac{1}{2}$ account for any state preparation and measurement errors. In our Ramsey experiment, the probability trace estimates shown in Fig.~\ref{fig:flowchart}C suggest that the state preparation, measurement and decoherence error rates are approximately time-independent, as the contrast is constant over time. So we define a time-resolved model that expands only  $\Omega$ into a time-dependent summation:
\begin{align}
\label{eq:tr-ramsey}
p_{l}(t) &= A + B \exp(- l/l_{0}) \sin(2\pi l t_{w} \Omega(t)),
\end{align}
where $\Omega(t) =  \gamma_0 + \sum_{\omega}  \gamma_{\omega} f_{\omega}(t)$. To select the set of frequencies in the summation, we observe that the dependence of the circuit probabilities on $\Omega$ is approximately linear for small $l$ (\eg, expand Eq.~\eqref{eq:tr-ramsey} around $lt_w\Omega(t)\approx 0$). Therefore, the oscillation frequencies in the model parameters necessarily appear in the circuit probabilities. So in our expansion of $\Omega$, we include all 13 frequencies detected in the circuit probabilities (\ie, the ones with power above the threshold in Fig.~\ref{fig:flowchart}C). The circuit probabilities will also contain sums, differences and harmonics of the frequencies in the true $\Omega$ --- Fig.~\ref{fig:flowchart}D shows clearly that the phase is wrapping around the Bloch sphere in the circuits with the longest wait times ($l \geq 2048$), so these harmonic contributions will be significant in our data. Therefore, this frequency selection strategy could result in erroneously including some of these harmonics in our model. We check for this using standard information-theoretic criteria \cite{akaike1974new}, and then discard any frequencies that should not be in the model (see Appendix~\ref{app:2}). This avoids over-fitting the data. Once the model is selected, we have a time-resolved parameterized model that we can directly fit to the time-series data. We do this with maximum likelihood estimation.

Fig.~\ref{fig:flowchart}E shows the estimated qubit detuning $\Omega(t)$ over time. It varies slowly between approximately $- \SI{0.5}{\hertz}$ and $\SI{0.5}{\hertz}$. The detuning is correlated with an ancillary measurement of the ambient laboratory temperature (the Spearman correlation coefficient magnitude is 0.92), which fluctuates by $\sim$1.5 C$^\circ$ over the course of the experiment. This suggests that temperature fluctuations are causing the drift in the qubit detuning (this conclusion is supported by further experiments: see later and the Methods). The detuning has been estimated to high precision, as highlighted by the $2\sigma$ confidence regions in Fig.~\ref{fig:flowchart}E. As with all standard confidence regions, these are in-model uncertainties, \ie, they do not account for any inadequacies in the model selection. However, we can confirm that the estimated detuning is reasonably consistent with the data by comparing the $p_l(t)$ predicted by the estimated model (dotted lines, Fig.~\ref{fig:flowchart}D) with the model-independent probability estimates obtained earlier (unbroken lines, Fig~\ref{fig:flowchart}D). These probabilities are in close agreement. 

\begin{figure*}[t!]
\includegraphics[width=18.cm]{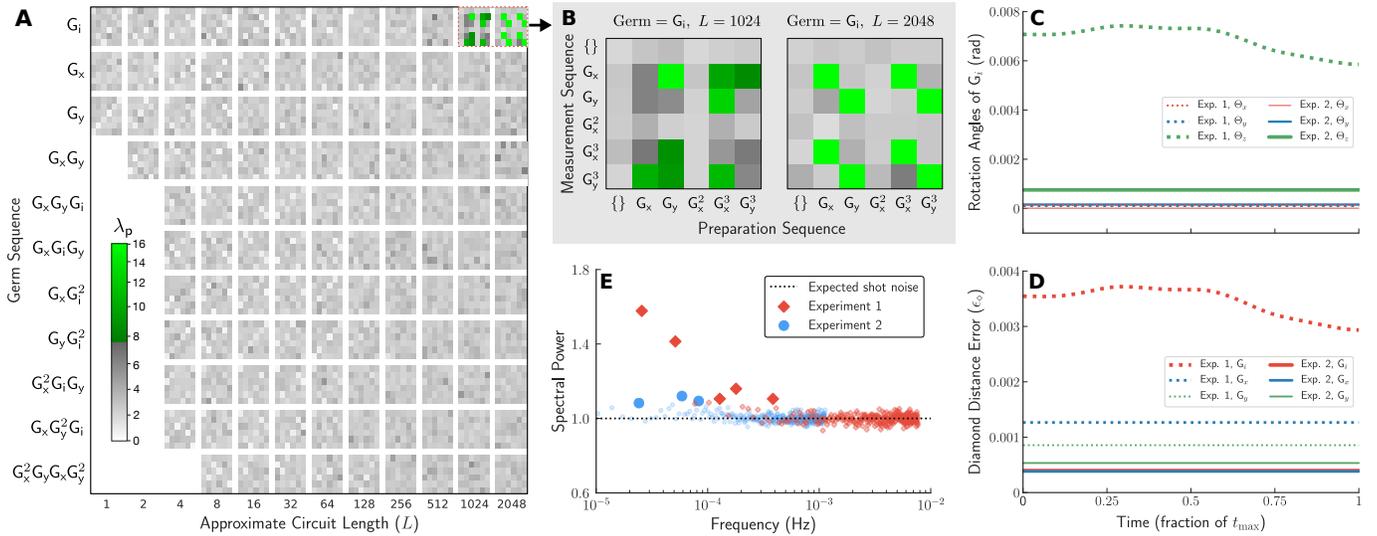}
\caption{{\bf Measuring qubit stability using time-resolved GST}. The results of two time-resolved GST experiments using the gates $\Gi$, $\Gx$ and $\Gy$, with drift compensation added for the second experiment.~{\bf A-B.} The evidence for instability in each circuit in the first experiment, quantified by $\lambda_{\mathsf{p}}  = -\log_{10}(\mathsf{p})$ where $\mathsf{p}$ is the p-value of the largest power in the spectrum for that circuit. A pixel is colored green when $\lambda_{\mathsf{p}}$ is large enough to be 5\% statistically significant, otherwise it is greyscale. Each circuit consists of repeating a ``germ'' sequence in between six initialization and pre-measurement sequences. The data is arranged by germ and approximate circuit length $L$, and then separated into the $6\times 6$ different preparation and measurement sequence pairs, as shown on the axes of B (``$\{\}$'' denotes the null sequence). Only long circuits containing repeated applications of $\Gi$ exhibit evidence of drift. In the second experiment none of the $\lambda_{\mathsf{p}}$ are statistically significant (data not shown).~{\bf C-D.} Time-resolved tomographic reconstructions of the gates in each experiment, summarized by the diamond distance error of each gate, and the decomposition of the coherent errors in $\Gi$ into rotation angles around $\hat{x}$, $\hat{y}$ and $\hat{z}$ ($t_{\max} \approx 5.5$ hours and $t_{\max} \approx 40$ hours for the first and second experiment, respectively).~{\bf E.} The power spectrum for each experiment obtained by averaging the individual power spectra for the different circuits, with filled points denoting power above the 5\% significance thresholds (the thresholds are not shown).}
\label{fig:gst-exp}
\end{figure*}

\vspace{0.2cm}
\noindent
{\bf Demonstration on simulated data.}~RB \cite{knill2008randomized, magesan2011scalable, proctor2018direct, magesan2012efficient, cross2016scalable, barends2014rolling, carignan2015characterizing, gambetta2012characterization} and GST \cite{blume2016certifying,merkel2013self} are two of the most popular methods for characterizing a QIP. Both methods are robust to state preparation and measurement errors; RB is fast and simple, whereas GST provides detailed diagnostic information about the types of errors afflicting the QIP. We now demonstrate time-resolved RB and GST on simulated data, using the general methodology introduced above. The number of circuits and circuit repetitions in these simulated experiments are in line with standard practice for these techniques, so they demonstrate that our techniques can be applied to RB and GST without additional experimental effort. 

We simulated data from 2000 rasters through 100 randomly sampled RB circuits \cite{knill2008randomized, magesan2011scalable, proctor2018direct} on two qubits. The error model consisted of 1\% depolarization on each qubit and a time-dependent coherent $\hat{z}$-rotation that is shown in inset of Fig.~\ref{fig:tr-rb}A (see Appendix~\ref{app:2} for details). The general instability analysis was implemented on this simulated data, after converting the 4-outcome data to the standard ``success''/``fail'' format of RB. This analysis yielded a time-dependent success probability for each circuit. Following our non-intrusive framework, instantaneous success probabilities at each time of interest were then fed into the standard RB data analysis (fitting an exponential) as shown for three times in Fig.~\ref{fig:tr-rb}B. The instantaneous RB error rate estimate is then (up to a constant \cite{proctor2018direct}) the decay rate of the fitted exponential at that time. The resultant time-resolved RB error rate is shown in Fig.~\ref{fig:tr-rb}A. It closely tracks the true error rate. 

GST is a method for high-precision tomographic reconstruction of a set of time-independent gates, state preparations and measurements \cite{blume2016certifying,merkel2013self}. We consider GST on a gate set comprised of standard $\hat{z}$-axis preparation and measurement, and three gates $\Gx$, $\Gy$, and $\Gi$. Here $\G_{\mathsf{x/y}}$ are $\nicefrac{\pi}{2}$ rotations around the $\hat{x}/\hat{y}$ axes and $\Gi$ is the idle gate. The GST circuits have the form $\mathsf{S}_{\text{prep}}\mathsf{S}_{\text{germ}}^{k}\mathsf{S}_{\text{meas}}$ (circuits are written in operation order where the leftmost operation occurs first). In this circuit: $\mathsf{S}_{\text{prep}}$ and $\mathsf{S}_{\text{meas}}$ are each one of six short sequences chosen to generate tomographically complete state preparations and measurements; $\mathsf{S}_{\text{germ}}$ is one of twelve short ``germ'' sequences, chosen so that powers (repetitions) of these germs amplify all coherent, stochastic and amplitude-damping errors; $k$ runs over an approximately logarithmically spaced set of integers, given by $k = \lfloor L/|\mathsf{S}_{\text{germ}}| \rfloor$ where $|\mathsf{S}_{\text{germ}}|$ is the length of the germ and $L =2^0,2^1,2^2,\dots,L_{\max}$ for some maximum germ power $L_{\max}$. 

We simulated data from 1000 rasters through these GST circuits (with $L_{\max} = 128$). The error model consisted of 0.1\% depolarization on each gate. Additionally, $\Gx$ and $\Gy$ are subject to over/under-rotation errors that oscillate both quickly and slowly, while $\Gi$ is subject to slowly varying $\hat{z}$-axis coherent errors. We used our intrusive approach to time-resolved tomography: the general instability analysis was implemented on this simulated data, the results were used to select a time-resolved model for the gates, and this model was then fit to the time-series data using maximum likelihood estimation (see Appendix~\ref{app:2} for details). The resulting time-resolved estimates of the gate rotation angles are shown in Fig.~\ref{fig:tr-rb}C-D. The estimates closely track the true values. 

\vspace{0.2cm}
\noindent
{\bf Demonstration on experimental data.}~Having verified that our methods are compatible with data from GST circuits, we now demonstrate time-resolved GST on two sets of experimental data, using the three gates $\Gx$, $\Gy$ and  $\Gi$. These experiments comprehensively quantify the stability of our $^{171}$Yb$^{+}$ qubit, because the GST circuits are tomographically complete and they amplify all standard types of error in the gates. The $\Gx$ and $\Gy$ gates were implemented with BB1 compensated pulses \cite{wimperis1994broadband,merrill2012progress}, and $\Gi$ was implemented with a dynamical decoupling $X_{\pi}Y_{\pi}X_{\pi}Y_{\pi}$ sequence \cite{khodjasteh2009dynamical}, where $X_{\pi}$ and $Y_{\pi}$ represent $\pi$ pulses around the $\hat{x}$ and $\hat{y}$ axes. The first round of data collection included the GST circuits to a maximum germ power of $L_{\text{max}} = 2048$ (resulting in 3889 circuits). These circuits were rastered  300 times over approximately $5.5$ hours. 

Fig.~\ref{fig:gst-exp}A-B summarizes the results of our general instability assessment on this data, using a representation that is tailored to GST circuits. Each pixel in this plot corresponds to a single circuit, and summarizes the evidence for instability by $\lambda_{\mathsf{p}}  = -\log_{10}(\mathsf{p})$ where $\mathsf{p}$ is the p-value of the largest power in the spectrum for that circuit ($\lambda_{\mathsf{p}}$  is 5\% significant when it is above the multi-test adjusted threshold $\lambda_{\mathsf{p},\text{threshold}} \approx 7$). The only circuits that displayed detectable instability are those that contain many sequential applications of $\Gi$. Fig.~\ref{fig:gst-exp}B further narrows this down to generalized Ramsey circuits, whereby the qubit is prepared on the equator of the Bloch sphere, active idle gates are applied, and then the qubit is measured on the equator of the Bloch sphere. These circuits amplify erroneous $\hat{z}$-axis rotations in $\Gi$. Other GST circuits amplify all other errors, but none of those circuits exhibit detectable drift. This is conclusive evidence that the angle of these $\hat{z}$-axis rotations is varying over the course of the experiment. 

The instability in $\Gi$ can be quantified by implementing time-resolved GST, with the $\hat{z}$-axis error in $\Gi$ expanded into a summation of Fourier coefficients (see Appendix~\ref{app:2} for details). The results are summarized in Fig.~\ref{fig:gst-exp}C-D (dotted lines). Fig.~\ref{fig:gst-exp}D shows the diamond distance error rate ($\epsilon_{\diamond}$) \cite{aharonov1998quantum} in the three gates over time. It shows that $\Gi$ is the worst performing gate, and that the error rate of $\Gi$ drifts substantially over the course of the experiment ($\epsilon_{\diamond}$ varies by $\sim$$25\%$). The gate infidelities are an order of magnitude smaller (see Table~\ref{table:error-rates} in Appendix~\ref{app:2}). Fig.~\ref{fig:gst-exp}C shows the coherent component of the $\Gi$ gate over time, resolved into rotation angles $\theta_{x}$, $\theta_{y}$ and $\theta_{z}$ around the three Bloch sphere axes $\hat{x}$, $\hat{y}$ and $\hat{z}$. The varying $\hat{z}$-axis component is the dominant source of error. 

This first round of experiments revealed instability, so we changed the experimental setup. Changes included the addition of periodic recalibration of the microwave drive frequency, the $\pi$-pulse duration, and the pointing of the detection laser (details in the Methods). We then repeated this GST experiment. To increase sensitivity to any instability, we collected more data, over a longer time period, and we included longer circuits. We ran the GST circuits out to a maximum germ power of $L_{\text{max}} = 16384$, rastering $328$ times through this set of 5041 circuits over approximately 40 hours. The purpose of running such a comprehensive experiment was to maximize sensitivity --- our methods need much fewer experimental resources for useful results (see below). Repeating the above analysis on this data, we found that none of the $\lambda_{\mathsf{p}}$ were statistically significant, \ie, no instability was detected in any circuit, including circuits containing over $10^5$ sequential $\Gi$ gates. Again, we performed time-resolved GST. Since no time dependence was detected, this reduces to standard time-independent GST. The results are summarized in Fig.~\ref{fig:gst-exp}C-D (unbroken lines). The gate error rates have been substantially suppressed ($\epsilon_{\diamond}$ decreased by $\sim$$10 \times$ for $\Gi$), and the $\hat{z}$-axis coherent error in $\Gi$ reduced and stabilized. This is a comprehensive demonstration that the recalibrations are stabilizing the qubit. Furthermore, the recalibrated parameters versus time are strongly correlated with ambient laboratory temperature (see the Methods), suggesting temperature stabilization as an alternative route to qubit stabilization, and supporting the conclusions of our Ramsey experiments.

No individual circuit exhibited signs of drift in this second GST experiment, but we can also perform a collective test for instability on the clickstreams from all the circuits. In particular, we can average the per-circuit power spectra, and look for statistically significant peaks in this single spectrum. This suppresses the shot noise inherent in each individual clickstream, so it can reveal low-power drift that would otherwise be hidden in the noise (see Appendix~\ref{app:1}). This average spectrum is shown for both experiments in Fig.~\ref{fig:gst-exp}E. The power at low frequencies decreases substantially from the first to the second experiment, further demonstrating that our drift compensation is stabilizing the qubit. However, there is power above the $5\%$ significance threshold for both experiments. So there is still some residual instability after the experimental improvements. But this residual drift is no longer a significant source of errors, as demonstrated by the low and stable error rates shown in Fig.~\ref{fig:gst-exp}D. 

\vspace{0.2cm}
\noindent
{\bf Experiment design.}~Our method is an efficient way to identify time dependence in the outcome probability distribution of any quantum circuit. In its most basic application, it can verify the stability of application or benchmarking data. No special-purpose circuits are required, as the drift detection can be applied to data that is already being taken. The analysis will then be sensitive to any drifting errors that impact this application, in proportion to their effect on the application.

As we have demonstrated, our method can also be used to create dedicated drift characterization protocols. This mode requires a carefully chosen set of quantum circuits that are sensitive to the specific parameters under study. Without \emph{a priori} knowledge about what may be drifting, this circuit set should be sensitive to all of the parameters of a gate set. The GST circuits are a good choice. However, if only a few parameters are expected to drift, a smaller set of circuits sensitive only to these parameters  can be used, resulting in a more efficient experiment. For example, Ramsey circuits serve as excellent probes of time variation in qubit phase rotation rates. Many of the most sensitive circuits, such as those used in GST, Ramsey spectroscopy, and robust phase estimation \cite{kimmel2015robust}, are periodic and extensible. These circuits achieve $\mathcal{O}(1/L)$ precision scaling, with $L$ the maximum circuit length, up until decoherence dominates. So, by choosing a suitably large $L$, very high-precision drift tracking can be achieved, as in our experiments. 

Interleaving dedicated drift characterization circuits with application circuits combines the two use cases for our methods --- dedicated drift characterization and auxiliary analysis. This reduces the data acquisition rate for both the application and characterization circuits, but it directly probes whether time variation in a parametric model is correlated with drift in the outcomes of an application circuit. While this reduces sensitivity to high-frequency instabilities, much of the drift seen in the laboratory is on timescales that are long compared to the data acquisition rate. As a simple demonstration of this, we note that discarding 80\% of our Ramsey data --- keeping only every fifth bit for each circuit --- still yields a high-precision time-resolved phase estimate, as shown in Fig.~\ref{fig:flowchart}E (gray dashed line). 

The sensitivity of our analysis depends on both the number of times a circuit is repeated ($N$) and the sampling rate ($t_{\rm gap}$). As in all signal analysis techniques, the sampling rate sets the Nyquist limit --- the highest frequency the analysis is sensitive to without aliasing --- while $(N-1)t_{\rm gap}$ sets the lowest frequency drift that will be visible. While the sensitivity of our methods increases with more data, statistically significant results can be achieved without dedicating hours or days to data collection. For example, both the simulated GST and RB experiments (see Fig.~\ref{fig:tr-rb}) used a number of circuits and repetitions consistent with standard practices. Further details relating the sampling parameters and the analysis sensitivity are provided in Appendix~\ref{app:3}.

\vspace{-0.2cm}
\section*{\normalsize Discussion}
Reliable quantum computation demands stable hardware. But current standards for characterizing QIPs assume stability --- they cannot verify that a QIP is stable,  nor can they quantify any instabilities. This is becoming a critical concern as stable sources of errors are steadily reduced. For example, drift significantly impacted the recent tomographic experiments of Wan \emph{et al.}~\cite{wan2019quantum} but this was only verified using a complex analysis. In this article we have introduced a general, flexible, and powerful methodology for diagnosing instabilities in a QIP. We have applied these methods to a trapped-ion qubit, demonstrating both time-resolved phase estimation and time-resolved tomographic reconstructions of logic gates. Using these tools, we were able to identify the most unstable gate, confirm that periodic recalibration stabilized the qubit to an extent that drift is no longer a significant source of error, and isolate the probable source of the instabilities (temperature changes).

Our methods are widely-applicable, platform-independent, and do not require special-purpose experiments. This is because the core techniques are applicable to data from any set of quantum circuits --- as long as it is recorded as a time series --- and the data analysis is fast and simple (speed is limited only by the fast Fourier transform). These techniques  enable routine stability analysis on data gathered primarily for other purposes, such as data from algorithm, benchmarking or error correction circuits. These techniques are even applicable outside of the context of quantum computing --- they could be used for time-resolved quantum sensing. We have incorporated these tools into an open-source software package~\cite{pygstiversion0.9.9.1, nielsen2020probing}, making it easy to check any time-series QIP data for signs of instability. Because of the disastrous impact of drift on characterization protocols \cite{dehollain2016optimization, epstein2014investigating, van2013quantum, fong2017randomized, fogarty2015nonexponential, chow2009randomized, wan2019quantum}, its largely unknown impact on QIP applications, and the minimal overhead required to implement our methods, we hope to see this analysis broadly and quickly adopted. 

\vspace{-0.1cm}
\section*{\normalsize Methods}
\begin{small}
\noindent
{\bf Experiment details.} We trap a single $^{171}$Yb$^{+}$ ion $\sim\SI{34}{\micro\meter}$ above a Sandia multi-layer surface ion trap with integrated microwave antennae, shown in Fig.~\ref{fig:ion-trap}. The radial trapping potential is formed with $\SI{170}{\volt}$ of rf-drive at $\SI{88}{\mega\hertz}$; the axial field is generated by up to $\SI{2}{\volt}$ on the segmented dc control electrodes. This yields secular trap frequencies of $\SI{0.7}{\mega\hertz}$, $\SI{5}{\mega\hertz}$, and $\SI{5.5}{\mega\hertz}$ for the axial and radial modes, respectively. An electromagnetic coil aligned with its axis perpendicular to the trap surface creates the quantization field of approximately $5\,\textrm{G}$ at the ion. The field magnitude is calibrated using the qubit transition frequency, which has a second-order dependence on the magnetic field of $f = \SI{12.642812118}{\giga\hertz} + 310.8B^2\SI{}{\hertz}$, where $B$ is the externally applied magnetic field in Gauss \cite{fisk1997accurate}. The qubit is encoded in the hyperfine clock states of the $^2\textrm{S}_{1/2}$ ground state of $^{171}$Yb$^{+}$, with logical 0 and 1 defined as $\ket{F = 0, m_F =0}$ and $\ket{F = 1, m_F =0}$, respectively.

Each run of a quantum circuit consists of four steps: cooling the ion, preparing the input state, performing the gates, and then measuring the ion. First, using an adaptive length Doppler cooling scheme, we verify the presence of the ion. The ion is Doppler cooled for $\SI{1}{\milli\second}$, during which fluorescence events are counted. If the number of detected photons is above a threshold ($\sim 85\%$ of the average fluorescence observed for a cooled ion) Doppler cooling is complete, otherwise, the cooling is repeated. If the threshold is not reached after 300 repetitions, the experiment is halted to load a new ion. This ensures that an ion is present in the trap and that it is approximately the same temperature for each run. After cooling and verifying the presence of the ion, it is prepared in the $\ket{F = 0, m_F =0}$ ground state using an optical pumping pulse \cite{olmschenk2007manipulation}. All active gates are implemented by directly driving the $\SI{12.6428}{\giga\hertz}$ hyperfine qubit transition, using a near-field antenna integrated into the trap (see Fig.~\ref{fig:ion-trap}). The methods used for generating the microwave radiation are discussed in Ref.~\cite{blume2016certifying}. A standard state fluorescence technique \cite{olmschenk2007manipulation} is used to measure the final state of the qubit.

The gates we use are $\Gx$, $\Gy$ and $\Gi$, which are $\nicefrac{\pi}{2}$ rotations around the $\hat{x}$- and $\hat{y}$-axes, and an idle gate. The $\Gx$ and $\Gy$ gates, used in both the Ramsey and GST experiments, are implemented using BB1 pulse sequences \cite{wimperis1994broadband,merrill2012progress}. The $\Gi$ gate, used only in the GST experiments, is a second-order compensation sequence: $\Gi = X_{\pi}Y_{\pi}X_{\pi}Y_{\pi}$, where $X_{\pi}$ and $Y_{\pi}$ denote $\pi$ pulses about the $\hat{x}$- and $\hat{y}$-axis, respectively \cite{khodjasteh2009dynamical}. To maintain a constant power on the microwave amplifier and reduce the errors from finite on/off times, active gates are performed gapless, \ie, we transition from one pulse to the next by adjusting the phase of the microwave signal without changing the amplitude of the microwave radiation.  In the first GST experiment the Rabi frequency was $\SI{119}{\kilo\hertz}$.

\begin{figure}[t!]
\includegraphics[width=\columnwidth]{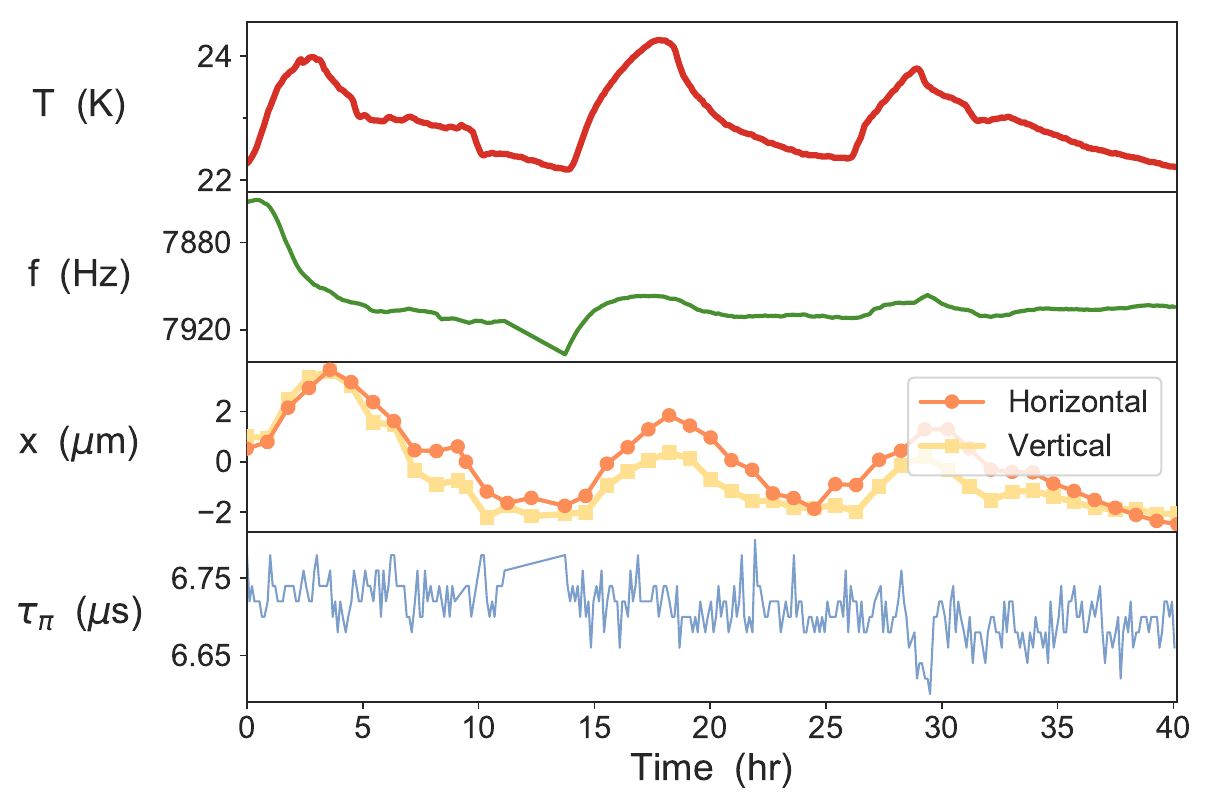}
\caption{{\bf Ancillary measurements in the stabilized GST experiment}. The microwave drive frequency $f$, the horizontal and vertical detection beam offsets $x$, and microwave $\pi$-pulse time $\tau_\pi$ were periodically recalibrated and tracked over the course of the experiment. The Spearman correlation coefficient ($\rho$) confirms that the ambient temperature $T$ is well correlated with the drive frequency ($\rho=0.64$) and the horizontal and vertical beam offsets ($\rho=0.86$ and $\rho=0.84$), but it is not well correlated with the $\pi$-pulse time ($\rho$=0.05). }
\label{fig:ancillary}
\end{figure}

 Changing the phase in the analog output signal takes $\sim \SI{5}{\nano\second}$ and this causes errors, because the pulse sequences are performed gapless. These errors are larger for shorter $\pi$-pulse times. To reduce this error, in this second GST experiment the Rabi frequency was decreased to $\SI{74}{\kilo\hertz}$. To compensate for the drift that we observed in both the Ramsey experiment and the first GST experiment, in the second GST experiment we incorporated three forms of active drift control. The detection laser position was recalibrated every 45 minutes, and both the $\pi$-time ($\tau_{\pi}$) and the microwave drive frequency ($f$) were updated based on the results of interleaved calibration circuits. After every $4^{\rm th}$ circuit, a circuit consisting of a $10.5\pi$ pulse was performed. If the outcome was 0 (resp., 1) then $\SI{1.25}{\nano\second}$ was added to $\tau_{\pi}$ (resp., subtracted from $\tau_{\pi}$). The applied $\pi$-time is $\tau_{\pi}$ rounded to an integer multiple of $\SI{20}{\nano\second}$, so only consistently bright or dark measurements result in changes of the pulse time. After every $16^{\rm th}$ circuit a $\SI{10}{\milli\second}$ wait Ramsey circuit was performed. If the outcome was 0 (resp., 1)  $\SI{10}{\milli\hertz}$ is added to $f$ (resp., subtracted from $f$).

Fig.~\ref{fig:ancillary} shows the detection beam position, $\tau_{\pi}$, $f$, and ambient laboratory temperature over the course of the second GST experiment. The calibrated $f$ is correlated with the ambient temperature. This is consistent with the observed correlation between the ambient temperature and the estimated detuning in the Ramsey experiment (see Fig.~\ref{fig:flowchart}E). The temperature is also strongly correlated with the calibrated detection beam location points, suggesting that thermal expansion is a plausible underlying cause of the frequency shift. 

\vspace{0.2cm}
\noindent
{\bf Data analysis details.} To generate a power spectrum from a clickstream we use the Type-II discrete cosine transform with an orthogonal normalization. This is the matrix $F$ with elements
\begin{align}
F_{\omega i} =\sqrt{\frac{2^{1-\delta_{\omega,0}}}{N}} \cos\left(\frac{\omega \pi}{N}\left( i + \frac{1}{2}\right)\right), 
\label{eq:dct}
\end{align}
where $\omega,i=0,\dots,N-1$ \cite{ahmed1974discrete}. However, note that the exact transform used is not important: we only require that $F$ is an orthogonal and Fourier-like matrix (see Appendix~\ref{app:1}). Our hypothesis testing is all at a statistical significance of 5\%, and uses a Bonferroni correction to maintain this significance when implementing many hypothesis tests (see Appendix~\ref{app:1}). All data fitting uses maximum likelihood estimation, except for the $p(t)$ estimation in the time-resolved RB simulations. In that case, we use a simple form of signal filtering (see Appendix~\ref{app:1}), so that the entire analysis chain maintains the speed and simplicity inherent to RB. When choosing between multiple time-resolved models, as in the time-resolved Ramsey tomography and GST analyses, we use the Akaike information criteria \cite{akaike1974new} to avoid overfitting (see Appendix~\ref{app:2}). Further details on these methods, and a supporting theory, is provided in Appendices~\ref{app:1}-\ref{app:3}.

\section*{Data availability} 
All experimental and simulated data presented in this paper are available at http://doi.org/10.5281/zenodo.4033077.
\section*{Code availability} 
Code for implementing the general drift charactierization methods introduced in this paper has been incorporated into the open-source Python package \texttt{pyGSTi} \cite{pygstiversion0.9.9.1, nielsen2020probing}. The \texttt{pyGSTi}-based Python scripts and notebooks used for the data analysis reported in this paper are available at http://doi.org/10.5281/zenodo.4033077.

\section*{Acknowledgments}
This paper describes objective technical results and analysis. Any subjective views or opinions that might be expressed in the paper do not necessarily represent the views of the U.S. Department of Energy or the United States Government. This work was supported by the U.S. Department of Energy, Office of Science, Office of Advanced Scientific Computing Research Quantum Testbed Program; the Office of the Director of National Intelligence (ODNI), Intelligence Advanced Research Projects Activity (IARPA); and the Laboratory Directed Research and Development program at Sandia National Laboratories. Sandia National Laboratories is a multi-program laboratory managed and operated by National Technology and Engineering Solutions of Sandia, LLC., a wholly owned subsidiary of Honeywell International, Inc., for the U.S. Department of Energy's National Nuclear Security Administration under contract DE-NA-0003525. All statements of fact, opinion or conclusions contained herein are those of the authors and should not be construed as representing the official views or policies of IARPA, the ODNI, the U.S. Department of Energy, or the U.S. Government.

\end{small}

\appendix
\section{Drift characterization using time-series data}\label{app:1}
Here we provide supporting theory and information for the methods that we propose for detecting and characterizing instability in the outcome distribution of arbitrary quantum circuits. This is the methods summarized in steps 1-3 of the flowchart in Fig.~\ref{fig:flowchart}.
\subsection{Power spectra}
To begin, we detail how a clickstream is transformed into the frequency domain and then converted into a power spectrum (step 1 in the flowchart of Fig.~\ref{fig:flowchart}A). To transform into the frequency domain we apply a Fourier-like transform $F$ to rescaled data. All of the data analysis in this article uses the Type-II discrete cosine transform (DCT), with an orthogonal normalization. It is define by the matrix $F^{(\textsc{dct})}$ with the elements
\begin{align}
F^{(\textsc{dct})}_{\omega i} =\sqrt{\frac{2^{1-\delta_{\omega,0}}}{N}} \cos\left(\frac{\omega \pi}{N}\left( i + \frac{1}{2}\right)\right), 
\label{eq:dct}
\end{align}
where $\omega,i=0,\dots,N-1$ \cite{rao2014discrete,ahmed1974discrete}. The Type-II DCT typically results in a sparser representation of signals with differing values at the boundaries than many of the alternative Fourier-like transforms \cite{rao2014discrete,ahmed1974discrete}. Functions of this sort are typical in our problem, and so this is our motivation for using this transform.

\begin{figure}[t!]
\includegraphics[width=8cm]{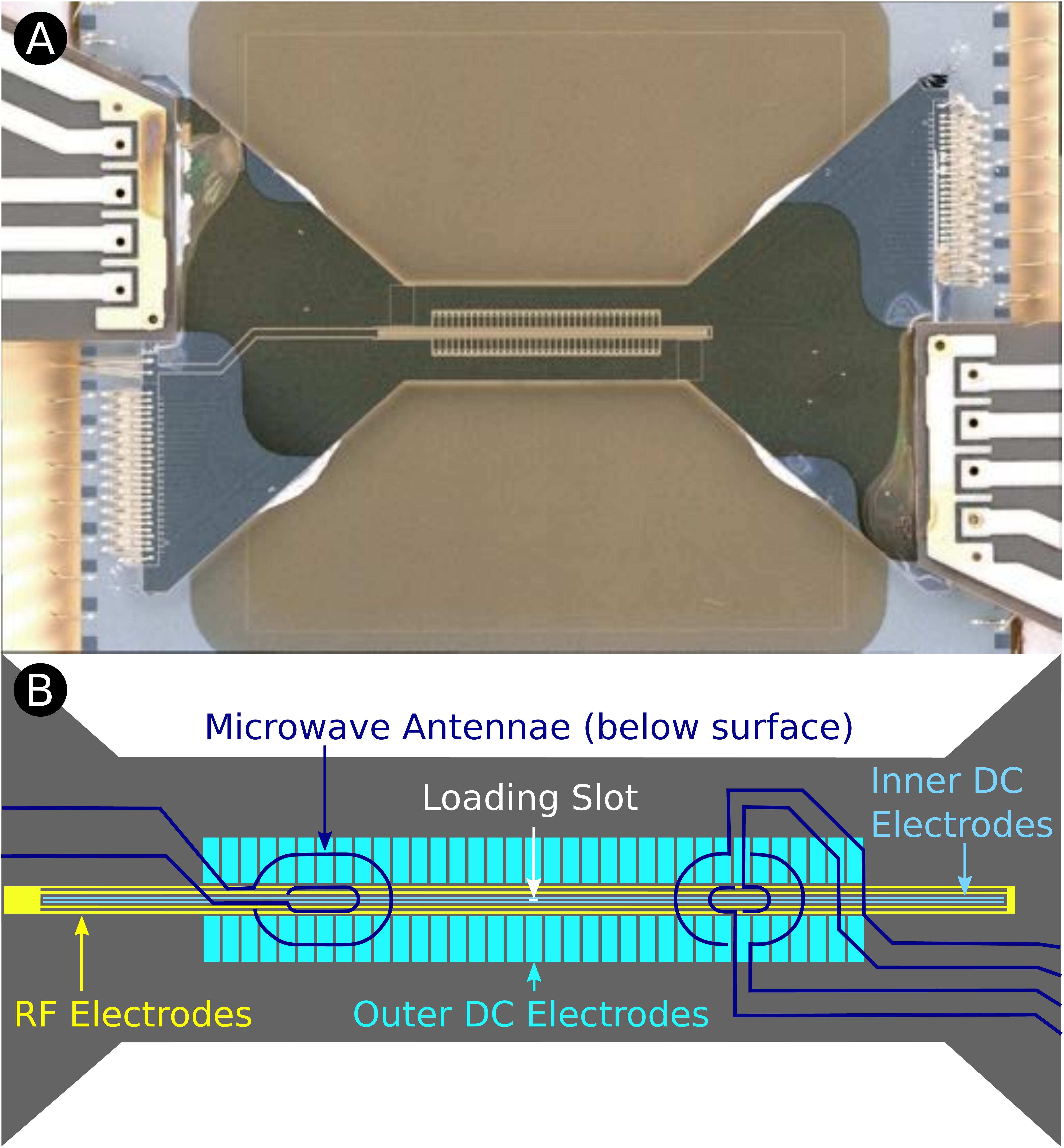}
\caption{{\bf The Sandia ion trap}. A $^{171}$Yb$^{+}$ ion is loaded in the center of the trap, by photo-ionizing neutral ytterbium vapor introduced through the loading slot in the device. It is then shuttled to the center of the microwave antenna on the left. {\bf A.} Flex cable microwave waveguides (upper left and lower right) are solder-die attached to the trap die to circumvent the trap package. This limits unintended impedance changes and losses, and optimizes the microwave power reaching the antenna. {\bf B.} A schematic of the trap showing the antennae locations with respect to the dc control and rf electrodes. The leads for the microwave antennae attach directly to the flex cable waveguides. The antennae are located on the lowest metal layer underneath the trap electrodes. While the generated magnetic field is attenuated by the metal layers above it, the gaps between electrodes make this design viable and the magnetic field above the trap is about half of the field strength generated if the antennae would not be covered by trap electrodes.  In the Ramsey experiment the ion was situated at the center of the trap, above the loading slot. In both GST experiments the ion was located above the center of the left microwave antenna.}
\label{fig:ion-trap}
\end{figure}

Although all our data analysis uses the Type-II DCT, our methods and our theory (except where stated) are all directly applicable for any transform matrix $F$ that  satisfies the following criteria: (i) $F$ is orthogonal, (ii) every matrix element of $F$ satisfies $F_{\omega i}^2 \leq b_{F}/N$ for a small constant $b_{F}$, and (iii) the top row of $F$ is proportional to the vector of all ones. This includes, \eg, sine transforms, other cosine transforms, and the Walsh-Hadamard transform \cite{feldman1987fast,rao2014discrete,ahmed1974discrete} (all with the appropriate normalizations). Note that the standard discrete Fourier transform does not satisfy these requirements (with an appropriate normalization it is unitary, but not orthogonal). However, the methods described here are compatible with the discrete Fourier transform as long as some minor adjustments are made to guarantee that, \eg, the probability trace estimates are real. Throughout the following theory, we refer to the indices of the $F$-domain vector as frequencies, but note that (i) when we report frequencies for experimental data these indices have been converted to Hertz, and (ii) for a generic transformation an index should only be interpreted as indicating the corresponding basis function of $F$, and it may or may not constitute a frequency.

As discussed in the main text, the motivation for using spectral analysis is that many physically plausible $p(t)$ are sparse in the frequency domain. But note that the relevant quantity to the data analysis is the discrete-time probability trajectory $\vec{p}$, where $p_i = p(t_i)$ and $t_i$ is the $i^{\rm th}$ data collection time for the circuit in question. Moreover, as we are concerned with deviations from time-independence, it is the ``signal'' vector 
\begin{equation}
\vec{s} = \vec{p} - \bar{p} \vec{1},
\end{equation}
that is of most relevance, where $\bar{v}$ denotes the mean of the vector $\vec{v}$ and $\vec{1} = (1,1,\dots, 1)^{\mathsf{T}}$. It is sparsity of $\tilde{\vec{s}}$ that is important for our data analysis. Sparsity in the frequency domain of the continuous-time probability $p(t)$ implies that $\tilde{\vec{s}}=F\vec{s}$ is a sparse vector if: (i) a sensible choice is made for the transformation $F$, such as the Type-II DCT, (ii) the number of sample times $N$ is sufficiently high, and (iii) the sample times $t_{0},t_1,t_2,\dots,t_{N-1}$ for the circuit in question are spaced reasonably uniformly. 

All the experiments and simulations herein consist of rastering through a set of circuits --- \ie, running each circuit once, and then looping through this until each circuit has been run $N$  times. So the time between sequential runs of a circuit is exactly constant (and the same for all the circuits) in an idealized experiment. Violations of this ideal are unavoidable, but if they are small (as they are in our experiments) they only mildly degrade the effectiveness of our techniques. This can be understood by noting that small perturbations on equally-spaced times (or a small number of erroneously long time-steps of any size) will only slightly reduce the sparsity of $\tilde{\vec{s}}$, for any Fourier-sparse $p(t)$. Large deviations from the equally-spaced-times ideal can substantial degrade the sensitivity of the methods we presented in the main text (although this will not cause instability to be spuriously detected). In this case, we can instead generate power spectra using a technique that explicitly takes into account the sample times. One such option is the floating-mean Lomb-Scargle periodogram \cite{zechmeister2009generalised,vanderplas2018understanding}, and all our methods can be adapted to this spectral analysis technique. As there are technical differences in this case, and this method is not needed to analyze our experimental data, we do not include the details here.

Before transforming a clickstream ($\vec{x}$) to the frequency domain, we first subtract its mean and divide by its variance. This simplifies the statistics of the power spectrum, but the spectrum is not well-defined if the clickstream consists of only zeros or only ones (because then the variance of $\vec{x}$ is zero). So we define the frequency-domain amplitudes by:
\begin{equation}
\tilde{\vec z} = \begin{cases}  \frac{F(\vec{x} - \bar{x}\vec{1})}{\sqrt{\bar{x}(1- \bar{x})}} \quad &\text{if}\,\,\, 0<\bar{x} <1, \\ (0,1,1,\dots)^{\mathsf{T}} \quad& \text{else}.\end{cases}
\label{eq:def:z}
\end{equation}
The power at frequency $\omega$ is then given by $|\tilde{z}_{\omega}|^2$. This convention for the constant clickstream case is convenient, particularly when averaging the power spectra from multiple clickstreams, as there are then fewer situations in which we need to explicitly take this special case into account. Moreover, $(0,1,1,\dots)$ roughly corresponds to the $N\to \infty$ limit of $\tilde{\vec z}$ for a clickstream containing a single 1 or 0 (the correspondence is not precise, because the spectrum of a delta function is not completely flat for a general transform $F$). 

\subsection{Statistical hypothesis testing} 
All of our methods are built upon statistical hypothesis testing \cite{lehmann2006testing,shaffer1995multiple}. We now review the relevant aspects of this field and formulate our hypothesis testing problem. Consider some set of random variables $\{A_1,A_2,\dots\}$ with a joint distribution that is parameterized by $\theta \in \mathcal{H}$ for some space $\mathcal{H}$. A statistical hypothesis is the conjecture that $\theta \in \mathcal{H}_0 \subset \mathcal{H}$. In this article, data consists of a clickstream of $N$ bits for each of $C\geq 1$ circuits. Generalizing the notation used in the main text so that we explicitly denote the circuit index, let $x_{c,i}$ be the bit output by the $i^{\rm th}$ repetition of the $c^{\rm th}$ circuit, and denote the corresponding Bernoulli random variable from which that datum is drawn by $X_{c,i}$. The set of random variables about which we state statistical hypotheses is $\{X_{c,i}\}$. The random variable $X_{c,i}$ has some unknown probability $p_{c,i}$ of taking the value 1, so the set of random variables $\{X_{c,i}\}$ is parameterized by the space 
\begin{equation}
\mathcal{H}=\{p_{c,i}\in[0,1] \}.
\end{equation} 

Statistical hypothesis testing starts from some set of null hypotheses $\{\mathcal{H}_{0,i}\}$ and uses statistical hypothesis tests on data drawn from the random variables to attempt to reject one or more of these null hypotheses (and/or intersections of these null hypotheses). Our methods look for evidence that $|\tilde{p}_{\omega,c}| > 0$ at each non-zero frequency $\omega$ and for each circuit $c$. This is formalized by the set of null hypotheses $\{\mathcal{H}_{0,c,\omega}\}$ where $\mathcal{H}_{0,c,\omega}$ is the conjecture that $\tilde{p}_{c,\omega} = 0$, with $\omega=1,2,\dots,N-1$ and $c=1,2,\dots,C$. For our purposes, testing a null hypothesis consists of the following four steps: (i) pick a significance level $\alpha \in (0,1)$, with $\alpha = 5\%$ a common choice; (ii) select a test statistic $\Lambda$, which is a function from data to $\mathbb{R}$; (iii) find a threshold such that the probability of observing $\Lambda$ larger than this threshold if the null hypothesis is true is at most $\alpha$; (iv) collect data, evaluate $\Lambda$, and reject the null hypothesis if and only if $\Lambda$ is larger than the threshold. This procedure guarantees that the null hypothesis is falsely rejected with probability at most $\alpha$. Our test statistic for testing the null hypothesis $\mathcal{H}_{0,c,\omega}$ is simply the power of the normalized data at the frequency $\omega$, \ie, $|\tilde{z}_{c,\omega}|^2$.

Implementing multiple hypothesis tests that use the above procedure will typically cause the probability of falsely rejecting at least one null hypothesis --- the family-wise error rate (FWER) --- to increase quickly with the number of tests (for $T$ tests $\text{FWER} \leq T\alpha$). The standard approach when implementing multiple tests is to adapt the testing procedure to, at a minimum, maintain weak control of the FWER. Weak control of the FWER means that if all the null hypotheses are true then, for some pre-specified global significance $\alpha$, the probability of falsely rejecting one or more null hypotheses is at most $\alpha$. It is common to choose a testing procedure that also maintains strong control of the FWER. Strong control of the FWER means that the probability of falsely rejecting one or more true null hypotheses is at most $\alpha$ and this holds even if one or more of the null hypotheses are false. We seek to maintain strong control of the FWER, so that whenever we reject $\mathcal{H}_{0,c,\omega}$ we can be $(1-\alpha)$ confident that $\omega$ is a frequency with a non-zero amplitude in the Fourier decomposition of $\vec{p}_{c}$.

Strong control of the FWER at a global significance of $\alpha$ can be maintained by implementing the $i^{\rm th}$ of $T$ tests at a local significance of $\alpha w_i$ for any $w_i\geq0$ satisfying $\sum_i w_i = 1$. This is known as the (generalized) Bonferroni correction \cite{lehmann2006testing,shaffer1995multiple}. The Bonferroni correction can be conservative; there are other methods for maintaining strong control of the FWER that are more powerful than the Bonferroni correction, such as the Holms procedure \cite{lehmann2006testing,shaffer1995multiple}. Moreover, strong control of the FWER is not the only way to ensure ``confidence'' in the results --- \eg, we could instead choose to control the false discovery rate, which is the expected ratio of the number of falsely rejected null hypotheses to the total number of rejected null hypotheses \cite{benjamini1995controlling}. Variants of our techniques that control the false discovery rate and/or use a different multi-test correction procedure are possible, and essentially involve setting a slightly lower power significance threshold in the analysis.

\subsection{Statistics of the power spectra}
We now derive the distribution of the power spectrum of the data, both when $p(t)$ is and is not constant. We consider the frequency-domain data $\tilde{z}_{\omega}$, defined in Eq.~\eqref{eq:def:z}, for an arbitrary non-zero frequency $\omega$, and we derive an approximation to the distribution of the corresponding random variable,
\begin{equation}
\tilde{Z}_{\omega} =  \frac{\tilde{X}_{\omega}}{\sqrt{\bar{X}(1- \bar{X})}},
\end{equation}
where $\omega > 0$ (we are again dropping the circuit indexing, and we have ignored the edge-case of the constant clickstream instance in this equation). Note that $\bar{X} = \frac{1}{N}\sum_i X_i$ is the average of random variables, and so it is itself a random variable. In the main text we asserted that $\tilde{Z}^2_{\omega}$ is approximately $\chi_1^2$ distributed whenever the probability trajectory is a constant. This is a sub-case of what we derive below, but we also study the marginal distribution of $\tilde{Z}_{\omega}^2$ when $\tilde{p}_{\omega}=0$ but the probability trajectory is not necessarily constant (\ie, there is perhaps power at one or more other non-zero frequencies). We do so in order to be able to claim that we are maintaining strong control of the FWER. By itself, the fact that $\tilde{Z}_{\omega}^2$ is approximately $\chi_1^2$ distributed when the probability trajectory is constant can only be used to ensure weak control of the FWER.

The random variable $\tilde{Z}_{\omega}$ is rather complex to directly analyze, because it depends on both $\tilde{X}_{\omega}$ and $\bar{X}$. So we first make an approximation. As $\bar{X}$ is the mean of the $X_i$ we have that $\mathbb{E}[\bar{X}] = \bar{p}$ and $\mathbb{V}[\bar{X}] = O(\nicefrac{1}{N})$, where $\mathbb{E}(A)$ and $\mathbb{V}(A)$ denote the expectation value and variance of the random variable $A$, respectively. Therefore 
\begin{equation}
Z_i \approx Y_{i} = \frac{X_{i} - \bar{p}}{\sqrt{\bar{p}(1- \bar{p})}},
\end{equation}
where we assume that $0 <\bar{p} < 1$. So we now derive the distribution of $\tilde{Y}_{\omega}$, as an approximation to the distribution of $\tilde{Z}_{\omega}$.

The random variable $\tilde{Y}_{\omega}$ is the sum of $N$ independent random variables:
\begin{equation}
\tilde{Y}_{\omega}=\sum_i F_{\omega i}Y_i.
\end{equation}
A heuristic appeal to the central limit theorem therefore suggests that $\tilde{Y}_{\omega}$ is normally distributed. That is, $\tilde{Y}_{\omega}$ is approximately $\mathcal{N}(\mu_{\omega},\nu_{\omega})$ distributed for $N \gg 1$, with some mean $\mu_{\omega}$ and variance $\nu_{\omega}$. Later we formally apply the central limit theorem to $\tilde{Y}_{\omega}$, but first we derive formulae for $\mu_{\omega}$ and $\nu_{\omega}$. For any non-zero $\omega$, $\mu_{\omega}$ is simply a rescaling of the component of the probability trajectory at this frequency:
\begin{equation}
\mu_{\omega} \equiv \mathbb{E}(\tilde{Y}_{\omega}) = \sum_i\mathbb{E}\left[\frac{F_{\omega i}(X_{i}-\bar{p})}{\sqrt{\bar{p}(1-\bar{p})}}\right]= \frac{\tilde{p}_{\omega}}{\sqrt{\bar{p}(1-\bar{p})}}.
\end{equation}
 So, $\mu_{\omega}=0$ under the null hypothesis that $\tilde{p}_{\omega} = 0$. The variance $\nu_{\omega}$ has a more subtle dependence on the probability trajectory. By noting that $\tilde{Y}_{\omega}$ is a summation of independent Bernoulli random variables multiplied by constants, and by noting that the variance of a Bernoulli random variable with bias $b$ is $b(1-b)$, we obtain 
 \begin{equation}
 \nu_{\omega} \equiv \mathbb{V}(\tilde{Y}_{\omega}) = \frac{1}{\bar{p}(1- \bar{p})}\sum_{i} F_{\omega i}^2p_i(1-p_i).
 \end{equation}
  It then follows from the orthogonality of $F$ that $\nu_{\omega} =1$ when the probability trajectory is constant. So $\tilde{Z}_{\omega}$ is approximately $\mathcal{N}(0,1)$ distributed when the probability trajectory is constant. Therefore, calculating significance thresholds using the model that $\tilde{Z}_{\omega}^2$ is $\chi^2_1$ distributed under the null hypothesis will maintain weak control of the FWER.

 Using the relation $p_i = s_i + \bar{p}$, where $\vec{s}$ is the signal vector, we can rewrite the variance as 
 \begin{equation}
 \nu_{\omega} = 1 + \frac{\Delta_{\omega}}{\bar{p}(1-\bar{p})},
 \end{equation}
  where 
\begin{equation}
\Delta_{\omega} = \sum_{i=0}^{N-1} F_{\omega i}^2\left( s_i[1 - 2\bar{p}]- s_i ^2 \right). \label{eq:nu-delta}
\end{equation}
So far, $F$ is a general Fourier-like transformation that is an orthogonal matrix (see earlier). For some transformation, such as the Walsh-Hadamard transformation (but not the DCT), $F_{\omega i}^2= 1/N$ and so Eq.~\eqref{eq:nu-delta} simplifies to
\begin{equation}
\Delta_{\omega} = -\frac{\| \vec{s}\|_2^2}{N},
\end{equation}
because $\sum_i s_i = 0$. Here
 \begin{equation}
 \|\vec{v}\|_2 = \left(\sum_i v_i^2\right)^{\nicefrac{1}{2}},
 \end{equation}
  is the 2-norm.  That is, the variance is upper-bounded by 1, and it decreases in proportion to the signal power per time-step (which is equal to the variance of $\vec{p}$). So the maximum variance over the null hypothesis space of $\tilde{p}_{\omega} = 0$ is $\nu_{\omega}=1$. This means that modeling $\tilde{Z}_{\omega}^2$ with a $\chi_1^2$ distribution is sufficient for strong control of the FWER if the power spectra are generated using, \eg, the Walsh-Hadamard transformation. 

For a general transformation the situation is more complicated. In particular, even with $p_{\omega}=0$ it is possible to obtain $\nu_{\omega} > 1$ for a matrix $F$ containing elements of magnitude greater than $\nicefrac{1}{\sqrt{N}}$. Because all of our data analysis uses the Type-II DCT, defined in Eq.~\eqref{eq:dct},
at this point we specialize to this transformation. In this case, simple algebra can be used to show that
\begin{equation}
\Delta_{\omega} =  \frac{1}{\sqrt{2N}}\left(\tilde{s}_{2\omega} [1 - 2\bar{p}]- \tilde{q}_{2\omega} \right)  - \frac{\|\vec{s}\|_2^2}{N} ,
\end{equation}
where $\vec{q} = (s_0^2,s_1^2,s_2^2,\dots)$ is the vector of time-domain signal powers, and we are allowing the frequency index to extend outside its range as specified in the definition of the DCT. This equation implies that the variance at $\omega$ is increased by signal power at $2\omega$ if the oscillations are not centered on $\nicefrac{1}{2}$. This can only result in a variance above 1 if the amplitude at $2\omega$ outweighs the decrease in the variance caused by the total power per time-step in $\vec{s}$. So the variance at $\omega$ is maximized by a pure-tone probability trajectory oscillating between $\nicefrac{1}{2}$ and 1 or 0 at a frequency of $2\omega$, resulting in the saturable bound $ \nu_{\omega} \leq 1 + \nicefrac{1}{6}$. Therefore the maximal variance over the null hypothesis space of $\tilde{p}_{\omega} = 0$ is $\nu_{\omega} =\nicefrac{7}{6}$. So, ignoring the two approximations made so far, there exists a probability trajectory within the null hypothesis subspace defined by $\tilde{p}_{\omega} = 0$ such that $\tilde{Z}_{\omega}$ is $\mathcal{N}(0,\nicefrac{7}{6})$ distributed. This implies that calculating significance thresholds using a $\chi_1^2$ does not strictly provide strong control of the FWER. 

We could guarantee strong control of the FWER by calculating thresholds using a rescaled $\chi_1^2$ distribution (or, equivalently, by normalizing the data differently). But we choose not to do this, because for a broad class of probability trajectories $ \nu_{\omega} \leq  1$ for all $\omega$. So accounting for the possibility of $\nu_{\omega} > 1$ would result in a reduction in test power --- \ie, higher significance thresholds --- to gain statistical rigor in some edge cases. In our opinion this is not a good tradeoff, and we instead settle for maintaining weak control of the FWER and ``almost'' maintaining strong control of the FWER. This is true in the sense that although there are some probability trajectories for which the FWER is slightly above the desired value, (i) the increase in the FWER is small, (ii) it can only occur for extremal points in the parameter space, and (iii) it is counteracted by the conservative nature of the Bonferroni correction that we use when calculating thresholds from a $\chi^2_1$ distribution (see later). All further statements about strong control of the FWER should be understood to implicitly contain this minor caveat.

The above reasoning was all based on the assumption that $\tilde{Y}_{\omega}$ is approximately $\mathcal{N}(\mu_{\omega},\nu_{\omega})$ distributed for large $N$. So, we now prove that the $N \to \infty$ limiting distribution of $(\tilde{Y}_{\omega}-\mu_{\omega})/\sqrt{\nu_{\omega}}$ is $\mathcal{N}(0,1)$, under a reasonable notion of the limit of a probability trajectory. In particular, we assume that $\epsilon_1 <\bar{p} < 1 - \epsilon_1$ and $\nu_{\omega}> \epsilon_2$ for some constants $\epsilon_1,\epsilon_2 > 0$ and all $N$. The proof will use Lyapunov's central limit theorem \cite{billingsley2008probability}, which we now state. Let $\{A_1,A_2,\dots, A_n\}$ be a sequence of independent random variables where $A_i$ has a finite expected value $\mu_i$ and variance $\nu_i$, and define $s_n^2 = \sum_{i}\nu_i$. Then, if Lyapunov's condition
\begin{equation}
\lim_{n \to \infty} \frac{1}{s_n^{2+\delta}}\sum_{i=1}^n \mathbb{E}\left[ |A_i -\mu_i|^{2+\delta} \right] =0,
\end{equation}
holds for some $\delta>0$, the distribution of $A = \sum_{i=1}^{n} (A_i-\mu_i)/s_n$ converges to $\mathcal{N}(0,1)$ as $N \to \infty$. The random variable $\tilde{Y}_{\omega}$ may be written in the format of this central limit theorem: $\tilde{Y}_{\omega} = \sum_{i}Q_{\omega i}$ where $Q_{\omega i} = F_{\omega i}Y_{i}$, and the expected value and variance of every $Q_{\omega i}$ is bounded (due to our assumptions). It remains to prove that Lyapunov's condition holds. We have that 
\begin{equation}
\mathbb{E}\left[\left| Q_{\omega i} - \mathbb{E}[Q_{\omega i}]\right|^{3}\right] \leq \mathcal{O}( | F_{\omega i}|^3),
\end{equation}
because $Q_{\omega i} = F_{\omega i }Y_{i}$ and the moments of $Y_{i}$ are bounded from above by a constant (as $\epsilon_1 < \bar{p} < 1 - \epsilon_1$).  Therefore 
\begin{align}
\frac{1}{s_n^{3}}\sum_{i=0}^{N-1} \mathbb{E}\left[ |Q_{\omega i} -\mathbb{E}[Q_{\omega i} ]|^3 \right] &\leq  \frac{1}{s_n^{3}} \sum_{i=0}^{N-1} \mathcal{O}(| F_{\omega i}|^3)   ,\\
&= \mathcal{O}\left(\frac{1}{\sqrt{N}}\right),
\end{align}
with the equality holding because $s_n^3=\nu_{\omega}^{\nicefrac{3}{2}}$ and $\nu_{\omega} > \epsilon_2$. So Lyapunov's condition holds for $\delta = 1$.

In addition to performing hypothesis testing on the power spectrum for each circuit, we also test the single power spectrum obtained by averaging over the power spectra for the different circuits, \eg, see Fig.~\ref{fig:gst-exp}. The power at frequency $\omega$ in this spectrum is given by 
\begin{equation}
\tilde{z}_{\text{avg},\omega}^2= \frac{1}{C} \sum_{c=1}^C \tilde{z}_{c,\omega}^2,
\end{equation}
 where $C$ is the number of circuits. The clickstreams for different circuits are independent, and so the corresponding random variable $\tilde{Z}_{\text{avg},\omega}^2$ is the average of independent random variables  that are approximately $\chi_1^2$ distributed under the (intersection) null hypothesis 
 \begin{equation}
 \mathcal{H}_{0,\omega} = \cap_c \mathcal{H}_{0,c,\omega},
 \end{equation}
  that $\tilde{p}_{c,\omega} = 0$, for all $c$. Therefore $\tilde{Z}_{\text{avg},\omega}^2$ is approximately $\nicefrac{\chi^2_{C}}{C}$ under this null hypothesis. 

\subsection{Hypothesis testing of the power spectra}
Here we use the statistical model developed above to calculate the power significance threshold. We argued above that $\tilde{Z}_{c,\omega}^2$ can be modeled as a $\chi^2_1$ random variable under the null hypothesis that $\tilde{p}_{c,\omega} = 0$ (up to minor caveats). So if we are only testing a single circuit $c$ and a single frequency $\omega$, then the $\alpha$-significance power threshold is simply
\begin{equation}
T = \mathsf{CDF}^{-1}_1(1-\alpha),
\label{eq:1-test}
\end{equation}
where $\mathsf{CDF}_{k}$ denotes the cumulative distribution function for the $\chi^2_k$ distribution. However, we test every circuit at every non-zero frequency, and we also test the average power spectrum.  So, to maintain strong control of the FWER with a global significance of $\alpha$ in this set of $(N-1)(C+1)$ tests, we use a Bonferroni correction. In general, this means setting the local significance of the test on circuit $c$ at frequency $\omega$ to $w_{c,\omega}\alpha$, and setting the local significance of the test on the average power spectrum at frequency $\omega$ to $w_{\omega}\alpha$ for some non-negative weightings $w_{c,\omega}$ and $w_{\omega}$ satisfying $\sum_{\omega}(w_{\omega}+\sum_{c}w_{c,\omega})=1$. The test weightings can be optimized for different usages. 

Often it is equally important to identify any drift at every frequency and in every circuit. A one-parameter subclass of weightings that is useful under this circumstance is given by $w_{c, \omega} = (1-w)/(C(N-1))$ and $w_{\omega} = w/(N-1)$ for some weighting factor $w$. This choice of weightings results in a single $\omega$-independent significance threshold for the power in the individual power spectra for the $C$ circuits  of
\begin{equation}
T_{\mathsf{individual}} =  \mathsf{CDF}^{-1}_1\left[1-\frac{(1-w)\alpha}{(N-1)C}\right],
\label{eq:test-individual}
\end{equation}
and an $\omega$-independent significance threshold for the power in the averaged spectrum of
\begin{equation}
T_{\mathsf{average}} =  \frac{1}{C} \mathsf{CDF}^{-1}_C \left[1-\frac{w\alpha}{N-1}\right].
\label{eq:test-average}
\end{equation}

The weighting factor $w$ controls the proportion of the test statistical significance that is allocated to the test on the averaged power spectrum versus the significance allocated to the individual power spectra. The averaged power spectrum does not contain information about which circuits (if any) exhibit drift, but has (under most circumstances) increased sensitivity to any drift. So the best choice for this weighting depends on the relative importance of detecting any drift versus being able to assigning it to a particular circuit. All our data analysis uses the thresholds given in Eqs.~(\ref{eq:test-individual}-\ref{eq:test-average}), with $\alpha = 5\%$ and $w = \nicefrac{1}{2}$, with the exception of time-resolved RB, wherein $w = 1$.

\subsection{Estimating probability trajectories}
We now explain our methods for estimating the circuit outcome probability trajectories (step 3, Fig.~\ref{fig:flowchart}A). The methods are based on model selection, to avoid overfitting. The general form of the model that we use for a probability trajectory is:
\begin{equation}
 p(\vec{\gamma},t) =  \gamma_{0} + \sum_{\omega \in \mathbb{W}} \gamma_{\omega}f_{\omega}(t),
\end{equation}
where $f_{\omega}(t)$ is the $\omega^{\rm th}$ basis function of the chosen Fourier transform, the summation is over some set of frequencies $\mathbb{W}$ that are to be chosen using model selection, and the $\gamma_{\omega}$ are parameters constrained only so that $ p(\vec{\gamma},t)$ is a valid probability at all data acquisition times (the condition that $p(\vec{\gamma},t)$ is within $[0,1]$ for all real-valued $t$ can cause overly severe compression of the amplitudes). In the case of the Type-II DCT the basis functions are
\begin{equation}
f_{\omega}(t) = \cos\left(\frac{\omega \pi}{N}\left[\frac{t - t_{0}}{t_{\text{step}}} + \frac{1}{2}\right] \right),
\end{equation}
where, as throughout, $t_{i}$ is the $i^{\rm th}$ data collection time for the circuit in question, and
\begin{equation}
t_{\text{step}} = \frac{t_{N-1} - t_{0}}{N-1},
\end{equation}
 is the size of the time-step.

The first step is model selection to identify a set of frequencies ($\mathbb{W}$). These frequencies can be chosen using the results of the hypothesis testing on the data spectra. It is useful to allow flexibility in exactly how they are chosen from the hypothesis testing results, because this allows the techniques to be adapted to different applications. A good general-purpose option, which results in estimates with a simple statistical meaning, is to set $\mathbb{W}$ to contain those and only those frequencies that are found to be statistically significant in the power spectrum for the circuit in question. With this method we are only including frequencies that contribute a nonzero component to the true $p(t)$ with confidence $(1-\alpha)$, where $\alpha$ is the chosen global significance level. All of the data analysis in this article uses this method, with the exception of the model selection in time-resolved RB (see later). 

It remains to estimate the model parameters, \ie, the amplitudes $\vec{\gamma}$. Maximum likelihood estimation (MLE) is a statistically sound methodology. But MLE requires a numerical optimization, which can sometimes be slow. So we also provide an optimization-free method. The data amplitude $\tilde{x}_{\omega}$ is a well-motivated estimate of the probability trajectory amplitude at $\omega$, since $\mathbb{E}(\tilde{X}_{\omega}) = \tilde{p}_{\omega}$. In terms of the rescaled amplitudes $\tilde{\vec z}$, this suggests setting $\gamma_{\omega}= S(\vec{x}) \tilde{z}_{\omega}$ where $S(\vec{x})$ is a rescaling factor that inverts the normalization of the $\tilde{\vec z}$ amplitudes by the data standard deviation (and also rescales for the chosen normalization of the basis functions). This is intuitive but flawed: the resulting probability trajectory is not necessarily a valid probability at all times. To address this, we add a regularization to this estimator. We map each amplitude via
\begin{equation}
\gamma_{\omega} \to \gamma_{\omega} - \delta\,\text{sign}(\gamma_{\omega}),
\end{equation}
 with $\delta$ the smallest constant such that the estimated probability trajectory $p(\vec{\gamma},t)$ is within $[\epsilon, 1-\epsilon]$ at all data acquisition times  $t_i$, for some chosen $\epsilon \geq 0$ (\eg, $\epsilon = 0$ if estimates on the parameter space boundary are considered acceptable). We refer to this estimation method as the ``Fourier filter,'' because it is a form of signal filtering. The Fourier filter returns very similar estimates to MLE under most conditions, although note that if $\epsilon=0$ it is possible for the Fourier filter to return estimates that have a likelihood of zero.

All data analysis in this paper uses MLE, except the time-resolved RB analysis. The motivation for using the Fourier filter in time-resolved RB is that then the speed of the analysis is limited only by the fast Fourier transform and curve fitting at the times of interest. This is particularly relevant to RB, as one motivation for using RB is its speed and simplicity. Moreover, note that for all the data herein the difference between the MLE and the Fourier filter estimates is negligible.

\subsection{Analyzing data from circuits with more than two outcomes}
Throughout this article we have only considered analysis of circuits with two possible outcomes. This is sufficient for all the experiments and simulations herein. However, the techniques we present can be easily generalized to $M$-outcome circuits, as we now briefly outline. In this more general setting, the aim is still to detect and quantify time variation in the probability distribution over outcomes for each circuit. However $M$ can be very large in commonly encountered circumstances, as $M=2^n$ for an $n$-qubit circuit terminated with the standard measurement.
Even accurate estimation of a time-independent distribution over a very large number of outcomes is infeasible with reasonable quantities of data, so it will not be possible to accurately estimate a time-dependent distribution with many outcomes. For this reasons, it is useful to allow for a coarse-graining of the outcomes into $\check{M} \leq M$ bins (\eg, we can marginalize over some subset of the qubits in an $n$-qubit circuit or bin the outcomes into two sets). The analysis will then interrogate time-dependence in the coarse-grained probability distribution. So coarse-graining can make the analysis feasible with a reasonable amount of data, but the analysis will be insensitive to any time dependence that disappears under the chosen coarse-graining.

Each coarse-grained outcome has a bit-string time-series associated with it, corresponding to whether or not that measurement outcome was observed at each time. A power spectrum can be calculated for each such bit-string, and exactly the same hypothesis testing performed on each such power spectrum as in the 0/1 outcomes cases (with an appropriate correction for the increased number of hypothesis tests). Moreover, the spectrum obtained by averaging the spectra for a circuit over course-grained measurement outcomes can also be tested (which will be approximately $\chi^2_{\check{M}-1}$ distributed under the null hypothesis), or it can be tested instead of the per-outcome spectra --- to avoid significance dilution by implementing too many hypothesis tests. The probability trajectory estimation (step 3, Fig.~\ref{fig:flowchart}A) then proceeds exactly as in the 0/1 outcome case, except that now the estimation is of $\check{M}$ trajectories that must sum to 1 at all times. Any time-resolved parameter estimation proceeds exactly as in the 0/1 outcome case.

\section{Time-resolved tomography and benchmarking}\label{app:2}
Here we provide supporting theory and information for our time-resolved tomography and benchmarking methods. This is the methods summarized in steps 4 and 5 of the flowchart in Fig.~\ref{fig:flowchart}.

\subsection{Model selection in time-resolved tomography}
We now explain the model selection step in our ``intrusive'' approach to time-resolved tomography and benchmarking (step 5a, Fig.~\ref{fig:flowchart}A). Our methods  for time-resolved tomography and benchmarking start from a time-independent $\{\gamma_j\}$-parameterized model $\mathcal{M}$ that predicts circuit outcomes. In the intrusive approach we expand each parameter $\gamma_j$ into the time-dependent form
\begin{equation}
\gamma_j(t) = \gamma_{j,0}+ \sum_{\omega  \in \mathbb{S}_j} \gamma_{j,\omega} f_{\omega}(t),
\end{equation}
 where the $\gamma_{j,\omega}$ are restricted only so that they satisfy any constraints on $\gamma_j$, the summation is over some set of non-zero frequencies $\mathbb{S}_j$, and $j$ indexes all the parameters in the time-independent model. So, we obtain a new model $\mathcal{M}_{\rm{tr}}$ containing 
 \begin{equation}
 k = \sum_j(|\mathbb{S}_{j}|+1),
 \end{equation}
 free parameters, where $|\mathbb{S}|$ denotes the size of the set $\mathbb{S}$. The parameterized model is entirely defined by the set $\mathbb{M} = \{ \mathbb{S}_j \}$, and the model selection consists of choosing this set.

For the related problem of choosing the frequencies in the model for a probability trajectory (see above) there is a clear solution: our instability detection methods are directly testing hypotheses about the frequencies in $p(t)$, and so there is a clear statistical justification for choosing the frequencies in our model for $p(t)$ using the results of these tests. In contrast, the parameters in the model $\mathcal{M}$ do not necessarily have a simple relationship to the circuit probabilities --- for an arbitrary model the relationship is entirely arbitrary. So, the relationship between the frequencies detected in the probability trajectories and the frequencies in each $\gamma_j$ is also, for a general model, entirely arbitrary. However, for typical models associated with tomographic methods and circuits, under some approximation there is a simple relationship between the model parameters and the predicted probabilities. 

To illustrate this, we turn to the Ramsey model with a time-dependent phase that was introduced in the main text:
\begin{align}
p_{l}(t) = A + B \exp(- l/l_{0}) \sin(2\pi l t_{w} \Omega(t)).
\end{align}
If we expand around $lt_{w} \Omega(t) = 0$ we obtain
\begin{equation}
p_l(t) = A + 2 \pi B \exp(-l/l_0) l t_{w} \Omega(t) + O((lt_{w} \Omega(t))^2).
\end{equation}
 So, if $l t_{w} \Omega(t)$ is small then the dominant frequencies in $p_l(t)$ are also necessarily frequencies in $\Omega(t)$. Equivalent expansions around the zero-error parameter values hold for the general process matrix model of GST --- and most other tomographic techniques use a model that is a restriction of that in GST. Therefore, we can use the results of the hypothesis tests on the data from sufficiently short circuits to select the frequencies for the model parameters. But longer circuits are more sensitive to small variations in the model parameters, so we do not want to entirely discount that data in the model selection. To address this problem we can pick a small set of candidate models and then choose between them using the Akaike information criterion (AIC) \cite{akaike1974new}. For a parameterized model $\mathcal{M}$ with $k$ free parameters, the AIC score for that model is 
\begin{equation}
\mathsf{AIC}(\mathcal{M}) = 2k - 2\log(\mathcal{L}_{\max}),
\end{equation}
 where $\mathcal{L}_{\max}$ is the maximum of the likelihood function for that model. For a set of candidate parameterized models, the preferred model is the one with the minimum AIC score. Moreover, $\mathsf{AIC}(\mathcal{M}_a) -\mathsf{AIC}(\mathcal{M}_b)$ is twice the log relative likelihood of model $\mathcal{M}_b$ with respect to model $\mathcal{M}_a$, so this difference quantifies how much one model is preferred over another.  This allows us to decide between multiple time-resolved models.

For the Ramsey experiment presented in the main text we select the expansion
\begin{equation}
\Omega(t) =\gamma_0 + \sum_{\omega \in \mathbb{W}}\gamma_{\omega} f_{\omega}(t),
\end{equation}
 where $\mathbb{W}= \{1,2,3,4,5,6,7,9,11,12,13,14,15 \}$ is the set of all the frequency indices found to be significant in any of the power spectra.  We then check that the data justifies including the higher frequencies in the model --- because the lowest frequencies appear for circuits where $l$ is small enough so that $lt_{w}\Omega(t)\approx 0$, and so they are almost certainly components in the true $\Omega(t)$. In particular, we select the 10 candidate parameterized models defined by the frequency sets $\mathbb{W}_{0} = \{1,2,3,4\}$, $\mathbb{W}_{1}  = \mathbb{W}_0 \cup \{ 5 \}$, $\mathbb{W}_{2} = \mathbb{W}_1 \cup \{6\}$ through to $\mathbb{W}_9 = \mathbb{W}$, and compare their AIC scores, $\mathsf{AIC}_{k}$ for $k=1,2,\dots,9$. The values of $\mathsf{AIC}_k-\mathsf{AIC}_{9}$ are all well above zero (ranging from around 10 to around 1000). So the model containing all the frequencies is clearly favored according to the AIC. The detuning estimate reported in the main text is the MLE over this parameterized model.

There is no guarantee that $\mathbb{W}$ contains all of the frequencies that are components in the true $\Omega(t)$. We could also select some larger models to consider, but we do not do this because it would be \emph{ad hoc}: there are $N -14= 5986$ remaining frequencies that we could also include, and there is no clear method for choosing the next frequency to add. We could calculate the AIC for every possible model, but this is clearly impractical. In general, within this framework of time-resolved models based on Fourier decompositions there are $2^{Nk-k}$ possible time-resolved models, where $k$ is the number of parameters in the time-independent model. Moreover, this would not be a reasonable use of the AIC. When adding an erroneous parameter to a model sometimes the AIC score will decrease slightly due to random fluctuations in the data slightly favoring this model, and if we consider adding many different parameters this will almost certainly occur in at least one case. 

In the main text we also presented the results of time-resolved Ramsey tomography after we had discarded 80\% of our data (\ie, we keep every $5^{\rm th}$ click for each circuit). This analysis followed the same procedure as the analysis on all the data. In particular, note that we implemented independent model selection, \ie, we did not ``cheat'' by using the time-resolved model selection that was performed using all the data. In particular, using only 80\% of the data resulted in a smaller model whereby $\mathbb{W} = \{1,2,3,5,6,7,9\}$. We therefore would not expect the two estimates of $\Omega(t)$ to be equal up to statistical uncertainty (which they are not --- see Fig.~\ref{fig:flowchart}E), as both these statistical uncertainties are standard in-model uncertainties and the two estimates are fits to different models.

\subsection{Time-resolved RB}
RB is a set of protocols for benchmarking a set of gates on $n$ qubits \cite{knill2008randomized, magesan2011scalable, proctor2018direct}. The core protocols all follow the same general procedure:
\begin{itemize}
\item[(i)] For a range of lengths $m$, run $K$ circuits sampled from some $\Phi_m$ distribution whereby (a) the average length of a sampled circuit scales linearly with $m$, and (b) each circuit will deterministically output some $n$-bit string if implemented perfectly.
\item[(ii)] Process the data to obtain the average success probability ($P_m$) at each length $m$, meaning the probability that the expected bit-string was successfully output.
\item[(iii)] Fit $P_m$ to $P_m = A + B\lambda^m$, and convert $\lambda$ to an error rate via $r=\mathcal{N}(1-\lambda)$ where $\mathcal{N}$ is a constant that we set to $\mathcal{N} = (4^n-1)/4^n$ where $n$ is the number of qubits.
\end{itemize}
Within this framework there are a range of different RB protocols, \eg, those of Refs.~\cite{knill2008randomized, magesan2011scalable, proctor2018direct}, that vary in how the random circuits are chosen.  Our methods can be used to upgrade any of these RB protocols to time-resolved techniques. 

We focus on the non-intrusive approach to time-resolved RB, which is the following method:
\begin{enumerate}
\item Obtain data by rastering through the sampled circuits.
\item Use the general stability analysis to estimate the time-resolved success probability $p(t)$ for each circuit.
\item Average the time-resolved success probabilities for all circuits of the same RB length $m$ to obtain time-resolved average success probabilities $P_m(t)$.
\item Apply the standard RB curve fitting analysis [step (iii) above] to $P_m(t_i)$ at every time of interest $t_i$.
\end{enumerate}
The general instability analysis step in this process has some useful flexibility. It has two aspects that can be optimized to a particular task: the exact statistical hypothesis testing on the power spectra, and the selection of the frequencies to include in the time-resolved $p(t)$ models. 

Because the RB circuits are random, they are broadly all sensitive to the same physical parameters, with the exact sensitivities depending on the precise details of the random circuit. So the frequencies in $p(t)$ will typically be the same in all the sampled circuits, even if the power in each frequency will generally vary between circuits. In this circumstance the averaged power spectrum is more powerful for detecting these frequencies, because the averaging suppresses measurement shot noise and amplifies signal if the same frequencies appear in most of the spectra. Moreover, here there is little penalty for erroneously including a frequency in the success probability trajectory in a few circuits --- we are implementing a global analysis of the data, rather than trying to extract detailed diagnostic information. So we only perform hypothesis testing on the average power spectrum (\ie, we set $w = 1$ in Eqs.~(\ref{eq:test-individual}--\ref{eq:test-average})), and then we include every statistically significant frequency in the model for each success probability.

\subsection{Time-resolved RB simulation} 
In the main text we demonstrate time-resolved RB on simulated data. This demonstration is of time-resolved direct randomized benchmarking (DRB) \cite{proctor2018direct} on two qubits, with a gate set consisting of a \textsc{cphase} gate between the qubits (\ie, a controlled $\sigma_z$), rotations around the $\hat{x}$ and $\hat{y}$ axes by $\pm \frac{\pi}{2}$ and $\pi$ on each qubit, and the Hadamard and idle gates on each qubit. The details of how DRB circuits are sampled (\ie, the $\Phi_m$ of DRB) can be found in Ref.~\cite{proctor2018direct} --- they are omitted here as they are not relevant to assessing the performance of time-resolved RB. 

We simulated gates with time-dependent coherent phase errors and time-independent depolarizing errors. Each gate is modeled as the perfect gate followed by the error map 
\begin{equation}
\mathcal{E}(t)=\mathcal{D}_{\gamma}\circ\mathcal{R}_{z}(\theta(t)),
\end{equation}
 on each qubit, where 
\begin{equation}
\mathcal{D}_{\gamma}[\rho] = \gamma\rho + (1-\gamma)\mathds{1}/2,
\end{equation}
is a one-qubit depolarizing map, and 
\begin{equation}
\mathcal{R}_{z}(\phi)[\rho] = \exp\left(-i \frac{\phi}{2} \sigma_z\right)\rho\exp\left(+i \frac{\phi}{2} \sigma_z\right).
\end{equation}
We set $\gamma = 1 - 0.04/3 $, making the probability of Pauli error 1\%. The rotation angle is set to
\begin{equation}
\theta(t) = at + b\sin(\phi t) + c_{t},
\end{equation}
 where $a = 2\pi \times 10^{-5}$, $b=1.5\times 10^{-2}$, $\phi =2\pi \times 10^{-2}$, and $c_{t}$ is a Brownian stochastic process given by $c_{0}=0$ and $c_{t+1} = c_{t} +\mathcal{N}(0,\nu)$ where $\nu = \nicefrac{3}{200}$.  The parameters in this error model have been chosen to give physically plausible phase trajectories, and to not be artificially easy for our methods (\eg, they do not induce $p(t)$ that are exceptionally sparse in the Type-II DCT basis). They have no further significance. The time $t$ starts at 0 and increases by 1 after a single run of any circuit.

\begin{figure*}[t!]
\includegraphics[width=18.cm]{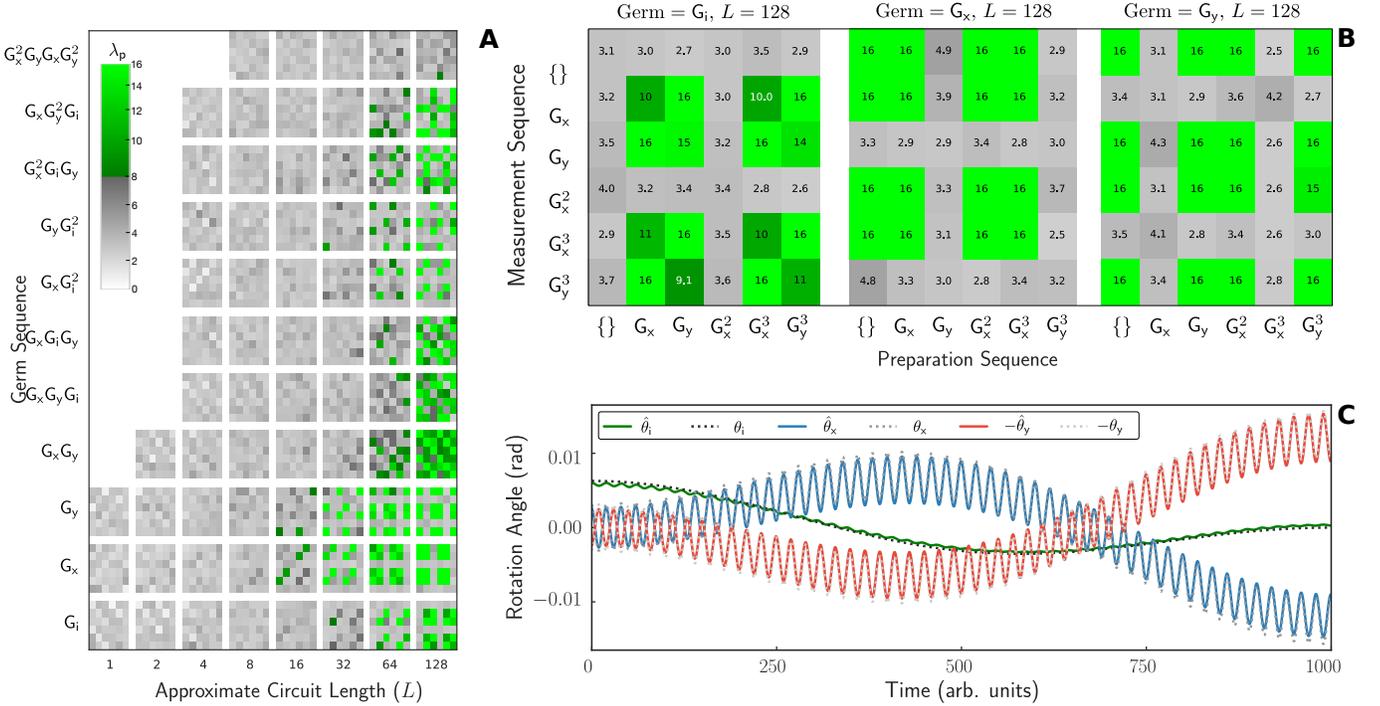}
\caption{{\bf Model selection in time-resolved GST on simulated data}. {\bf A-B.} The evidence for instability in the outcomes of each circuit in simulated GST data, as quantified by the test statistic $\lambda_{\mathsf{p}}$. A pixel is colored when $\lambda_{\mathsf{p}}$ is large enough to be statistically significant, otherwise it is greyscale. The arrangement of the data is as in Fig.~\ref{fig:gst-exp}. The pattern of significant $\lambda_{\mathsf{p}}$ is used to decide which model parameters to expand into a sum of Fourier coefficients. {\bf C.} The results of time-resolved GST when we select a time-resolved parameterized model using a simple frequency selection technique that erroneously includes high-frequency components in the $\hat{z}$-axis rotation angle of $\Gi$. The parameter estimates still closely track the true values, demonstrating that time-resolved GST can be successful even with simple model selection methods.}
\label{fig:gst-sim-boxplot}
\end{figure*}

In Fig.~\ref{fig:tr-rb}A the estimated time-resolved error rate is compared to the true time-dependent $r(t)$. The value of $r(t)$ is obtained by fitting the exact instantaneous average success probabilities for the set of 100 sampled circuits. This $r(t)$ is the target that time-resolved DRB with this particular set of random circuits is aiming for --- it does not correspond to the large-sample-limit error rate that finite-sampling DRB is estimating. We choose this comparison because analysis of time-series data from one particular set of $K$ randomly sampled circuits cannot correct for any finite sampling error introduced when sampling this set of circuits. By defining $r(t)$ in this way, we avoid conflating this finite-circuit-sampling error with any inaccuracies that are intrinsic to time-resolved benchmarking.

\subsection{Gate set tomography}
GST is a method for jointly reconstructing a set of gates, a state preparation and a measurement \cite{blume2016certifying,merkel2013self}. GST is predicated on the standard Markovian model for errors in QIPs, whereby the prepared state is a fixed density operator, the gates are fixed completely positive and trace preserving (CPTP) maps  --- which are the linear transformations that map density operators to density operators --- and the measurement is a fixed positive-operator valued measure. For our implementation of GST, we parameterize a CPTP gate by $\mathsf{G} = \Lambda \circ \mathsf{G}_{\mathsf{target}}$ where $ \mathsf{G}_{\mathsf{target}}$ is the fixed target map and $\Lambda$ is a CPTP map encompassing the errors in the gate. This error map is then parameterized as $\Lambda = \exp(G)$ with $G$ defined by
\begin{equation}
G[\rho] =\sum_{k} \theta_{k} H_{k}[\rho] +\sum_{j,k} s_{jk} S_{jk}[\rho],
\end{equation}
where the summations are over $x$, $y$ and $z$, and
\begin{align}
H_{k}[\rho] &= \frac{i}{2}(\rho\sigma_k-\sigma_k\rho),\\
S_{jk}[\rho] &= \sigma_j\rho\sigma_k - \frac{1}{2}(\rho \sigma_k\sigma_j + \sigma_k\sigma_j \rho).
\end{align}
Here, $\sigma_{x}$, $\sigma_{y}$ and $\sigma_{z}$ are the Pauli operators. The $\{H_{k}\}$ maps generate all unitary errors. If $(\theta_x,\theta_y,\theta_z) = \phi \hat{v}$ where $\hat{v}$ is a unit vector then the generated unitary is a rotation around the $\hat{v}$ axis of the Bloch sphere by an angle $\phi$. So $\theta_x$, $\theta_y$ and $\theta_z$ are the components of the unitary error along the $\hat{x}$, $\hat{y}$ and $\hat{z}$ axes in the standard sense. The $\{S_{ij}\}$ maps generate all non-unitary Markovian errors, so the matrix $s$ dictates the stochastic error rates and the size of any amplitude damping errors. In this parameterization, the CPTP constraint is that $s$ is positive semi-definite. Finally, note that this parameterization of $\Lambda$ includes only infinitesimally-generated CPTP maps; it does not strictly permit every possible CPTP map.

\subsection{Time-resolved GST simulation}
In the main text, time-resolved GST is demonstrated on simulated data. Here we detail the error model that was simulated, and use this example to explain how we apply our methods to implement time-resolved GST. The simulation is for GST circuits out to $L=128$ containing the gates $\Gi$, $\Gx$ and $\Gy$ with an error model that, when expressed in the parameterization above, consists of:
\begin{itemize}
\item[(i)] Setting the $s$ matrix for all three gates so that if there are no coherent errors $\Lambda$ is a uniform depolarizing channel with a 1\% error rate (this is $s\propto \mathds{1}$).
\item[(ii)] Setting all the coherent error parameters to zero, except 
\begin{align*}
\theta_{z}(\Gi) &= \frac{2\pi}{10^{3}}\left[f_{1}(t)+f_{2}(t)\right],\\
\theta_{x}(\Gx) &= \theta_{y}(\Gy) = \frac{4\pi}{10^3}\left[f_{1}(t)-f_{2}(t)+0.5f_{100}(t)\right],
\end{align*}
where $\theta_{k}(\G)$ is the $H_{k}$ error rate for gate $\G$, and $f_k(t)$ is the $k^{\rm th}$ basis function of the Type-II DCT.
\end{itemize}
Time starts at 0 and increments by 1 after each raster. The trajectories of the time-dependent parameters over the 1000 simulated rasters are shown in Fig.~\ref{fig:tr-rb}D (and are also included in Fig~\ref{fig:gst-sim-boxplot}C).

To implement time-resolved GST on this simulated data we first apply the general stability analysis. The resulting 
\begin{equation}
\lambda_{\mathsf{p}}=-\log_{10}(\mathsf{p}),
\end{equation}
 statistic for every circuit is shown in Fig.~\ref{fig:gst-sim-boxplot}A, where $\mathsf{p}$ is the p-value of the largest peak in the power spectrum for that circuit. (Due to numerical accuracy limitations, the minimal p-value is $\mathsf{p} = 10^{-16}$). The value of $\lambda_{\mathsf{p}}$ varies strongly between circuits of the same approximate length $L$. The strong variations in $\lambda_{\mathsf{p}}$ imply that the time-varying errors are coherent, as stochastic errors have a fairly uniform impact across the GST circuits. Moreover, by inspecting the values of $\lambda_{\mathsf{p}}$ for the simple germs $\Gi$, $\Gx$ and $\Gy$ it becomes clear that $\theta_{z}(\Gi)$, $\theta_{x}(\Gx)$ and $\theta_y(\Gy)$ are varying. For example, with the $\Gi$ germ it is the generalized Ramsey circuits for which $\lambda_{\mathsf{p}}$ is statistically significant, as highlighted in Fig.~\ref{fig:gst-sim-boxplot}B. These circuits amplify  $\theta_{z}(\Gi)$, so this parameter is likely varying over time. Similarly, it is generalized Rabi sequences that exhibit statistically significant $\lambda_{\mathsf{p}}$ for the $\Gx$ and $\Gy$ germs, and these circuits amplify over/under-rotation errors in these gates, \ie, the parameters $\theta_{x}(\Gx)$ and $\theta_y(\Gy)$. So we conclude that $\theta_{z}(\Gi)$, $\theta_{x}(\Gx)$ and $\theta_{y}(\Gy)$ should be expanded into a summation of Fourier coefficients. Having decided on the set of parameters that are to be expanded into a time-dependent functions, we now use the short-hand $\theta_{\mathsf{i}}  \equiv \theta_{z}(\Gi)$, $\theta_{\mathsf{x}}  \equiv \theta_{x}(\Gx)$ and $\theta_{\mathsf{y}} \equiv \theta_{y}(\Gy)$.

Deciding on the frequencies to include in the expansion of each of these parameters is more involved. One simple option is to use the frequencies that are statistically significant in the global power spectrum --- the set $\mathbb{W} = \{1, 2, 3, 4, 97, 98, 99, 100, 101, 102\}$ --- as the frequency set for all three parameters. This parameterized model $(\mathcal{M}_1)$ includes all of the frequencies in the true model, but also sums and differences of these frequencies, as well as high-frequencies in the expansion of $\Gi$ that are not present in the true model. Yet, implementing MLE over this parameterized model still provides an accurate time-resolved estimate of $\theta_{\mathsf{i}}$, $\theta_{\mathsf{x}}$ and $\theta_{\mathsf{y}}$, as shown in Fig.~\ref{fig:gst-sim-boxplot}C. 

A more sophisticated model selection strategy is to select the frequencies to include in the expansion for $\theta_{\mathsf{i}}$ (the set  $\mathbb{W}_{\mathsf{i}}$) to be those frequencies that are statistically significant in the circuits in which $\Gi$ is a germ, and similarly for $\theta_{\mathsf{x}}$ and $\theta_{\mathsf{y}}$. This results in the sets $\mathbb{W}_{\mathsf{i}}=\{1,2,3\}$, $\mathbb{W}_{\mathsf{x}}=\{1,2,3,100,103\}$ and $\mathbb{W}_{\mathsf{y}}=\{1,2,3,5,97,100,101,103\}$, and constitutes a smaller parameterized model ($\mathcal{M}_{2}$). Even this smaller model still contains frequencies that are not in the true model, but each frequency set now only contains sums and differences of the frequencies that appear in the corresponding parameter. The MLE over this parameterized model is what is reported in the main text. The difference in the AIC scores of these two candidate parameterized models is $\mathsf{AIC}(\mathcal{M}_{1})-\mathsf{AIC}(\mathcal{M}_{2}) \gg 1$. So the AIC favors discarding the erroneous parameters in $\mathcal{M}_1$. 

\begin{figure}[t!]
\includegraphics[width=8.8cm]{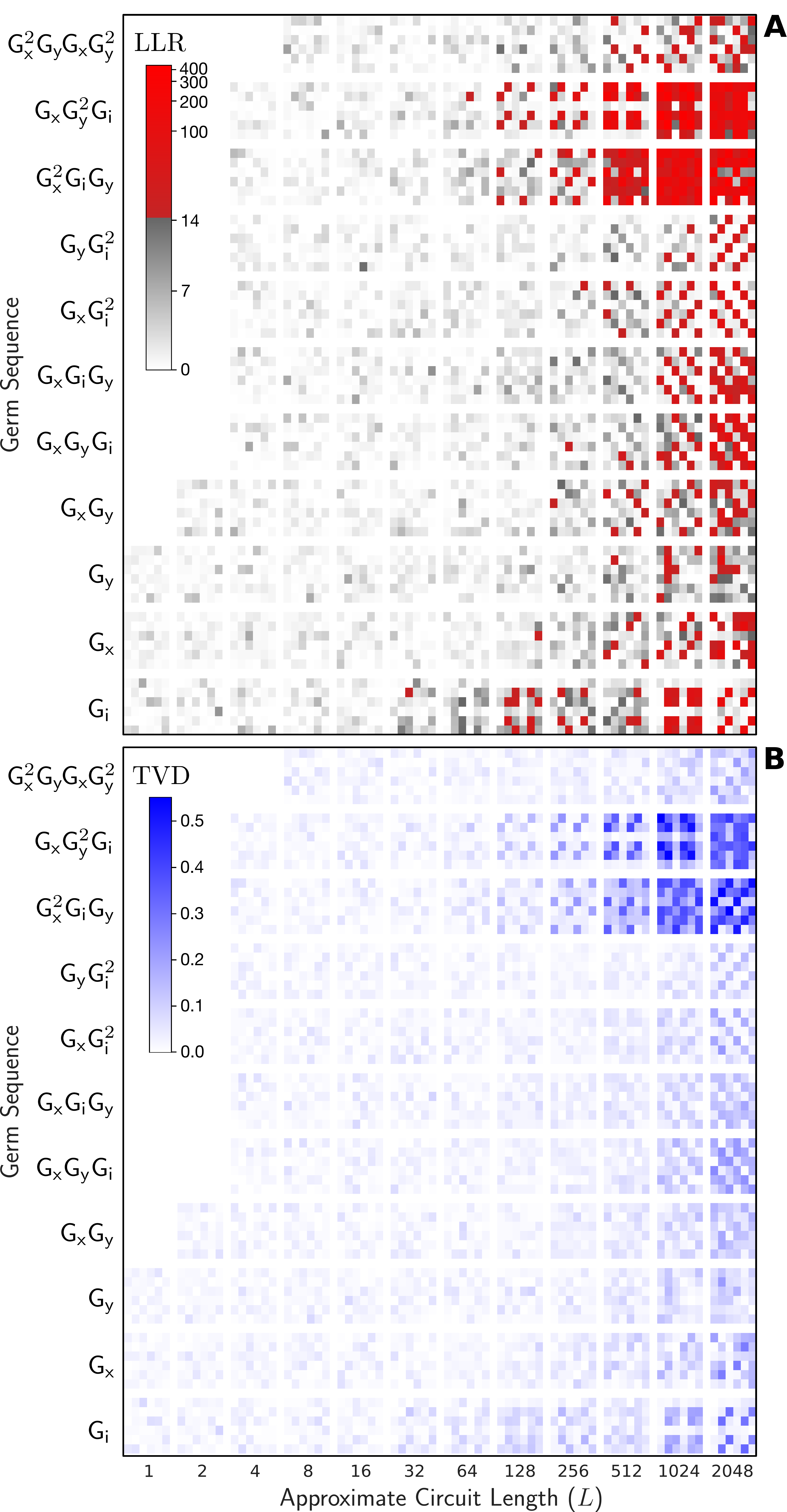}
\caption{{\bf Model violation in standard GST on experimental data}. The discrepancy between the predictions of the time-independent GST estimate and the data, for the first GST experiment. {\bf A.}~The log-likelihood ratio (LLR) statistics of the GST estimate. A pixel is colored if and only if the LLR for that circuit is so large that it is inconsistent with the model of standard GST, at a global significance of $5\%$. {\bf B.}~The total variational distance (TVD) between the probabilities predicted by the GST estimate and the observed frequencies. The TVD is not being used as a statistic in a hypothesis test and so there is no differentiation between significant and insignificant TVDs.}
\label{fig:gst-exp-modelviolation}
\end{figure}

\subsection{Time-resolved GST on experimental data}
We now detail the exact approach used to implement time-resolved GST on the data from the first of our two GST experiments (as discussed in the main text, we implemented standard time-independent GST on the data from the second experiment). As discussed in the main text, we use the $\lambda_{\mathsf{p}}$ statistics to motivate only expanding the $\hat{z}$-axis rotation angle of the $\Gi$ gate --- \ie, using the same notation as above, the parameter $\theta_{z}  \equiv \theta_z(\Gi)$ --- into a sum of Fourier components. Specifically, we expand $\theta_{z}$ into
\begin{equation}
\theta_{z} = \gamma_{0} + \sum_{\omega \in \mathbb{W}} \gamma_{\omega} f_{\omega}(t),
\end{equation}
 where $\mathbb{W}=\{1,2,5,7,15\}$ is the set of significant frequencies in the average power spectrum (shown in Fig.~\ref{fig:gst-exp}D). This method for the frequency selection is motivated by noting that, as we have decided that only $\theta_{z}$ exhibits time-dependence, all the frequencies in the averaged power spectrum must be attributable to variations in $\theta_z$. So far this is exactly the same type of analysis as with the time-resolved GST on simulated data. However, fitting the model requires more care with experimental data.

First we implemented standard time-independent GST on all the data. The predictions of this GST estimate are compared to the aggregated counts data in Fig.~\ref{fig:gst-exp-modelviolation}. This summarizes the discrepancy between the data and the predictions of the model estimated by standard GST using (A) the log-likelihood ratio (LLR) and (B) the total variational distance (TVD). The LLR test statistic for a circuit is 
\begin{equation}
\mathsf{LLR} = -2\log\left(\mathcal{L}_{\mathsf{gst}}/\mathcal{L}_{\mathsf{freq}}\right).
\end{equation}
Here $\mathcal{L}_{\mathsf{gst}}$ is the likelihood, given the data, of the outcome probabilities for that circuit that are predicted by the GST estimate. Conversely, $\mathcal{L}_{\mathsf{freq}}$ is the likelihood of the outcome probabilities that are equal to the observed frequencies (\ie, for each outcome, set the corresponding probability equal to the number of times that particular outcome was observed divided by the total number of outcome counts for the circuit).  Under the null hypothesis that the model of standard GST is true, Wilks' theorem tells us that the LLR is asymptotically $\chi^2_1$ distributed.  This is because, since we have many circuits, the degrees of freedom per circuit is approximately zero for the GST model (used to compute $\mathcal{L}_{\mathsf{gst}}$) and one for the model maximized over to compute $\mathcal{L}_{\mathsf{freq}}$.The TVD for a circuit is 
\begin{equation}
\mathsf{TVD} = \left| p_{\mathsf{model}} - \frac{1}{N}\sum_i x_i\right|,
\end{equation}
 where $p_{\mathsf{model}}$ is the probability for the circuit to output 1 predicted by the estimated model and $\vec{x}$ is the clickstream for the circuit. We do not use the TVD as a test static (it has less convenient properties for hypothesis testing than the LLR), but the TVD is useful as --- unlike the LLR --- it quantifies the distance between the observed frequencies and the predicted probabilities. There are statistically significant LLRs in Fig.~\ref{fig:gst-exp-modelviolation}A, so the data are inconsistent with the model of standard GST.

\begin{table}
\begin{center}
\setlength{\tabcolsep}{6pt}
\renewcommand{\arraystretch}{1.5}
 \begin{tabular}{ r | c  c  c  c  c  c }
Error rate & \multicolumn{2}{c}{$\epsilon_{\diamond} \times 10^3$} &  \multicolumn{2}{c}{$ \epsilon_{\textsf{f}} \times 10^3$} &  \multicolumn{2}{c}{$\epsilon_{\mathsf{spec},\textsf{f}} \times 10^3$}    \\
\cline{1-7}
 Exp. \# &  1 &  2 &  1 &  2 &  1 &  2 \\
\cline{1-7}
 $\Gi$ & 3.45 & 0.40 & 0.02 &  0.07 &  0.02 &  0.07\\  
$\Gx$ & 1.25 & 0.38  & 0.03 &  0.06 & 0.03&   0.06\\ 
$\Gy$ & 0.82 & 0.53  & 0.15  &  0.04 & 0.15 &  0.04
\end{tabular}
\end{center}
\caption{{\bf Estimated gate error rates}. Error rates for the three gates in the two GST experiments, obtained using time-resolved GST. In the first experiment the error rates of the $\Gi$ gate vary over time (see Fig.~\ref{fig:gst-exp}C of the main text). The values reported here are the average over the experiment duration. All other error rates are static over the course of each experiment. See the text for definitions of these three different error rates.}
\label{table:error-rates}
\end{table}

From the general instability analysis, we know that there is some instability in the qubit (see Fig.~\ref{fig:gst-exp}A). So we did not expect the time-independent model of standard GST to be consistent with the data. But, if drift was the only effect outside of GST's model that is needed to explain the data, we would expect the predictions of the estimated GST model to be inconsistent with the data only for those circuits that exhibit drift. This is not the case; comparing Fig.~\ref{fig:gst-exp-modelviolation} to Fig.~\ref{fig:gst-exp}A shows that the inconsistency between the estimate of standard GST and the data cannot all be attributed to detectable drift. In particular, the data from many of the long circuits containing germs other than $\Gi$ are irreconcilable with the model of GST, but there is no evidence for drift in these circuits. So there is some other non-Markovian effect that our time-resolved parameterized model does not take into account. 
 
This sort of ``model violation'' interferes with the model selection and data fitting techniques that we rely on. If we fit our time-resolved parameterized model to all the data, the data that are inconsistent with any set of parameter values can pollute the estimates and render them meaningless. So we first estimate most of the time-independent parameters in the model --- specifically, the state preparation, measurement, $\Gx$, $\Gy$ and the non-unitary errors matrix $s$ for $\Gi$ ---  by implementing ordinary GST on the data from circuits with $L \leq 64$ (these circuits are short enough for the model of GST to be reasonably consistent with the data). Only the two time-independent coherent error parameters in $\Gi$ and $\theta_z$ --- which is a summation of 6 Fourier coefficients --- are not fixed by this initial fit. Using MLE, we then fit this 8-parameter time-resolved model to the time-series data for all circuits with $L \leq 64$ and all the circuits where $\Gi$ is a germ with $L \leq 1024$. This allows us to extract information about the time-dependent parameter without polluting the results with data that does not fit this model.

\subsection{Estimated gate error rates}
In the main text, we report the diamond distance error rate ($\epsilon_{\diamond}$) of the estimated (time-resolved) gate process matrices in the two GST experiments. Table~\ref{table:error-rates} includes additional information on the gate error rates.  In addition to $\epsilon_{\diamond}$, this table reports the entanglement infidelity ($\epsilon_{\mathsf{f}}$) and spectral entanglement infidelity ($\epsilon_{\mathsf{spec},\mathsf{f}}$) for each gate. The diamond distance error rate for process matrix $\mathsf{G}$ is 
\begin{equation}
\epsilon_{\diamond}(\mathsf{G},\mathsf{G}_{\text{target}}) = \frac{1}{2}\| \mathsf{G} - \mathsf{G}_{\text{target}}\|_{\diamond},
\end{equation}
 where $\|\cdot\|_{\diamond}$ is the diamond norm \cite{watrous2005notes} and $\mathsf{G}_{\text{target}}$ is the target process matrix that $\mathsf{G}$ ideally should be. The entanglement infidelity is
 \begin{equation}
 \epsilon_{\textsf{f}}(\mathsf{G},\mathsf{G}_{\text{target}}) = 1- \langle \varphi |\left( \mathds{1} \otimes \mathsf{G}\mathsf{G}^{\dagger}_{\mathsf{target}}\right)(\varphi)|\varphi \rangle,
 \end{equation}
where $\varphi$ is any maximally entangled state \cite{nielsen2002simple}. Although the entanglement infidelity is defined with respect to a maximally entangled state, note that it is well-defined for one-qubit gates and it does not imply that any entangling operations are being performed. Another commonly used gate infidelity is the average gate infidelity ($\bar{\epsilon}$). The entanglement infidelity is related to the average gate infidelity via a linear rescaling. Specifically, $\epsilon_{\textsf{f}} = (d + 1) \bar{\epsilon}/2$  \cite{nielsen2002simple} where $d$ is the dimension, so here $d=2$. We use the entanglement infidelity because $\epsilon_{\diamond}=\epsilon_{\textsf{f}}$ for depolarizing errors, so the difference between $\epsilon_{\diamond}$ and $\epsilon_{\textsf{f}}$ is a rough measure of how structured the errors are, \eg, for coherent errors $\epsilon_{\diamond} >\epsilon_{\textsf{f}}$.

The spectral entanglement infidelity ($\epsilon_{\mathsf{spec},\mathsf{f}}$) is defined to be
\begin{equation}
\epsilon_{\mathsf{spec},\mathsf{f}} =  \epsilon_{\textsf{f}}\left(\mathsf{Spec}(\mathsf{G}),\mathsf{Spec}(\mathsf{G}_{\mathsf{target}})\right),
\end{equation}
where $\mathsf{Spec}(\mathsf{A})$ is the diagonal matrix that has the eigenvalues of the matrix $\mathsf{A}$ on the diagonal (and the relative ordering of the eigenvalues in the two matrices is chosen to minimize $\epsilon_{\mathsf{spec},\mathsf{f}}$). The spectral entanglement infidelity is an alternative to the standard entanglement infidelity that is invariant under transformation of the ``GST gauge'' (see Ref.~\cite{blume2016certifying} for discussions of this gauge).

Looking at Table~\ref{table:error-rates} we see that all three gates have a smaller diamond distance error rate in the second experiment, with a decrease from $\epsilon_{\diamond} \sim0.5 \%$ to $\epsilon_{\diamond} \sim0.05\%$. The infidelity also decreases for $\Gy$, but, in contrast, for $\Gi$ and $\Gx$ the infidelity increases from $\epsilon_{\textsf{f}} \sim 0.003\%$ to $\epsilon_{\textsf{f}} \sim 0.006\%$. The diamond distance error rate is sensitive at order $\theta$ to a coherent error with a rotation angle of $\theta$, whereas the infidelity is sensitive to coherent errors only at order $\theta^2$. So, a decrease in $\epsilon_{\diamond}$ and an increase in $\epsilon_{\textsf{f}}$ is a sign that the magnitude of the coherent errors (\ie, calibration errors) has been reduced in the second experiment, but that the rate of stochastic errors has slightly increased.

\section{Experiment Design}\label{app:3}
In this section we discuss how to design experiments to efficiently detect and characterize instability. This includes details on the relationship between the amount of data taken and the sensitivity of our methods.

\begin{figure}[t!]
\includegraphics[width=8cm]{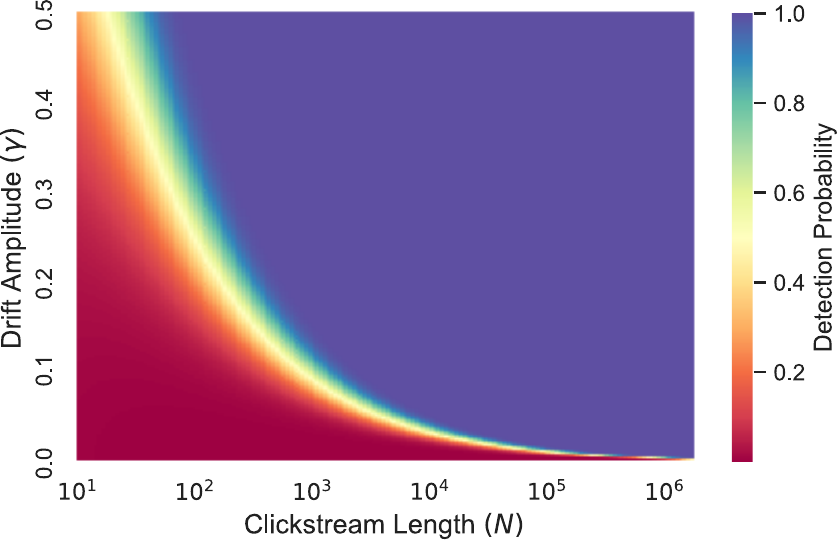}
\caption{{\bf Instability detection sensitivity}. The probability of detecting that $\vec{p}$ has a nonzero component at $\omega$, from a clickstream of length $N$ and with an amplitude of $\gamma$ at any non-zero basis frequency $\omega$, using tests of the power spectrum at a statistical significance of $5\%$. Here $\gamma$ corresponds to the amplitude in an unnormalized DCT expansion of $\vec{p}$ (\ie,  the expansion of Eq.~\eqref{si:dct-exp}), meaning that if this is the only non-zero frequency with a non-zero amplitude in this expansion of $\vec{p}$ then $\vec{p}$ oscillates between $\bar{p} \pm \gamma$.
}
\label{fig:stats-1}
\end{figure}

\subsection{How much data do we need?}
Consider looking for instability in a given set of circuits. For example, we could be running an algorithm, or syndrome extraction circuits, or the circuits from some benchmarking routine. Furthermore, assume that we are going to collect data by rastering through this set of circuits $N$ times. It then only remains to choose $N$, the length of the clickstream for each circuit. In this circumstance, it is important to understand how the sensitivity of our methods varies with $N$. 

Each circuit has some discrete-time probability trajectory $\vec{p}= (p_{0}, p_{1}, \dots, p_{N-1})$. We can always express this as a summation over the basis vectors of $F$. As we use the Type-II DCT for all of our explicit data analysis, we will use this basis for clarity. In this basis, we write
 \begin{equation}
  p_i = \bar{p} + \sum_{\omega=1}^{N-1} \gamma_{\omega} \cos \left(\frac{\omega \pi}{N} \left( i + \frac{1}{2}\right)\right), \label{si:dct-exp}
  \end{equation}
for some $\{\gamma_{\omega}\}$ amplitudes. We can use our statistical model for the frequency-domain data (see earlier) to approximately calculate the probability of successfully detecting that some $\gamma_{\omega}$ is non-zero, as a function of $\gamma_{\omega}$, $N$ and the test significance $\alpha$. In particular, the probability of detecting that $\vec{p}$ has a nonzero component at $\omega$, from a clickstream of length $N$ and with tests at a statistical significance of $\alpha$, is approximately
\begin{equation}
P_{\text{det}}(N, \gamma_{\omega},  \alpha) = 1 - \frac{1}{2} \left[ \text{erf} \left(\frac{\delta_{+}}{\sqrt{2}} \right) +  \text{erf} \left(\frac{\delta_{-}}{\sqrt{2}} \right)\right].
\end{equation}
Here, $\text{erf}(\cdot)$ is the error function,
\begin{equation}
\delta_{\pm} = \sqrt{T(\alpha,N)} \pm  \gamma_{\omega} \sqrt{\frac{N}{2\bar{p}(1-\bar{p})}},
\end{equation}
and $T(\alpha,N)$ is the power significance threshold for the hypothesis test at that frequency. In the case of testing a single clickstream, and splitting the test significance equally over all $N-1$ non-zero frequencies (as we do for all our data analysis, but which is not necessarily always desirable), this is given by
\begin{equation}
T(\alpha,N)=  \mathsf{CDF}^{-1}_1\left[1-\frac{\alpha}{(N-1)}\right],
\end{equation}
where $\mathsf{CDF}_k$ is the $\chi^2_k$ cumulative distribution function. If testing $C$ clickstreams, we let $\alpha \to \alpha/C$ in this equation.

\begin{figure}[t!]
\includegraphics[width=7.45cm]{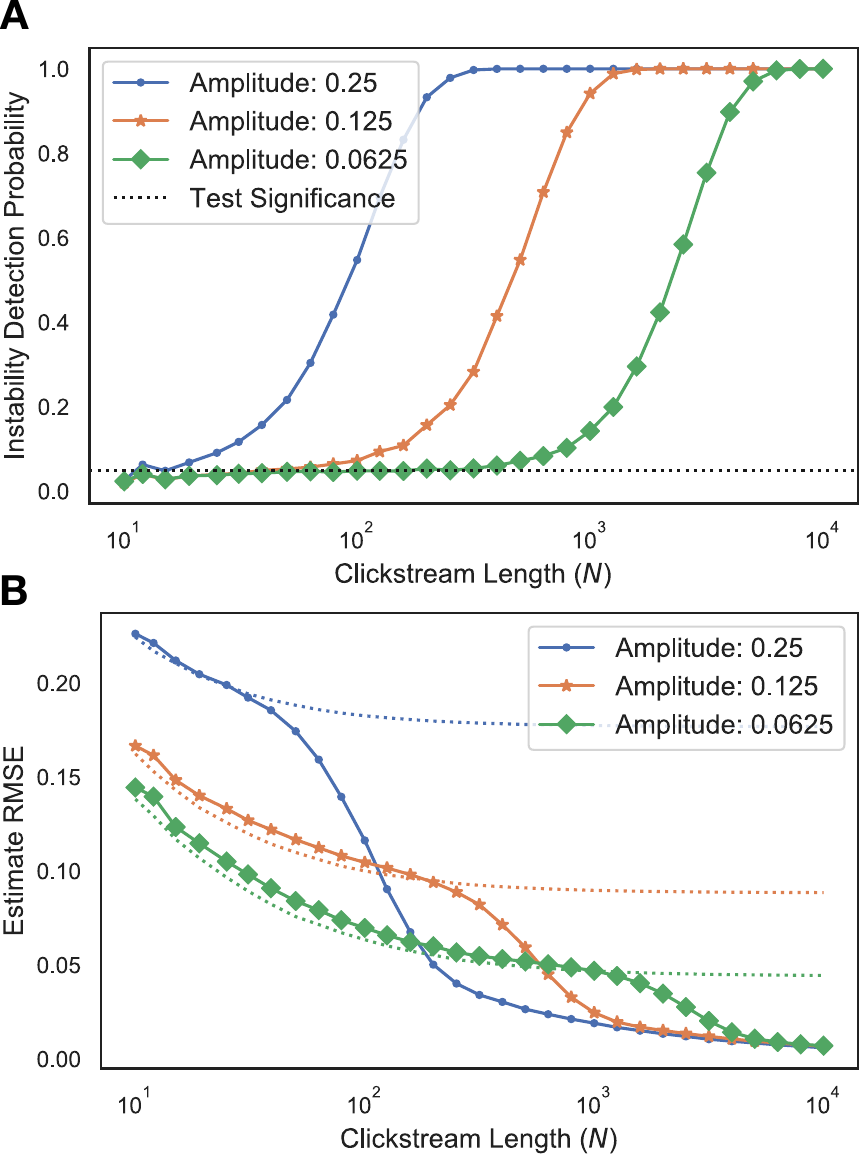}
\caption{{\bf Detection probability and estimation error}.~\textbf{A.}~The detection probability versus clickstream length for data sampled from a pure-tone probability $p(t) = 0.5 + \gamma \cos(\omega t + \tau)$ for three values of the amplitude $\gamma$, where $\omega$ and $\tau$ are such that this is a pure-tone in the DCT basis. Note that this is the probability of detecting significant power at any non-zero $\omega$, so the low-$N$ limit is the false detection probability of approximately the test significance $\alpha$, set here to $5\%$.~\textbf{B.}~The root mean squared error (RMSE) of our estimate of $p(t)$. The dotted lines show the RMSE of the constant-probability estimates, whereby we estimate $p(t)$ to be the mean of the clickstream. All quantities are calculated by averaging over the analysis results from $10^5$ simulated clickstream instances.
}
\label{fig:stats-2}
\end{figure}

Fig.~\ref{fig:stats-1} shows a heatmap of $P_{\text{det}}(N, \gamma, 0.05)$ as $N$ and $\gamma$ are varied. From this plot we can read off the clickstream length required to reliably detect an instability of a given amplitude. It shows that only moderate amounts of data are required to detect relatively small instabilities, \eg, an instability of amplitude $\gamma \approx 0.1$ can be reliably detected with $N \approx 1000$. This quantity of data is routinely taken to estimate static circuit outcome distributions.

As well as detecting instabilities in probability trajectories, our methods also estimate $p(t)$. To quantify the accuracy of our probability trajectory estimation we use simulations. We simulated clickstream data sampled from a pure-tone probability trajectory $\vec{p}(\gamma)$:
 \begin{equation}
  p_i(\gamma) = 0.5 + \gamma \cos \left(\frac{\omega \pi}{N} \left( i + \frac{1}{2}\right)\right).
  \end{equation}
Fig.~\ref{fig:stats-2} shows both the instability detection probability and the error in our estimator. Fig.~\ref{fig:stats-2}B shows the root mean squared error (RMSE) of our estimator for three values of $\gamma$ (unbroken lines), alongside the corresponding baselines set by estimating $p(t)$ as a constant given by the mean of the clickstream (dotted lines). This is the mean RMSE, obtained in $10^5$ simulations, and note that this simulation used the ``Fourier filter'' variant of our estimator (as it is faster to calculate than the MLE version). In the regime where $N$ is so low that instability is almost never detected, the RMSE of our estimator is essentially equal to the RMSE estimator of the baseline clickstream-mean estimate. This is because our estimate is the mean of the clickstream unless instability is detected. In fact, when $N$ is so low that instability is almost never detected (as shown in Fig.~\ref{fig:stats-2}A) our estimator actually performs worse than the baseline, on average. This is because in approximately $5\%$ of instances (for testing at 5\% statistical significance) our detection routine will spuriously detect that there is a component of $\vec{p}$ at some random frequency, and so we will spuriously add a component at that frequency into our estimator. Comparing Fig.~\ref{fig:stats-2}A and \ref{fig:stats-2}B, we see that the RMSE of our estimator rapidly improves once $N$ moves into the regime where instability at frequency $\omega$ is successfully detected with high probability. Once the detection probability saturates at essentially 1, the RMSE then follows $O(\nicefrac{1}{\sqrt{N}}$) scaling. This is because the RMSE in the estimate of each amplitude in our selected model is essentially shot-noise limited.

\begin{figure}[t!]
\includegraphics[width=7.6cm]{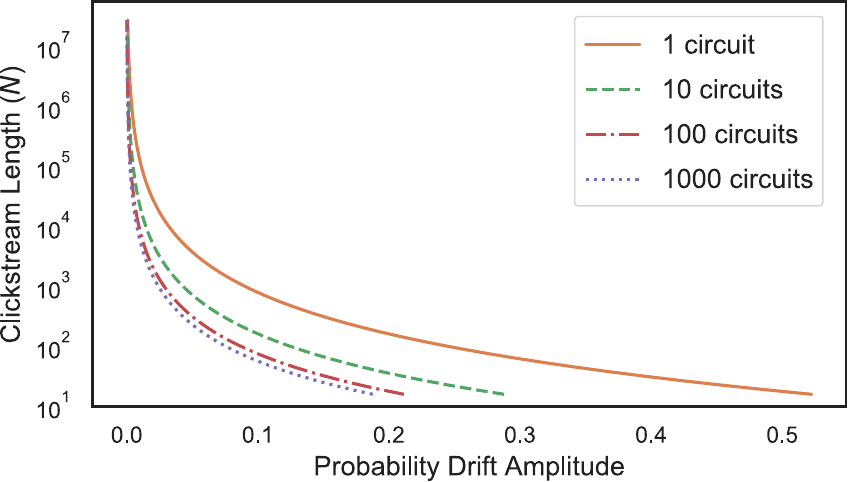}
\caption{{\bf Detection probability with circuit number}. The minimum clickstream length $N$ required for a 50\% instability detection probability when averaging spectra from 1, 10, 100 and 1000 circuits. The clickstreams for all circuits are sampled from the same probability trajectory $p(t) = 0.5 + \gamma \cos(\omega t + \tau)$, where $\omega$ and $\tau$ are such that this is a pure-tone in the DCT basis.
}
\label{fig:stats-3}
\end{figure}

So far we have considered how long a single clickstream needs to be to reliably detect instabilities of a given magnitude. But our analysis can also jointly test the data from multiple circuits, by calculating and then testing an averaged power spectrum. This allows for much greater sensitivity for a fixed $N$, where $N$ is the clickstream length for each circuit. Fig.~\ref{fig:stats-3} demonstrates this enhancement, by plotting the minimum $N$ required for a 50\% detection probability ($N_{\min}$) when averaging spectra from 1, 10, 100 and 1000 circuits. This is for the case of clickstreams that are all sampled from the same pure-tone $\vec{p}(\gamma)$. This case is chosen as it provides the maximal enhancement from averaging spectra, and  --- although it is unlikely that multiple circuits will have identical $p(t)$ --- there are many circumstances under which many or all of a set of circuits would contain drift at the same frequencies.

Fig.~\ref{fig:stats-3} shows that increasing the number of circuits provides a substantial reduction in $N_{\min}$. For example, if $\gamma=0.1$ then $N_{\min} \approx 100$ for 100 clickstreams whereas $N_{\min} \approx 1000$ for a single clickstream. But note that the sensitivity enhancement from increasing the number of circuits is smaller than that from increasing the clickstream length. As illustrated in the above example, to obtain a 50\% detection probability at $\gamma =0.1$ we need 10 times more bits in total in the 100 clickstream case than in the single clickstream case. Essentially, this is due to the difference between the $O(\nicefrac{1}{\sqrt{C}})$ shot-noise suppression obtained when averaging $C$ spectra with independent statistical fluctuations verses the linear growth with $N$ of the power in $\vec{p}$ (for $\vec{p}$ with an $N$-independent variance).

\subsection{Choosing the circuit repetition rate}
Our methods can be used for both dedicated drift characterization and for auxiliary data analysis. For dedicated drift characterization we need to carefully chose a set of quantum circuits that are as sensitive as possible to the specific parameters under study, as we discussed in the main text. Greater sensitivity results in large fluctuations in the circuits' outcome probabilities, for a given instability on the parameters of interest, meaning that fewer data are needed to reliably detect and characterize the drift --- as quantified above (see Fig.~\ref{fig:stats-1}-\ref{fig:stats-3}). But there is also another important aspect for designing  dedicated drift characterization experiments: the rate that the circuits need to be repeated. If the dedicated drift characterization circuits can be run sufficiently infrequently then it becomes possible to interleave them with application circuits. Furthermore, understanding the effect of reducing the repetition rate is important even in experiments run only for the purpose of drift characterization --- as we will often have to decide between running more circuits, or repeating fewer circuits at a faster rate. For example, in the case of both GST and Ramsey experiments, we must choose the maximum circuit length, and this directly impacts the rate at which we can raster through the full set of circuits.

Reducing the repetition rate of the drift characterization circuits --- \ie, increasing the time between each run of a circuit, $t_{\rm gap}$ --- will have two effects. Firstly, for a fixed experiment duration we will obtain fewer data points, \ie, the clickstreams will be shorter. The impact of the clickstream length on the sensitivity of our methods has already been discussed in detail earlier (see Fig.~\ref{fig:stats-1}-\ref{fig:stats-3}). But note that, for dedicated drift characterization, we are free to choose the circuits that maximally amplify the parameters we expect might be drifting --- so that the oscillations in the $p(t)$ need only be small if they are small in every possible circuit (under the assumption that we have correctly identified all the parameters that might be drifting). Therefore, relatively low $N$ will typically still allow for sensitive drift characterization. In the main text we demonstrated this, by discarding 80\% of our Ramsey data (giving $N=1200$) and showing that we still obtain a high-precision time-resolved estimate of the qubit detuning (see Fig.~\ref{fig:flowchart}E). Moreover, much more of that data could be discarded if we only wanted to detect drift and obtain a rough time-resolved estimate of the detuning.

The second impact of increasing $t_{\rm gap}$ is that the sensitivity to high-frequency drift is reduced. As with all signal analysis techniques that use regularly sampled data, there is a Nyquist limit --- \ie, oscillations at a frequency of $\nicefrac{1}{2t_{\rm gap}}$ are undetectable, and oscillations at higher frequencies are either aliased, or are undetectable if they are integer multiples of this frequency. The repetition rate should be chosen to maintain sensitivity to the frequency range of interest. As illustrated by our analysis of the Ramsey data, this means that we can have a slow repetition rate if there is only low-frequency drift.


\begin{thebibliography}{56}%
\makeatletter
\providecommand \@ifxundefined [1]{%
 \@ifx{#1\undefined}
}%
\providecommand \@ifnum [1]{%
 \ifnum #1\expandafter \@firstoftwo
 \else \expandafter \@secondoftwo
 \fi
}%
\providecommand \@ifx [1]{%
 \ifx #1\expandafter \@firstoftwo
 \else \expandafter \@secondoftwo
 \fi
}%
\providecommand \natexlab [1]{#1}%
\providecommand \enquote  [1]{``#1''}%
\providecommand \bibnamefont  [1]{#1}%
\providecommand \bibfnamefont [1]{#1}%
\providecommand \citenamefont [1]{#1}%
\providecommand \href@noop [0]{\@secondoftwo}%
\providecommand \href [0]{\begingroup \@sanitize@url \@href}%
\providecommand \@href[1]{\@@startlink{#1}\@@href}%
\providecommand \@@href[1]{\endgroup#1\@@endlink}%
\providecommand \@sanitize@url [0]{\catcode `\\12\catcode `\$12\catcode
  `\&12\catcode `\#12\catcode `\^12\catcode `\_12\catcode `\%12\relax}%
\providecommand \@@startlink[1]{}%
\providecommand \@@endlink[0]{}%
\providecommand \url  [0]{\begingroup\@sanitize@url \@url }%
\providecommand \@url [1]{\endgroup\@href {#1}{\urlprefix }}%
\providecommand \urlprefix  [0]{URL }%
\providecommand \Eprint [0]{\href }%
\providecommand \doibase [0]{http://dx.doi.org/}%
\providecommand \selectlanguage [0]{\@gobble}%
\providecommand \bibinfo  [0]{\@secondoftwo}%
\providecommand \bibfield  [0]{\@secondoftwo}%
\providecommand \translation [1]{[#1]}%
\providecommand \BibitemOpen [0]{}%
\providecommand \bibitemStop [0]{}%
\providecommand \bibitemNoStop [0]{.\EOS\space}%
\providecommand \EOS [0]{\spacefactor3000\relax}%
\providecommand \BibitemShut  [1]{\csname bibitem#1\endcsname}%
\let\auto@bib@innerbib\@empty
%</preamble>
\bibitem [{\citenamefont {Rol}\ \emph {et~al.}(2017)\citenamefont {Rol},
  \citenamefont {Bultink}, \citenamefont {O'Brien}, \citenamefont {de~Jong},
  \citenamefont {Theis}, \citenamefont {Fu}, \citenamefont {Luthi},
  \citenamefont {Vermeulen}, \citenamefont {de~Sterke}, \citenamefont {Bruno}
  \emph {et~al.}}]{rol2017restless}%
  \BibitemOpen
  \bibfield  {author} {\bibinfo {author} {\bibfnamefont {MA}~\bibnamefont
  {Rol}}, \bibinfo {author} {\bibfnamefont {CC}~\bibnamefont {Bultink}},
  \bibinfo {author} {\bibfnamefont {TE}~\bibnamefont {O'Brien}}, \bibinfo
  {author} {\bibfnamefont {SR}~\bibnamefont {de~Jong}}, \bibinfo {author}
  {\bibfnamefont {LS}~\bibnamefont {Theis}}, \bibinfo {author} {\bibfnamefont
  {X}~\bibnamefont {Fu}}, \bibinfo {author} {\bibfnamefont {F}~\bibnamefont
  {Luthi}}, \bibinfo {author} {\bibfnamefont {RFL}\ \bibnamefont {Vermeulen}},
  \bibinfo {author} {\bibfnamefont {JC}~\bibnamefont {de~Sterke}}, \bibinfo
  {author} {\bibfnamefont {A}~\bibnamefont {Bruno}},  \emph {et~al.},\
  }\bibfield  {title} {\enquote {\bibinfo {title} {Restless tuneup of
  high-fidelity qubit gates},}\ }\href
  {https://journals.aps.org/prapplied/abstract/10.1103/PhysRevApplied.7.041001}
  {\bibfield  {journal} {\bibinfo  {journal} {Phys. Rev. Appl.}\ }\textbf
  {\bibinfo {volume} {7}},\ \bibinfo {pages} {041001} (\bibinfo {year}
  {2017})}\BibitemShut {NoStop}%
\bibitem [{\citenamefont {Otterbach}\ \emph {et~al.}(2017)\citenamefont
  {Otterbach}, \citenamefont {Manenti}, \citenamefont {Alidoust}, \citenamefont
  {Bestwick}, \citenamefont {Block}, \citenamefont {Bloom}, \citenamefont
  {Caldwell}, \citenamefont {Didier}, \citenamefont {Fried}, \citenamefont
  {Hong} \emph {et~al.}}]{otterbach2017unsupervised}%
  \BibitemOpen
  \bibfield  {author} {\bibinfo {author} {\bibfnamefont {JS}~\bibnamefont
  {Otterbach}}, \bibinfo {author} {\bibfnamefont {R}~\bibnamefont {Manenti}},
  \bibinfo {author} {\bibfnamefont {N}~\bibnamefont {Alidoust}}, \bibinfo
  {author} {\bibfnamefont {A}~\bibnamefont {Bestwick}}, \bibinfo {author}
  {\bibfnamefont {M}~\bibnamefont {Block}}, \bibinfo {author} {\bibfnamefont
  {B}~\bibnamefont {Bloom}}, \bibinfo {author} {\bibfnamefont {S}~\bibnamefont
  {Caldwell}}, \bibinfo {author} {\bibfnamefont {N}~\bibnamefont {Didier}},
  \bibinfo {author} {\bibfnamefont {E~Schuyler}\ \bibnamefont {Fried}},
  \bibinfo {author} {\bibfnamefont {S}~\bibnamefont {Hong}},  \emph {et~al.},\
  }\bibfield  {title} {\enquote {\bibinfo {title} {Unsupervised machine
  learning on a hybrid quantum computer},}\ }\href
  {https://arxiv.org/abs/1712.05771} {\bibfield  {journal} {\bibinfo  {journal}
  {Preprint at http://arxiv.org/abs/1712.05771}\ } (\bibinfo {year} {2017})}\BibitemShut
  {NoStop}%
\bibitem [{\citenamefont {Friis}\ \emph {et~al.}(2018)\citenamefont {Friis},
  \citenamefont {Marty}, \citenamefont {Maier}, \citenamefont {Hempel},
  \citenamefont {Holz{\"a}pfel}, \citenamefont {Jurcevic}, \citenamefont
  {Plenio}, \citenamefont {Huber}, \citenamefont {Roos}, \citenamefont {Blatt}
  \emph {et~al.}}]{friis2018observation}%
  \BibitemOpen
  \bibfield  {author} {\bibinfo {author} {\bibfnamefont {Nicolai}\ \bibnamefont
  {Friis}}, \bibinfo {author} {\bibfnamefont {Oliver}\ \bibnamefont {Marty}},
  \bibinfo {author} {\bibfnamefont {Christine}\ \bibnamefont {Maier}}, \bibinfo
  {author} {\bibfnamefont {Cornelius}\ \bibnamefont {Hempel}}, \bibinfo
  {author} {\bibfnamefont {Milan}\ \bibnamefont {Holz{\"a}pfel}}, \bibinfo
  {author} {\bibfnamefont {Petar}\ \bibnamefont {Jurcevic}}, \bibinfo {author}
  {\bibfnamefont {Martin~B}\ \bibnamefont {Plenio}}, \bibinfo {author}
  {\bibfnamefont {Marcus}\ \bibnamefont {Huber}}, \bibinfo {author}
  {\bibfnamefont {Christian}\ \bibnamefont {Roos}}, \bibinfo {author}
  {\bibfnamefont {Rainer}\ \bibnamefont {Blatt}},  \emph {et~al.},\ }\bibfield
  {title} {\enquote {\bibinfo {title} {Observation of entangled states of a
  fully controlled 20-qubit system},}\ }\href
  {https://journals.aps.org/prx/abstract/10.1103/PhysRevX.8.021012} {\bibfield
  {journal} {\bibinfo  {journal} {Phys. Rev. X}\ }\textbf {\bibinfo {volume}
  {8}},\ \bibinfo {pages} {021012} (\bibinfo {year} {2018})}\BibitemShut
  {NoStop}%
\bibitem [{\citenamefont {Arute}\ \emph {et~al.}(2019)\citenamefont {Arute},
  \citenamefont {Arya}, \citenamefont {Babbush}, \citenamefont {Bacon},
  \citenamefont {Bardin}, \citenamefont {Barends}, \citenamefont {Biswas},
  \citenamefont {Boixo}, \citenamefont {Brandao}, \citenamefont {Buell} \emph
  {et~al.}}]{arute2019quantum}%
  \BibitemOpen
  \bibfield  {author} {\bibinfo {author} {\bibfnamefont {Frank}\ \bibnamefont
  {Arute}}, \bibinfo {author} {\bibfnamefont {Kunal}\ \bibnamefont {Arya}},
  \bibinfo {author} {\bibfnamefont {Ryan}\ \bibnamefont {Babbush}}, \bibinfo
  {author} {\bibfnamefont {Dave}\ \bibnamefont {Bacon}}, \bibinfo {author}
  {\bibfnamefont {Joseph~C}\ \bibnamefont {Bardin}}, \bibinfo {author}
  {\bibfnamefont {Rami}\ \bibnamefont {Barends}}, \bibinfo {author}
  {\bibfnamefont {Rupak}\ \bibnamefont {Biswas}}, \bibinfo {author}
  {\bibfnamefont {Sergio}\ \bibnamefont {Boixo}}, \bibinfo {author}
  {\bibfnamefont {Fernando~GSL}\ \bibnamefont {Brandao}}, \bibinfo {author}
  {\bibfnamefont {David~A}\ \bibnamefont {Buell}},  \emph {et~al.},\ }\bibfield
   {title} {\enquote {\bibinfo {title} {Quantum supremacy using a programmable
  superconducting processor},}\ }\href
  {https://www.nature.com/articles/s41586-019-1666-5} {\bibfield  {journal}
  {\bibinfo  {journal} {Nature}\ }\textbf {\bibinfo {volume} {574}},\ \bibinfo
  {pages} {505--510} (\bibinfo {year} {2019})}\BibitemShut {NoStop}%
\bibitem [{\citenamefont {Blume-Kohout}\ \emph {et~al.}(2017)\citenamefont
  {Blume-Kohout}, \citenamefont {Gamble}, \citenamefont {Nielsen},
  \citenamefont {Rudinger}, \citenamefont {Mizrahi}, \citenamefont {Fortier},\
  and\ \citenamefont {Maunz}}]{blume2016certifying}%
  \BibitemOpen
  \bibfield  {author} {\bibinfo {author} {\bibfnamefont {Robin}\ \bibnamefont
  {Blume-Kohout}}, \bibinfo {author} {\bibfnamefont {John~King}\ \bibnamefont
  {Gamble}}, \bibinfo {author} {\bibfnamefont {Erik}\ \bibnamefont {Nielsen}},
  \bibinfo {author} {\bibfnamefont {Kenneth}\ \bibnamefont {Rudinger}},
  \bibinfo {author} {\bibfnamefont {Jonathan}\ \bibnamefont {Mizrahi}},
  \bibinfo {author} {\bibfnamefont {Kevin}\ \bibnamefont {Fortier}}, \ and\
  \bibinfo {author} {\bibfnamefont {Peter}\ \bibnamefont {Maunz}},\ }\bibfield
  {title} {\enquote {\bibinfo {title} {Demonstration of qubit operations below
  a rigorous fault tolerance threshold with gate set tomography},}\ }\href
  {https://www.nature.com/articles/ncomms14485} {\bibfield  {journal} {\bibinfo
   {journal} {Nat. Commun.}\ }\textbf {\bibinfo {volume} {8}},\ \bibinfo
  {pages} {14485} (\bibinfo {year} {2017})}\BibitemShut {NoStop}%
\bibitem [{\citenamefont {Merkel}\ \emph {et~al.}(2013)\citenamefont {Merkel},
  \citenamefont {Gambetta}, \citenamefont {Smolin}, \citenamefont {Poletto},
  \citenamefont {C{\'o}rcoles}, \citenamefont {Johnson}, \citenamefont {Ryan},\
  and\ \citenamefont {Steffen}}]{merkel2013self}%
  \BibitemOpen
  \bibfield  {author} {\bibinfo {author} {\bibfnamefont {Seth~T}\ \bibnamefont
  {Merkel}}, \bibinfo {author} {\bibfnamefont {Jay~M}\ \bibnamefont
  {Gambetta}}, \bibinfo {author} {\bibfnamefont {John~A}\ \bibnamefont
  {Smolin}}, \bibinfo {author} {\bibfnamefont {Stefano}\ \bibnamefont
  {Poletto}}, \bibinfo {author} {\bibfnamefont {Antonio~D}\ \bibnamefont
  {C{\'o}rcoles}}, \bibinfo {author} {\bibfnamefont {Blake~R}\ \bibnamefont
  {Johnson}}, \bibinfo {author} {\bibfnamefont {Colm~A}\ \bibnamefont {Ryan}},
  \ and\ \bibinfo {author} {\bibfnamefont {Matthias}\ \bibnamefont {Steffen}},\
  }\bibfield  {title} {\enquote {\bibinfo {title} {Self-consistent quantum
  process tomography},}\ }\href
  {https://journals.aps.org/pra/abstract/10.1103/PhysRevA.87.062119} {\bibfield
   {journal} {\bibinfo  {journal} {Phys. Rev. A}\ }\textbf {\bibinfo {volume}
  {87}},\ \bibinfo {pages} {062119} (\bibinfo {year} {2013})}\BibitemShut
  {NoStop}%
\bibitem [{\citenamefont {Knill}\ \emph {et~al.}(2008)\citenamefont {Knill},
  \citenamefont {Leibfried}, \citenamefont {Reichle}, \citenamefont {Britton},
  \citenamefont {Blakestad}, \citenamefont {Jost}, \citenamefont {Langer},
  \citenamefont {Ozeri}, \citenamefont {Seidelin},\ and\ \citenamefont
  {Wineland}}]{knill2008randomized}%
  \BibitemOpen
  \bibfield  {author} {\bibinfo {author} {\bibfnamefont {Emanuel}\ \bibnamefont
  {Knill}}, \bibinfo {author} {\bibfnamefont {D}~\bibnamefont {Leibfried}},
  \bibinfo {author} {\bibfnamefont {R}~\bibnamefont {Reichle}}, \bibinfo
  {author} {\bibfnamefont {J}~\bibnamefont {Britton}}, \bibinfo {author}
  {\bibfnamefont {RB}~\bibnamefont {Blakestad}}, \bibinfo {author}
  {\bibfnamefont {JD}~\bibnamefont {Jost}}, \bibinfo {author} {\bibfnamefont
  {C}~\bibnamefont {Langer}}, \bibinfo {author} {\bibfnamefont {R}~\bibnamefont
  {Ozeri}}, \bibinfo {author} {\bibfnamefont {S}~\bibnamefont {Seidelin}}, \
  and\ \bibinfo {author} {\bibfnamefont {DJ}~\bibnamefont {Wineland}},\
  }\bibfield  {title} {\enquote {\bibinfo {title} {Randomized benchmarking of
  quantum gates},}\ }\href
  {https://journals.aps.org/pra/abstract/10.1103/PhysRevA.77.012307} {\bibfield
   {journal} {\bibinfo  {journal} {Phys. Rev. A}\ }\textbf {\bibinfo {volume}
  {77}},\ \bibinfo {pages} {012307} (\bibinfo {year} {2008})}\BibitemShut
  {NoStop}%
\bibitem [{\citenamefont {Magesan}\ \emph {et~al.}(2011)\citenamefont
  {Magesan}, \citenamefont {Gambetta},\ and\ \citenamefont
  {Emerson}}]{magesan2011scalable}%
  \BibitemOpen
  \bibfield  {author} {\bibinfo {author} {\bibfnamefont {Easwar}\ \bibnamefont
  {Magesan}}, \bibinfo {author} {\bibfnamefont {Jay~M}\ \bibnamefont
  {Gambetta}}, \ and\ \bibinfo {author} {\bibfnamefont {Joseph}\ \bibnamefont
  {Emerson}},\ }\bibfield  {title} {\enquote {\bibinfo {title} {Scalable and
  robust randomized benchmarking of quantum processes},}\ }\href
  {https://journals.aps.org/prl/abstract/10.1103/PhysRevLett.106.180504}
  {\bibfield  {journal} {\bibinfo  {journal} {Phys. Rev. Lett.}\ }\textbf
  {\bibinfo {volume} {106}},\ \bibinfo {pages} {180504} (\bibinfo {year}
  {2011})}\BibitemShut {NoStop}%
\bibitem [{\citenamefont {Proctor}\ \emph {et~al.}(2019)\citenamefont
  {Proctor}, \citenamefont {Carignan-Dugas}, \citenamefont {Rudinger},
  \citenamefont {Nielsen}, \citenamefont {Blume-Kohout},\ and\ \citenamefont
  {Young}}]{proctor2018direct}%
  \BibitemOpen
  \bibfield  {author} {\bibinfo {author} {\bibfnamefont {Timothy~J}\
  \bibnamefont {Proctor}}, \bibinfo {author} {\bibfnamefont {Arnaud}\
  \bibnamefont {Carignan-Dugas}}, \bibinfo {author} {\bibfnamefont {Kenneth}\
  \bibnamefont {Rudinger}}, \bibinfo {author} {\bibfnamefont {Erik}\
  \bibnamefont {Nielsen}}, \bibinfo {author} {\bibfnamefont {Robin}\
  \bibnamefont {Blume-Kohout}}, \ and\ \bibinfo {author} {\bibfnamefont
  {Kevin}\ \bibnamefont {Young}},\ }\bibfield  {title} {\enquote {\bibinfo
  {title} {Direct randomized benchmarking for multiqubit devices},}\ }\href
  {https://journals.aps.org/prl/abstract/10.1103/PhysRevLett.123.030503}
  {\bibfield  {journal} {\bibinfo  {journal} {Phys. Rev. Lett.}\ }\textbf
  {\bibinfo {volume} {123}} (\bibinfo {year} {2019})}\BibitemShut {NoStop}%
\bibitem [{\citenamefont {Magesan}\ \emph {et~al.}(2012)\citenamefont
  {Magesan}, \citenamefont {Gambetta}, \citenamefont {Johnson}, \citenamefont
  {Ryan}, \citenamefont {Chow}, \citenamefont {Merkel}, \citenamefont
  {da~Silva}, \citenamefont {Keefe}, \citenamefont {Rothwell}, \citenamefont
  {Ohki} \emph {et~al.}}]{magesan2012efficient}%
  \BibitemOpen
  \bibfield  {author} {\bibinfo {author} {\bibfnamefont {Easwar}\ \bibnamefont
  {Magesan}}, \bibinfo {author} {\bibfnamefont {Jay~M}\ \bibnamefont
  {Gambetta}}, \bibinfo {author} {\bibfnamefont {Blake~R}\ \bibnamefont
  {Johnson}}, \bibinfo {author} {\bibfnamefont {Colm~A}\ \bibnamefont {Ryan}},
  \bibinfo {author} {\bibfnamefont {Jerry~M}\ \bibnamefont {Chow}}, \bibinfo
  {author} {\bibfnamefont {Seth~T}\ \bibnamefont {Merkel}}, \bibinfo {author}
  {\bibfnamefont {Marcus~P}\ \bibnamefont {da~Silva}}, \bibinfo {author}
  {\bibfnamefont {George~A}\ \bibnamefont {Keefe}}, \bibinfo {author}
  {\bibfnamefont {Mary~B}\ \bibnamefont {Rothwell}}, \bibinfo {author}
  {\bibfnamefont {Thomas~A}\ \bibnamefont {Ohki}},  \emph {et~al.},\ }\bibfield
   {title} {\enquote {\bibinfo {title} {Efficient measurement of quantum gate
  error by interleaved randomized benchmarking},}\ }\href
  {https://journals.aps.org/prl/abstract/10.1103/PhysRevLett.109.080505}
  {\bibfield  {journal} {\bibinfo  {journal} {Phys. Rev. Lett.}\ }\textbf
  {\bibinfo {volume} {109}},\ \bibinfo {pages} {080505} (\bibinfo {year}
  {2012})}\BibitemShut {NoStop}%
\bibitem [{\citenamefont {Cross}\ \emph {et~al.}(2016)\citenamefont {Cross},
  \citenamefont {Magesan}, \citenamefont {Bishop}, \citenamefont {Smolin},\
  and\ \citenamefont {Gambetta}}]{cross2016scalable}%
  \BibitemOpen
  \bibfield  {author} {\bibinfo {author} {\bibfnamefont {Andrew~W}\
  \bibnamefont {Cross}}, \bibinfo {author} {\bibfnamefont {Easwar}\
  \bibnamefont {Magesan}}, \bibinfo {author} {\bibfnamefont {Lev~S}\
  \bibnamefont {Bishop}}, \bibinfo {author} {\bibfnamefont {John~A}\
  \bibnamefont {Smolin}}, \ and\ \bibinfo {author} {\bibfnamefont {Jay~M}\
  \bibnamefont {Gambetta}},\ }\bibfield  {title} {\enquote {\bibinfo {title}
  {Scalable randomised benchmarking of non-clifford gates},}\ }\href
  {http://www.nature.com/articles/npjqi201612} {\bibfield  {journal} {\bibinfo
  {journal} {NPJ Quantum Inf.}\ }\textbf {\bibinfo {volume} {2}},\ \bibinfo
  {pages} {16012} (\bibinfo {year} {2016})}\BibitemShut {NoStop}%
\bibitem [{\citenamefont {Barends}\ \emph {et~al.}(2014)\citenamefont
  {Barends}, \citenamefont {Kelly}, \citenamefont {Veitia}, \citenamefont
  {Megrant}, \citenamefont {Fowler}, \citenamefont {Campbell}, \citenamefont
  {Chen}, \citenamefont {Chen}, \citenamefont {Chiaro}, \citenamefont
  {Dunsworth} \emph {et~al.}}]{barends2014rolling}%
  \BibitemOpen
  \bibfield  {author} {\bibinfo {author} {\bibfnamefont {R}~\bibnamefont
  {Barends}}, \bibinfo {author} {\bibfnamefont {J}~\bibnamefont {Kelly}},
  \bibinfo {author} {\bibfnamefont {A}~\bibnamefont {Veitia}}, \bibinfo
  {author} {\bibfnamefont {A}~\bibnamefont {Megrant}}, \bibinfo {author}
  {\bibfnamefont {AG}~\bibnamefont {Fowler}}, \bibinfo {author} {\bibfnamefont
  {B}~\bibnamefont {Campbell}}, \bibinfo {author} {\bibfnamefont
  {Y}~\bibnamefont {Chen}}, \bibinfo {author} {\bibfnamefont {Z}~\bibnamefont
  {Chen}}, \bibinfo {author} {\bibfnamefont {B}~\bibnamefont {Chiaro}},
  \bibinfo {author} {\bibfnamefont {A}~\bibnamefont {Dunsworth}},  \emph
  {et~al.},\ }\bibfield  {title} {\enquote {\bibinfo {title} {Rolling quantum
  dice with a superconducting qubit},}\ }\href
  {https://journals.aps.org/pra/abstract/10.1103/PhysRevA.90.030303} {\bibfield
   {journal} {\bibinfo  {journal} {Phys. Rev. A}\ }\textbf {\bibinfo {volume}
  {90}},\ \bibinfo {pages} {030303} (\bibinfo {year} {2014})}\BibitemShut
  {NoStop}%
\bibitem [{\citenamefont {Carignan-Dugas}\ \emph {et~al.}(2015)\citenamefont
  {Carignan-Dugas}, \citenamefont {Wallman},\ and\ \citenamefont
  {Emerson}}]{carignan2015characterizing}%
  \BibitemOpen
  \bibfield  {author} {\bibinfo {author} {\bibfnamefont {Arnaud}\ \bibnamefont
  {Carignan-Dugas}}, \bibinfo {author} {\bibfnamefont {Joel~J}\ \bibnamefont
  {Wallman}}, \ and\ \bibinfo {author} {\bibfnamefont {Joseph}\ \bibnamefont
  {Emerson}},\ }\bibfield  {title} {\enquote {\bibinfo {title} {Characterizing
  universal gate sets via dihedral benchmarking},}\ }\href
  {https://journals.aps.org/pra/abstract/10.1103/PhysRevA.92.060302} {\bibfield
   {journal} {\bibinfo  {journal} {Phys. Rev. A}\ }\textbf {\bibinfo {volume}
  {92}},\ \bibinfo {pages} {060302} (\bibinfo {year} {2015})}\BibitemShut
  {NoStop}%
\bibitem [{\citenamefont {Gambetta}\ \emph {et~al.}(2012)\citenamefont
  {Gambetta}, \citenamefont {C{\'o}rcoles}, \citenamefont {Merkel},
  \citenamefont {Johnson}, \citenamefont {Smolin}, \citenamefont {Chow},
  \citenamefont {Ryan}, \citenamefont {Rigetti}, \citenamefont {Poletto},
  \citenamefont {Ohki} \emph {et~al.}}]{gambetta2012characterization}%
  \BibitemOpen
  \bibfield  {author} {\bibinfo {author} {\bibfnamefont {Jay~M}\ \bibnamefont
  {Gambetta}}, \bibinfo {author} {\bibfnamefont {AD}~\bibnamefont
  {C{\'o}rcoles}}, \bibinfo {author} {\bibfnamefont {Seth~T}\ \bibnamefont
  {Merkel}}, \bibinfo {author} {\bibfnamefont {Blake~R}\ \bibnamefont
  {Johnson}}, \bibinfo {author} {\bibfnamefont {John~A}\ \bibnamefont
  {Smolin}}, \bibinfo {author} {\bibfnamefont {Jerry~M}\ \bibnamefont {Chow}},
  \bibinfo {author} {\bibfnamefont {Colm~A}\ \bibnamefont {Ryan}}, \bibinfo
  {author} {\bibfnamefont {Chad}\ \bibnamefont {Rigetti}}, \bibinfo {author}
  {\bibfnamefont {S}~\bibnamefont {Poletto}}, \bibinfo {author} {\bibfnamefont
  {Thomas~A}\ \bibnamefont {Ohki}},  \emph {et~al.},\ }\bibfield  {title}
  {\enquote {\bibinfo {title} {Characterization of addressability by
  simultaneous randomized benchmarking},}\ }\href
  {https://journals.aps.org/prl/abstract/10.1103/PhysRevLett.109.240504}
  {\bibfield  {journal} {\bibinfo  {journal} {Phys. Rev. Lett.}\ }\textbf
  {\bibinfo {volume} {109}},\ \bibinfo {pages} {240504} (\bibinfo {year}
  {2012})}\BibitemShut {NoStop}%
\bibitem [{\citenamefont {Kimmel}\ \emph {et~al.}(2015)\citenamefont {Kimmel},
  \citenamefont {Low},\ and\ \citenamefont {Yoder}}]{kimmel2015robust}%
  \BibitemOpen
  \bibfield  {author} {\bibinfo {author} {\bibfnamefont {Shelby}\ \bibnamefont
  {Kimmel}}, \bibinfo {author} {\bibfnamefont {Guang~Hao}\ \bibnamefont {Low}},
  \ and\ \bibinfo {author} {\bibfnamefont {Theodore~J}\ \bibnamefont {Yoder}},\
  }\bibfield  {title} {\enquote {\bibinfo {title} {Robust calibration of a
  universal single-qubit gate set via robust phase estimation},}\ }\href
  {https://journals.aps.org/pra/abstract/10.1103/PhysRevA.92.062315} {\bibfield
   {journal} {\bibinfo  {journal} {Phys. Rev. A}\ }\textbf {\bibinfo {volume}
  {92}},\ \bibinfo {pages} {062315} (\bibinfo {year} {2015})}\BibitemShut
  {NoStop}%
\bibitem [{\citenamefont {Dehollain}\ \emph {et~al.}(2016)\citenamefont
  {Dehollain}, \citenamefont {Muhonen}, \citenamefont {Blume-Kohout},
  \citenamefont {Rudinger}, \citenamefont {Gamble}, \citenamefont {Nielsen},
  \citenamefont {Laucht}, \citenamefont {Simmons}, \citenamefont {Kalra},
  \citenamefont {Dzurak} \emph {et~al.}}]{dehollain2016optimization}%
  \BibitemOpen
  \bibfield  {author} {\bibinfo {author} {\bibfnamefont {Juan~P}\ \bibnamefont
  {Dehollain}}, \bibinfo {author} {\bibfnamefont {Juha~T}\ \bibnamefont
  {Muhonen}}, \bibinfo {author} {\bibfnamefont {Robin}\ \bibnamefont
  {Blume-Kohout}}, \bibinfo {author} {\bibfnamefont {Kenneth~M}\ \bibnamefont
  {Rudinger}}, \bibinfo {author} {\bibfnamefont {John~King}\ \bibnamefont
  {Gamble}}, \bibinfo {author} {\bibfnamefont {Erik}\ \bibnamefont {Nielsen}},
  \bibinfo {author} {\bibfnamefont {Arne}\ \bibnamefont {Laucht}}, \bibinfo
  {author} {\bibfnamefont {Stephanie}\ \bibnamefont {Simmons}}, \bibinfo
  {author} {\bibfnamefont {Rachpon}\ \bibnamefont {Kalra}}, \bibinfo {author}
  {\bibfnamefont {Andrew~S}\ \bibnamefont {Dzurak}},  \emph {et~al.},\
  }\bibfield  {title} {\enquote {\bibinfo {title} {Optimization of a
  solid-state electron spin qubit using gate set tomography},}\ }\href
  {http://iopscience.iop.org/article/10.1088/1367-2630/18/10/103018} {\bibfield
   {journal} {\bibinfo  {journal} {New J. Phys.}\ }\textbf {\bibinfo {volume}
  {18}},\ \bibinfo {pages} {103018} (\bibinfo {year} {2016})}\BibitemShut
  {NoStop}%
\bibitem [{\citenamefont {Epstein}\ \emph {et~al.}(2014)\citenamefont
  {Epstein}, \citenamefont {Cross}, \citenamefont {Magesan},\ and\
  \citenamefont {Gambetta}}]{epstein2014investigating}%
  \BibitemOpen
  \bibfield  {author} {\bibinfo {author} {\bibfnamefont {Jeffrey~M}\
  \bibnamefont {Epstein}}, \bibinfo {author} {\bibfnamefont {Andrew~W}\
  \bibnamefont {Cross}}, \bibinfo {author} {\bibfnamefont {Easwar}\
  \bibnamefont {Magesan}}, \ and\ \bibinfo {author} {\bibfnamefont {Jay~M}\
  \bibnamefont {Gambetta}},\ }\bibfield  {title} {\enquote {\bibinfo {title}
  {Investigating the limits of randomized benchmarking protocols},}\ }\href
  {https://journals.aps.org/pra/abstract/10.1103/PhysRevA.89.062321} {\bibfield
   {journal} {\bibinfo  {journal} {Phys. Rev. A}\ }\textbf {\bibinfo {volume}
  {89}},\ \bibinfo {pages} {062321} (\bibinfo {year} {2014})}\BibitemShut
  {NoStop}%
\bibitem [{\citenamefont {van Enk}\ and\ \citenamefont
  {Blume-Kohout}(2013)}]{van2013quantum}%
  \BibitemOpen
  \bibfield  {author} {\bibinfo {author} {\bibfnamefont {Steven~J}\
  \bibnamefont {van Enk}}\ and\ \bibinfo {author} {\bibfnamefont {Robin}\
  \bibnamefont {Blume-Kohout}},\ }\bibfield  {title} {\enquote {\bibinfo
  {title} {When quantum tomography goes wrong: drift of quantum sources and
  other errors},}\ }\href
  {http://iopscience.iop.org/article/10.1088/1367-2630/15/2/025024} {\bibfield
  {journal} {\bibinfo  {journal} {New J. Phys.}\ }\textbf {\bibinfo {volume}
  {15}},\ \bibinfo {pages} {025024} (\bibinfo {year} {2013})}\BibitemShut
  {NoStop}%
\bibitem [{\citenamefont {Fong}\ and\ \citenamefont
  {Merkel}(2017)}]{fong2017randomized}%
  \BibitemOpen
  \bibfield  {author} {\bibinfo {author} {\bibfnamefont {Bryan~H}\ \bibnamefont
  {Fong}}\ and\ \bibinfo {author} {\bibfnamefont {Seth~T}\ \bibnamefont
  {Merkel}},\ }\bibfield  {title} {\enquote {\bibinfo {title} {Randomized
  benchmarking, correlated noise, and ising models},}\ }\href
  {https://arxiv.org/abs/1703.09747v1} {\bibfield  {journal} {\bibinfo
  {journal} {Preprint at http://arxiv.org/abs/1703.09747}\ } (\bibinfo {year}
  {2017})}\BibitemShut {NoStop}%
\bibitem [{\citenamefont {Chow}\ \emph {et~al.}(2009)\citenamefont {Chow},
  \citenamefont {Gambetta}, \citenamefont {Tornberg}, \citenamefont {Koch},
  \citenamefont {Bishop}, \citenamefont {Houck}, \citenamefont {Johnson},
  \citenamefont {Frunzio}, \citenamefont {Girvin},\ and\ \citenamefont
  {Schoelkopf}}]{chow2009randomized}%
  \BibitemOpen
  \bibfield  {author} {\bibinfo {author} {\bibfnamefont {JM}~\bibnamefont
  {Chow}}, \bibinfo {author} {\bibfnamefont {Jay~M}\ \bibnamefont {Gambetta}},
  \bibinfo {author} {\bibfnamefont {Lars}\ \bibnamefont {Tornberg}}, \bibinfo
  {author} {\bibfnamefont {Jens}\ \bibnamefont {Koch}}, \bibinfo {author}
  {\bibfnamefont {Lev~S}\ \bibnamefont {Bishop}}, \bibinfo {author}
  {\bibfnamefont {Andrew~A}\ \bibnamefont {Houck}}, \bibinfo {author}
  {\bibfnamefont {BR}~\bibnamefont {Johnson}}, \bibinfo {author} {\bibfnamefont
  {L}~\bibnamefont {Frunzio}}, \bibinfo {author} {\bibfnamefont {Steven~M}\
  \bibnamefont {Girvin}}, \ and\ \bibinfo {author} {\bibfnamefont {Robert~J}\
  \bibnamefont {Schoelkopf}},\ }\bibfield  {title} {\enquote {\bibinfo {title}
  {Randomized benchmarking and process tomography for gate errors in a
  solid-state qubit},}\ }\href
  {https://journals.aps.org/prl/abstract/10.1103/PhysRevLett.102.090502}
  {\bibfield  {journal} {\bibinfo  {journal} {Phys. Rev. Lett.}\ }\textbf
  {\bibinfo {volume} {102}},\ \bibinfo {pages} {090502} (\bibinfo {year}
  {2009})}\BibitemShut {NoStop}%
\bibitem [{\citenamefont {Fogarty}\ \emph {et~al.}(2015)\citenamefont
  {Fogarty}, \citenamefont {Veldhorst}, \citenamefont {Harper}, \citenamefont
  {Yang}, \citenamefont {Bartlett}, \citenamefont {Flammia},\ and\
  \citenamefont {Dzurak}}]{fogarty2015nonexponential}%
  \BibitemOpen
  \bibfield  {author} {\bibinfo {author} {\bibfnamefont {MA}~\bibnamefont
  {Fogarty}}, \bibinfo {author} {\bibfnamefont {M}~\bibnamefont {Veldhorst}},
  \bibinfo {author} {\bibfnamefont {R}~\bibnamefont {Harper}}, \bibinfo
  {author} {\bibfnamefont {CH}~\bibnamefont {Yang}}, \bibinfo {author}
  {\bibfnamefont {SD}~\bibnamefont {Bartlett}}, \bibinfo {author}
  {\bibfnamefont {ST}~\bibnamefont {Flammia}}, \ and\ \bibinfo {author}
  {\bibfnamefont {AS}~\bibnamefont {Dzurak}},\ }\bibfield  {title} {\enquote
  {\bibinfo {title} {Nonexponential fidelity decay in randomized benchmarking
  with low-frequency noise},}\ }\href
  {https://journals.aps.org/pra/abstract/10.1103/PhysRevA.92.022326} {\bibfield
   {journal} {\bibinfo  {journal} {Phys. Rev. A}\ }\textbf {\bibinfo {volume}
  {92}},\ \bibinfo {pages} {022326} (\bibinfo {year} {2015})}\BibitemShut
  {NoStop}%
\bibitem [{\citenamefont {Wan}\ \emph {et~al.}(2019)\citenamefont {Wan},
  \citenamefont {Kienzler}, \citenamefont {Erickson}, \citenamefont {Mayer},
  \citenamefont {Tan}, \citenamefont {Wu}, \citenamefont {Vasconcelos},
  \citenamefont {Glancy}, \citenamefont {Knill}, \citenamefont {Wineland} \emph
  {et~al.}}]{wan2019quantum}%
  \BibitemOpen
  \bibfield  {author} {\bibinfo {author} {\bibfnamefont {Yong}\ \bibnamefont
  {Wan}}, \bibinfo {author} {\bibfnamefont {Daniel}\ \bibnamefont {Kienzler}},
  \bibinfo {author} {\bibfnamefont {Stephen~D}\ \bibnamefont {Erickson}},
  \bibinfo {author} {\bibfnamefont {Karl~H}\ \bibnamefont {Mayer}}, \bibinfo
  {author} {\bibfnamefont {Ting~Rei}\ \bibnamefont {Tan}}, \bibinfo {author}
  {\bibfnamefont {Jenny~J}\ \bibnamefont {Wu}}, \bibinfo {author}
  {\bibfnamefont {Hilma~M}\ \bibnamefont {Vasconcelos}}, \bibinfo {author}
  {\bibfnamefont {Scott}\ \bibnamefont {Glancy}}, \bibinfo {author}
  {\bibfnamefont {Emanuel}\ \bibnamefont {Knill}}, \bibinfo {author}
  {\bibfnamefont {David~J}\ \bibnamefont {Wineland}},  \emph {et~al.},\
  }\bibfield  {title} {\enquote {\bibinfo {title} {Quantum gate teleportation
  between separated qubits in a trapped-ion processor},}\ }\href
  {https://science.sciencemag.org/content/364/6443/875} {\bibfield  {journal}
  {\bibinfo  {journal} {Science}\ }\textbf {\bibinfo {volume} {364}},\ \bibinfo
  {pages} {875--878} (\bibinfo {year} {2019})}\BibitemShut {NoStop}%
\bibitem [{\citenamefont {Harris}\ \emph {et~al.}(2008)\citenamefont {Harris},
  \citenamefont {Johnson}, \citenamefont {Han}, \citenamefont {Berkley},
  \citenamefont {Johansson}, \citenamefont {Bunyk}, \citenamefont {Ladizinsky},
  \citenamefont {Govorkov}, \citenamefont {Thom}, \citenamefont {Uchaikin}
  \emph {et~al.}}]{harris2008probing}%
  \BibitemOpen
  \bibfield  {author} {\bibinfo {author} {\bibfnamefont {R}~\bibnamefont
  {Harris}}, \bibinfo {author} {\bibfnamefont {MW}~\bibnamefont {Johnson}},
  \bibinfo {author} {\bibfnamefont {S}~\bibnamefont {Han}}, \bibinfo {author}
  {\bibfnamefont {AJ}~\bibnamefont {Berkley}}, \bibinfo {author} {\bibfnamefont
  {J}~\bibnamefont {Johansson}}, \bibinfo {author} {\bibfnamefont
  {P}~\bibnamefont {Bunyk}}, \bibinfo {author} {\bibfnamefont {E}~\bibnamefont
  {Ladizinsky}}, \bibinfo {author} {\bibfnamefont {S}~\bibnamefont {Govorkov}},
  \bibinfo {author} {\bibfnamefont {MC}~\bibnamefont {Thom}}, \bibinfo {author}
  {\bibfnamefont {S}~\bibnamefont {Uchaikin}},  \emph {et~al.},\ }\bibfield
  {title} {\enquote {\bibinfo {title} {Probing noise in flux qubits via
  macroscopic resonant tunneling},}\ }\href
  {https://journals.aps.org/prl/abstract/10.1103/PhysRevLett.101.117003}
  {\bibfield  {journal} {\bibinfo  {journal} {Phys. Rev. Lett.}\ }\textbf
  {\bibinfo {volume} {101}},\ \bibinfo {pages} {117003} (\bibinfo {year}
  {2008})}\BibitemShut {NoStop}%
\bibitem [{\citenamefont {Bylander}\ \emph {et~al.}(2011)\citenamefont
  {Bylander}, \citenamefont {Gustavsson}, \citenamefont {Yan}, \citenamefont
  {Yoshihara}, \citenamefont {Harrabi}, \citenamefont {Fitch}, \citenamefont
  {Cory}, \citenamefont {Nakamura}, \citenamefont {Tsai},\ and\ \citenamefont
  {Oliver}}]{bylander2011noise}%
  \BibitemOpen
  \bibfield  {author} {\bibinfo {author} {\bibfnamefont {Jonas}\ \bibnamefont
  {Bylander}}, \bibinfo {author} {\bibfnamefont {Simon}\ \bibnamefont
  {Gustavsson}}, \bibinfo {author} {\bibfnamefont {Fei}\ \bibnamefont {Yan}},
  \bibinfo {author} {\bibfnamefont {Fumiki}\ \bibnamefont {Yoshihara}},
  \bibinfo {author} {\bibfnamefont {Khalil}\ \bibnamefont {Harrabi}}, \bibinfo
  {author} {\bibfnamefont {George}\ \bibnamefont {Fitch}}, \bibinfo {author}
  {\bibfnamefont {David~G}\ \bibnamefont {Cory}}, \bibinfo {author}
  {\bibfnamefont {Yasunobu}\ \bibnamefont {Nakamura}}, \bibinfo {author}
  {\bibfnamefont {Jaw-Shen}\ \bibnamefont {Tsai}}, \ and\ \bibinfo {author}
  {\bibfnamefont {William~D}\ \bibnamefont {Oliver}},\ }\bibfield  {title}
  {\enquote {\bibinfo {title} {Noise spectroscopy through dynamical decoupling
  with a superconducting flux qubit},}\ }\href
  {https://www.nature.com/articles/nphys1994} {\bibfield  {journal} {\bibinfo
  {journal} {Nature Physics}\ }\textbf {\bibinfo {volume} {7}},\ \bibinfo
  {pages} {565} (\bibinfo {year} {2011})}\BibitemShut {NoStop}%
\bibitem [{\citenamefont {Chan}\ \emph {et~al.}(2018)\citenamefont {Chan},
  \citenamefont {Huang}, \citenamefont {Yang}, \citenamefont {Hwang},
  \citenamefont {Hensen}, \citenamefont {Tanttu}, \citenamefont {Hudson},
  \citenamefont {Itoh}, \citenamefont {Laucht}, \citenamefont {Morello} \emph
  {et~al.}}]{chan2018assessment}%
  \BibitemOpen
  \bibfield  {author} {\bibinfo {author} {\bibfnamefont {KW}~\bibnamefont
  {Chan}}, \bibinfo {author} {\bibfnamefont {W}~\bibnamefont {Huang}}, \bibinfo
  {author} {\bibfnamefont {CH}~\bibnamefont {Yang}}, \bibinfo {author}
  {\bibfnamefont {JCC}\ \bibnamefont {Hwang}}, \bibinfo {author} {\bibfnamefont
  {B}~\bibnamefont {Hensen}}, \bibinfo {author} {\bibfnamefont {T}~\bibnamefont
  {Tanttu}}, \bibinfo {author} {\bibfnamefont {FE}~\bibnamefont {Hudson}},
  \bibinfo {author} {\bibfnamefont {Kohei~M}\ \bibnamefont {Itoh}}, \bibinfo
  {author} {\bibfnamefont {A}~\bibnamefont {Laucht}}, \bibinfo {author}
  {\bibfnamefont {A}~\bibnamefont {Morello}},  \emph {et~al.},\ }\bibfield
  {title} {\enquote {\bibinfo {title} {Assessment of a silicon quantum dot spin
  qubit environment via noise spectroscopy},}\ }\href
  {https://journals.aps.org/prapplied/pdf/10.1103/PhysRevApplied.10.044017}
  {\bibfield  {journal} {\bibinfo  {journal} {Phys. Rev. Appl.}\ }\textbf
  {\bibinfo {volume} {10}},\ \bibinfo {pages} {044017} (\bibinfo {year}
  {2018})}\BibitemShut {NoStop}%
\bibitem [{\citenamefont {Klimov}\ \emph {et~al.}(2018)\citenamefont {Klimov},
  \citenamefont {Kelly}, \citenamefont {Chen}, \citenamefont {Neeley},
  \citenamefont {Megrant}, \citenamefont {Burkett}, \citenamefont {Barends},
  \citenamefont {Arya}, \citenamefont {Chiaro}, \citenamefont {Chen} \emph
  {et~al.}}]{klimov2018fluctuations}%
  \BibitemOpen
  \bibfield  {author} {\bibinfo {author} {\bibfnamefont {PV}~\bibnamefont
  {Klimov}}, \bibinfo {author} {\bibfnamefont {Julian}\ \bibnamefont {Kelly}},
  \bibinfo {author} {\bibfnamefont {Z}~\bibnamefont {Chen}}, \bibinfo {author}
  {\bibfnamefont {Matthew}\ \bibnamefont {Neeley}}, \bibinfo {author}
  {\bibfnamefont {Anthony}\ \bibnamefont {Megrant}}, \bibinfo {author}
  {\bibfnamefont {Brian}\ \bibnamefont {Burkett}}, \bibinfo {author}
  {\bibfnamefont {Rami}\ \bibnamefont {Barends}}, \bibinfo {author}
  {\bibfnamefont {Kunal}\ \bibnamefont {Arya}}, \bibinfo {author}
  {\bibfnamefont {Ben}\ \bibnamefont {Chiaro}}, \bibinfo {author}
  {\bibfnamefont {Yu}~\bibnamefont {Chen}},  \emph {et~al.},\ }\bibfield
  {title} {\enquote {\bibinfo {title} {Fluctuations of energy-relaxation times
  in superconducting qubits},}\ }\href
  {https://journals.aps.org/prl/abstract/10.1103/PhysRevLett.121.090502}
  {\bibfield  {journal} {\bibinfo  {journal} {Phys. Rev. Lett.}\ }\textbf
  {\bibinfo {volume} {121}},\ \bibinfo {pages} {090502} (\bibinfo {year}
  {2018})}\BibitemShut {NoStop}%
\bibitem [{\citenamefont {Megrant}\ \emph {et~al.}(2012)\citenamefont
  {Megrant}, \citenamefont {Neill}, \citenamefont {Barends}, \citenamefont
  {Chiaro}, \citenamefont {Chen}, \citenamefont {Feigl}, \citenamefont {Kelly},
  \citenamefont {Lucero}, \citenamefont {Mariantoni}, \citenamefont {O'Malley}
  \emph {et~al.}}]{megrant2012planar}%
  \BibitemOpen
  \bibfield  {author} {\bibinfo {author} {\bibfnamefont {Anthony}\ \bibnamefont
  {Megrant}}, \bibinfo {author} {\bibfnamefont {Charles}\ \bibnamefont
  {Neill}}, \bibinfo {author} {\bibfnamefont {Rami}\ \bibnamefont {Barends}},
  \bibinfo {author} {\bibfnamefont {Ben}\ \bibnamefont {Chiaro}}, \bibinfo
  {author} {\bibfnamefont {Yu}~\bibnamefont {Chen}}, \bibinfo {author}
  {\bibfnamefont {Ludwig}\ \bibnamefont {Feigl}}, \bibinfo {author}
  {\bibfnamefont {Julian}\ \bibnamefont {Kelly}}, \bibinfo {author}
  {\bibfnamefont {Erik}\ \bibnamefont {Lucero}}, \bibinfo {author}
  {\bibfnamefont {Matteo}\ \bibnamefont {Mariantoni}}, \bibinfo {author}
  {\bibfnamefont {Peter~JJ}\ \bibnamefont {O'Malley}},  \emph {et~al.},\
  }\bibfield  {title} {\enquote {\bibinfo {title} {Planar superconducting
  resonators with internal quality factors above one million},}\ }\href
  {https://aip.scitation.org/doi/10.1063/1.4818710} {\bibfield  {journal}
  {\bibinfo  {journal} {Appl. Phys. Lett.}\ }\textbf {\bibinfo {volume}
  {100}},\ \bibinfo {pages} {113510} (\bibinfo {year} {2012})}\BibitemShut
  {NoStop}%
\bibitem [{\citenamefont {M{\"u}ller}\ \emph {et~al.}(2015)\citenamefont
  {M{\"u}ller}, \citenamefont {Lisenfeld}, \citenamefont {Shnirman},\ and\
  \citenamefont {Poletto}}]{muller2015interacting}%
  \BibitemOpen
  \bibfield  {author} {\bibinfo {author} {\bibfnamefont {Clemens}\ \bibnamefont
  {M{\"u}ller}}, \bibinfo {author} {\bibfnamefont {J{\"u}rgen}\ \bibnamefont
  {Lisenfeld}}, \bibinfo {author} {\bibfnamefont {Alexander}\ \bibnamefont
  {Shnirman}}, \ and\ \bibinfo {author} {\bibfnamefont {Stefano}\ \bibnamefont
  {Poletto}},\ }\bibfield  {title} {\enquote {\bibinfo {title} {Interacting
  two-level defects as sources of fluctuating high-frequency noise in
  superconducting circuits},}\ }\href
  {https://journals.aps.org/prb/abstract/10.1103/PhysRevB.92.035442} {\bibfield
   {journal} {\bibinfo  {journal} {Phys. Rev. B}\ }\textbf {\bibinfo {volume}
  {92}},\ \bibinfo {pages} {035442} (\bibinfo {year} {2015})}\BibitemShut
  {NoStop}%
\bibitem [{\citenamefont {Mei{\ss}ner}\ \emph {et~al.}(2018)\citenamefont
  {Mei{\ss}ner}, \citenamefont {Seiler}, \citenamefont {Lisenfeld},
  \citenamefont {Ustinov},\ and\ \citenamefont {Weiss}}]{meissner2018probing}%
  \BibitemOpen
  \bibfield  {author} {\bibinfo {author} {\bibfnamefont {Saskia~M}\
  \bibnamefont {Mei{\ss}ner}}, \bibinfo {author} {\bibfnamefont {Arnold}\
  \bibnamefont {Seiler}}, \bibinfo {author} {\bibfnamefont {J{\"u}rgen}\
  \bibnamefont {Lisenfeld}}, \bibinfo {author} {\bibfnamefont {Alexey~V}\
  \bibnamefont {Ustinov}}, \ and\ \bibinfo {author} {\bibfnamefont {Georg}\
  \bibnamefont {Weiss}},\ }\bibfield  {title} {\enquote {\bibinfo {title}
  {Probing individual tunneling fluctuators with coherently controlled
  tunneling systems},}\ }\href
  {https://journals.aps.org/prb/abstract/10.1103/PhysRevB.97.180505} {\bibfield
   {journal} {\bibinfo  {journal} {Phys. Rev. B}\ }\textbf {\bibinfo {volume}
  {97}},\ \bibinfo {pages} {180505} (\bibinfo {year} {2018})}\BibitemShut
  {NoStop}%
\bibitem [{\citenamefont {De~Graaf}\ \emph {et~al.}(2018)\citenamefont
  {De~Graaf}, \citenamefont {Faoro}, \citenamefont {Burnett}, \citenamefont
  {Adamyan}, \citenamefont {Tzalenchuk}, \citenamefont {Kubatkin},
  \citenamefont {Lindstr{\"o}m},\ and\ \citenamefont
  {Danilov}}]{de2018suppression}%
  \BibitemOpen
  \bibfield  {author} {\bibinfo {author} {\bibfnamefont {SE}~\bibnamefont
  {De~Graaf}}, \bibinfo {author} {\bibfnamefont {L}~\bibnamefont {Faoro}},
  \bibinfo {author} {\bibfnamefont {J}~\bibnamefont {Burnett}}, \bibinfo
  {author} {\bibfnamefont {AA}~\bibnamefont {Adamyan}}, \bibinfo {author}
  {\bibfnamefont {A~Ya}\ \bibnamefont {Tzalenchuk}}, \bibinfo {author}
  {\bibfnamefont {SE}~\bibnamefont {Kubatkin}}, \bibinfo {author}
  {\bibfnamefont {T}~\bibnamefont {Lindstr{\"o}m}}, \ and\ \bibinfo {author}
  {\bibfnamefont {AV}~\bibnamefont {Danilov}},\ }\bibfield  {title} {\enquote
  {\bibinfo {title} {Suppression of low-frequency charge noise in
  superconducting resonators by surface spin desorption},}\ }\href
  {https://www.nature.com/articles/s41467-018-03577-2} {\bibfield  {journal}
  {\bibinfo  {journal} {Nat. Commun.}\ }\textbf {\bibinfo {volume} {9}},\
  \bibinfo {pages} {1143} (\bibinfo {year} {2018})}\BibitemShut {NoStop}%
\bibitem [{\citenamefont {Merkel}\ \emph {et~al.}(2019)\citenamefont {Merkel},
  \citenamefont {Thirumalai}, \citenamefont {Tarlton}, \citenamefont
  {Sch{\"a}fer}, \citenamefont {Ballance}, \citenamefont {Harty},\ and\
  \citenamefont {Lucas}}]{merkel2018magnetic}%
  \BibitemOpen
  \bibfield  {author} {\bibinfo {author} {\bibfnamefont {B}~\bibnamefont
  {Merkel}}, \bibinfo {author} {\bibfnamefont {K}~\bibnamefont {Thirumalai}},
  \bibinfo {author} {\bibfnamefont {JE}~\bibnamefont {Tarlton}}, \bibinfo
  {author} {\bibfnamefont {VM}~\bibnamefont {Sch{\"a}fer}}, \bibinfo {author}
  {\bibfnamefont {CJ}~\bibnamefont {Ballance}}, \bibinfo {author}
  {\bibfnamefont {TP}~\bibnamefont {Harty}}, \ and\ \bibinfo {author}
  {\bibfnamefont {DM}~\bibnamefont {Lucas}},\ }\bibfield  {title} {\enquote
  {\bibinfo {title} {Magnetic field stabilization system for atomic physics
  experiments},}\ }\href {https://dx.doi.org/10.1063/1.5080093} {\bibfield
  {journal} {\bibinfo  {journal} {Rev. Sci. Instrum}\ }\textbf {\bibinfo
  {volume} {90}},\ \bibinfo {pages} {044702} (\bibinfo {year}
  {2019})}\BibitemShut {NoStop}%
\bibitem [{\citenamefont {Burnett}\ \emph {et~al.}(2019)\citenamefont
  {Burnett}, \citenamefont {Bengtsson}, \citenamefont {Scigliuzzo},
  \citenamefont {Niepce}, \citenamefont {Kudra}, \citenamefont {Delsing},\ and\
  \citenamefont {Bylander}}]{burnett2019decoherence}%
  \BibitemOpen
  \bibfield  {author} {\bibinfo {author} {\bibfnamefont {Jonathan}\
  \bibnamefont {Burnett}}, \bibinfo {author} {\bibfnamefont {Andreas}\
  \bibnamefont {Bengtsson}}, \bibinfo {author} {\bibfnamefont {Marco}\
  \bibnamefont {Scigliuzzo}}, \bibinfo {author} {\bibfnamefont {David}\
  \bibnamefont {Niepce}}, \bibinfo {author} {\bibfnamefont {Marina}\
  \bibnamefont {Kudra}}, \bibinfo {author} {\bibfnamefont {Per}\ \bibnamefont
  {Delsing}}, \ and\ \bibinfo {author} {\bibfnamefont {Jonas}\ \bibnamefont
  {Bylander}},\ }\bibfield  {title} {\enquote {\bibinfo {title} {Decoherence
  benchmarking of superconducting qubits},}\ }\href
  {https://dx.doi.org/10.1038/s41534-019-0168-5} {\bibfield  {journal}
  {\bibinfo  {journal} {NPJ Quantum Inf.}\ }\textbf {\bibinfo {volume} {5}}
  (\bibinfo {year} {2019})}\BibitemShut {NoStop}%
\bibitem [{\citenamefont {Cortez}\ \emph {et~al.}(2017)\citenamefont {Cortez},
  \citenamefont {Chantasri}, \citenamefont {Garc{\'\i}a-Pintos}, \citenamefont
  {Dressel},\ and\ \citenamefont {Jordan}}]{cortez2017rapid}%
  \BibitemOpen
  \bibfield  {author} {\bibinfo {author} {\bibfnamefont {Luis}\ \bibnamefont
  {Cortez}}, \bibinfo {author} {\bibfnamefont {Areeya}\ \bibnamefont
  {Chantasri}}, \bibinfo {author} {\bibfnamefont {Luis~Pedro}\ \bibnamefont
  {Garc{\'\i}a-Pintos}}, \bibinfo {author} {\bibfnamefont {Justin}\
  \bibnamefont {Dressel}}, \ and\ \bibinfo {author} {\bibfnamefont {Andrew~N}\
  \bibnamefont {Jordan}},\ }\bibfield  {title} {\enquote {\bibinfo {title}
  {Rapid estimation of drifting parameters in continuously measured quantum
  systems},}\ }\href
  {https://journals.aps.org/pra/abstract/10.1103/PhysRevA.95.012314} {\bibfield
   {journal} {\bibinfo  {journal} {Phys. Rev. A}\ }\textbf {\bibinfo {volume}
  {95}},\ \bibinfo {pages} {012314} (\bibinfo {year} {2017})}\BibitemShut
  {NoStop}%
\bibitem [{\citenamefont {Bonato}\ and\ \citenamefont
  {Berry}(2017)}]{bonato2017adaptive}%
  \BibitemOpen
  \bibfield  {author} {\bibinfo {author} {\bibfnamefont {Cristian}\
  \bibnamefont {Bonato}}\ and\ \bibinfo {author} {\bibfnamefont {Dominic~W}\
  \bibnamefont {Berry}},\ }\bibfield  {title} {\enquote {\bibinfo {title}
  {Adaptive tracking of a time-varying field with a quantum sensor},}\ }\href
  {https://journals.aps.org/pra/abstract/10.1103/PhysRevA.95.052348} {\bibfield
   {journal} {\bibinfo  {journal} {Physical Review A}\ }\textbf {\bibinfo
  {volume} {95}},\ \bibinfo {pages} {052348} (\bibinfo {year}
  {2017})}\BibitemShut {NoStop}%
\bibitem [{\citenamefont {Wheatley}\ \emph {et~al.}(2010)\citenamefont
  {Wheatley}, \citenamefont {Berry}, \citenamefont {Yonezawa}, \citenamefont
  {Nakane}, \citenamefont {Arao}, \citenamefont {Pope}, \citenamefont {Ralph},
  \citenamefont {Wiseman}, \citenamefont {Furusawa},\ and\ \citenamefont
  {Huntington}}]{wheatley2010adaptive}%
  \BibitemOpen
  \bibfield  {author} {\bibinfo {author} {\bibfnamefont {TA}~\bibnamefont
  {Wheatley}}, \bibinfo {author} {\bibfnamefont {DW}~\bibnamefont {Berry}},
  \bibinfo {author} {\bibfnamefont {H}~\bibnamefont {Yonezawa}}, \bibinfo
  {author} {\bibfnamefont {D}~\bibnamefont {Nakane}}, \bibinfo {author}
  {\bibfnamefont {H}~\bibnamefont {Arao}}, \bibinfo {author} {\bibfnamefont
  {DT}~\bibnamefont {Pope}}, \bibinfo {author} {\bibfnamefont {TC}~\bibnamefont
  {Ralph}}, \bibinfo {author} {\bibfnamefont {HM}~\bibnamefont {Wiseman}},
  \bibinfo {author} {\bibfnamefont {A}~\bibnamefont {Furusawa}}, \ and\
  \bibinfo {author} {\bibfnamefont {EH}~\bibnamefont {Huntington}},\ }\bibfield
   {title} {\enquote {\bibinfo {title} {Adaptive optical phase estimation using
  time-symmetric quantum smoothing},}\ }\href
  {https://journals.aps.org/prl/abstract/10.1103/PhysRevLett.104.093601}
  {\bibfield  {journal} {\bibinfo  {journal} {Phys. Rev. Lett.}\ }\textbf
  {\bibinfo {volume} {104}},\ \bibinfo {pages} {093601} (\bibinfo {year}
  {2010})}\BibitemShut {NoStop}%
\bibitem [{\citenamefont {Young}\ and\ \citenamefont
  {Whaley}(2012)}]{young2012qubits}%
  \BibitemOpen
  \bibfield  {author} {\bibinfo {author} {\bibfnamefont {Kevin~C}\ \bibnamefont
  {Young}}\ and\ \bibinfo {author} {\bibfnamefont {K~Birgitta}\ \bibnamefont
  {Whaley}},\ }\bibfield  {title} {\enquote {\bibinfo {title} {Qubits as
  spectrometers of dephasing noise},}\ }\href
  {https://journals.aps.org/pra/abstract/10.1103/PhysRevA.86.012314} {\bibfield
   {journal} {\bibinfo  {journal} {Phys. Rev. A}\ }\textbf {\bibinfo {volume}
  {86}},\ \bibinfo {pages} {012314} (\bibinfo {year} {2012})}\BibitemShut
  {NoStop}%
\bibitem [{\citenamefont {Gupta}\ and\ \citenamefont
  {Biercuk}(2018)}]{gupta2018machine}%
  \BibitemOpen
  \bibfield  {author} {\bibinfo {author} {\bibfnamefont {Riddhi~Swaroop}\
  \bibnamefont {Gupta}}\ and\ \bibinfo {author} {\bibfnamefont {Michael~J}\
  \bibnamefont {Biercuk}},\ }\bibfield  {title} {\enquote {\bibinfo {title}
  {Machine learning for predictive estimation of qubit dynamics subject to
  dephasing},}\ }\href
  {https://journals.aps.org/prapplied/abstract/10.1103/PhysRevApplied.9.064042}
  {\bibfield  {journal} {\bibinfo  {journal} {Physical Review Applied}\
  }\textbf {\bibinfo {volume} {9}},\ \bibinfo {pages} {064042} (\bibinfo {year}
  {2018})}\BibitemShut {NoStop}%
\bibitem [{\citenamefont {Granade}\ \emph {et~al.}(2016)\citenamefont
  {Granade}, \citenamefont {Combes},\ and\ \citenamefont
  {Cory}}]{granade2016practical}%
  \BibitemOpen
  \bibfield  {author} {\bibinfo {author} {\bibfnamefont {Christopher}\
  \bibnamefont {Granade}}, \bibinfo {author} {\bibfnamefont {Joshua}\
  \bibnamefont {Combes}}, \ and\ \bibinfo {author} {\bibfnamefont
  {DG}~\bibnamefont {Cory}},\ }\bibfield  {title} {\enquote {\bibinfo {title}
  {Practical bayesian tomography},}\ }\href
  {http://iopscience.iop.org/article/10.1088/1367-2630/18/3/033024/meta}
  {\bibfield  {journal} {\bibinfo  {journal} {New J. Phys.}\ }\textbf {\bibinfo
  {volume} {18}},\ \bibinfo {pages} {033024} (\bibinfo {year}
  {2016})}\BibitemShut {NoStop}%
\bibitem [{\citenamefont {Granade}\ \emph {et~al.}(2017)\citenamefont
  {Granade}, \citenamefont {Ferrie}, \citenamefont {Hincks}, \citenamefont
  {Casagrande}, \citenamefont {Alexander}, \citenamefont {Gross}, \citenamefont
  {Kononenko},\ and\ \citenamefont {Sanders}}]{granade2017qinfer}%
  \BibitemOpen
  \bibfield  {author} {\bibinfo {author} {\bibfnamefont {Christopher}\
  \bibnamefont {Granade}}, \bibinfo {author} {\bibfnamefont {Christopher}\
  \bibnamefont {Ferrie}}, \bibinfo {author} {\bibfnamefont {Ian}\ \bibnamefont
  {Hincks}}, \bibinfo {author} {\bibfnamefont {Steven}\ \bibnamefont
  {Casagrande}}, \bibinfo {author} {\bibfnamefont {Thomas}\ \bibnamefont
  {Alexander}}, \bibinfo {author} {\bibfnamefont {Jonathan}\ \bibnamefont
  {Gross}}, \bibinfo {author} {\bibfnamefont {Michal}\ \bibnamefont
  {Kononenko}}, \ and\ \bibinfo {author} {\bibfnamefont {Yuval}\ \bibnamefont
  {Sanders}},\ }\bibfield  {title} {\enquote {\bibinfo {title} {Qinfer:
  Statistical inference software for quantum applications},}\ }\href
  {https://quantum-journal.org/papers/q-2017-04-25-5/} {\bibfield  {journal}
  {\bibinfo  {journal} {Quantum}\ }\textbf {\bibinfo {volume} {1}},\ \bibinfo
  {pages} {5} (\bibinfo {year} {2017})}\BibitemShut {NoStop}%
\bibitem [{\citenamefont {Huo}\ and\ \citenamefont
  {Li}(2017)}]{huo2017learning}%
  \BibitemOpen
  \bibfield  {author} {\bibinfo {author} {\bibfnamefont {Ming-Xia}\
  \bibnamefont {Huo}}\ and\ \bibinfo {author} {\bibfnamefont {Ying}\
  \bibnamefont {Li}},\ }\bibfield  {title} {\enquote {\bibinfo {title}
  {Learning time-dependent noise to reduce logical errors: Real time error rate
  estimation in quantum error correction},}\ }\href
  {http://iopscience.iop.org/article/10.1088/1367-2630/aa916e} {\bibfield
  {journal} {\bibinfo  {journal} {New J. Phys.}\ }\textbf {\bibinfo {volume}
  {19}},\ \bibinfo {pages} {123032} (\bibinfo {year} {2017})}\BibitemShut
  {NoStop}%
\bibitem [{\citenamefont {Kelly}\ \emph {et~al.}(2016)\citenamefont {Kelly},
  \citenamefont {Barends}, \citenamefont {Fowler}, \citenamefont {Megrant},
  \citenamefont {Jeffrey}, \citenamefont {White}, \citenamefont {Sank},
  \citenamefont {Mutus}, \citenamefont {Campbell}, \citenamefont {Chen} \emph
  {et~al.}}]{kelly2016scalable}%
  \BibitemOpen
  \bibfield  {author} {\bibinfo {author} {\bibfnamefont {J}~\bibnamefont
  {Kelly}}, \bibinfo {author} {\bibfnamefont {R}~\bibnamefont {Barends}},
  \bibinfo {author} {\bibfnamefont {AG}~\bibnamefont {Fowler}}, \bibinfo
  {author} {\bibfnamefont {A}~\bibnamefont {Megrant}}, \bibinfo {author}
  {\bibfnamefont {E}~\bibnamefont {Jeffrey}}, \bibinfo {author} {\bibfnamefont
  {TC}~\bibnamefont {White}}, \bibinfo {author} {\bibfnamefont {D}~\bibnamefont
  {Sank}}, \bibinfo {author} {\bibfnamefont {JY}~\bibnamefont {Mutus}},
  \bibinfo {author} {\bibfnamefont {B}~\bibnamefont {Campbell}}, \bibinfo
  {author} {\bibfnamefont {Yu}~\bibnamefont {Chen}},  \emph {et~al.},\
  }\bibfield  {title} {\enquote {\bibinfo {title} {Scalable in situ qubit
  calibration during repetitive error detection},}\ }\href
  {https://journals.aps.org/pra/abstract/10.1103/PhysRevA.94.032321} {\bibfield
   {journal} {\bibinfo  {journal} {Phys. Rev. A}\ }\textbf {\bibinfo {volume}
  {94}},\ \bibinfo {pages} {032321} (\bibinfo {year} {2016})}\BibitemShut
  {NoStop}%
\bibitem [{\citenamefont {Huo}\ and\ \citenamefont
  {Li}(2018)}]{huo2018temporally}%
  \BibitemOpen
  \bibfield  {author} {\bibinfo {author} {\bibfnamefont {Mingxia}\ \bibnamefont
  {Huo}}\ and\ \bibinfo {author} {\bibfnamefont {Ying}\ \bibnamefont {Li}},\
  }\bibfield  {title} {\enquote {\bibinfo {title} {Self-consistent tomography
  of temporally correlated errors},}\ }\href {https://arxiv.org/abs/1811.02734}
  {\bibfield  {journal} {\bibinfo  {journal} {Preprint at http://arxiv.org/abs/1811.02734}\
  } (\bibinfo {year} {2018})}\BibitemShut {NoStop}%
\bibitem [{\citenamefont {Rudinger}\ \emph {et~al.}(2019)\citenamefont
  {Rudinger}, \citenamefont {Proctor}, \citenamefont {Langharst}, \citenamefont
  {Sarovar}, \citenamefont {Young},\ and\ \citenamefont
  {Blume-Kohout}}]{rudinger2018probing}%
  \BibitemOpen
  \bibfield  {author} {\bibinfo {author} {\bibfnamefont {Kenneth}\ \bibnamefont
  {Rudinger}}, \bibinfo {author} {\bibfnamefont {Timothy}\ \bibnamefont
  {Proctor}}, \bibinfo {author} {\bibfnamefont {Dylan}\ \bibnamefont
  {Langharst}}, \bibinfo {author} {\bibfnamefont {Mohan}\ \bibnamefont
  {Sarovar}}, \bibinfo {author} {\bibfnamefont {Kevin}\ \bibnamefont {Young}},
  \ and\ \bibinfo {author} {\bibfnamefont {Robin}\ \bibnamefont
  {Blume-Kohout}},\ }\bibfield  {title} {\enquote {\bibinfo {title} {Probing
  context-dependent errors in quantum processors},}\ }\href
  {https://journals.aps.org/prx/abstract/10.1103/PhysRevX.9.021045} {\bibfield
  {journal} {\bibinfo  {journal} {Phys. Rev. X}\ }\textbf {\bibinfo {volume}
  {9}},\ \bibinfo {pages} {021045} (\bibinfo {year} {2019})}\BibitemShut
  {NoStop}%
\bibitem [{\citenamefont {Donoho}(2006)}]{donoho2006compressed}%
  \BibitemOpen
  \bibfield  {author} {\bibinfo {author} {\bibfnamefont {David~L}\ \bibnamefont
  {Donoho}},\ }\bibfield  {title} {\enquote {\bibinfo {title} {Compressed
  sensing},}\ }\href {https://ieeexplore.ieee.org/document/1614066} {\bibfield
  {journal} {\bibinfo  {journal} {IEEE Trans. Inf. Theory}\ }\textbf {\bibinfo
  {volume} {52}},\ \bibinfo {pages} {1289--1306} (\bibinfo {year}
  {2006})}\BibitemShut {NoStop}%
\bibitem [{\citenamefont {Lehmann}\ and\ \citenamefont
  {Romano}(2006)}]{lehmann2006testing}%
  \BibitemOpen
  \bibfield  {author} {\bibinfo {author} {\bibfnamefont {Erich~L}\ \bibnamefont
  {Lehmann}}\ and\ \bibinfo {author} {\bibfnamefont {Joseph~P}\ \bibnamefont
  {Romano}},\ }\href@noop {} {\emph {\bibinfo {title} {Testing statistical
  hypotheses}}}\ (\bibinfo  {publisher} {Springer Science \& Business Media},\
  \bibinfo {year} {2006})\BibitemShut {NoStop}%
\bibitem [{\citenamefont {Shaffer}(1995)}]{shaffer1995multiple}%
  \BibitemOpen
  \bibfield  {author} {\bibinfo {author} {\bibfnamefont {Juliet~Popper}\
  \bibnamefont {Shaffer}},\ }\bibfield  {title} {\enquote {\bibinfo {title}
  {Multiple hypothesis testing},}\ }\href
  {https://www.annualreviews.org/doi/10.1146/annurev.ps.46.020195.003021}
  {\bibfield  {journal} {\bibinfo  {journal} {Annual review of psychology}\
  }\textbf {\bibinfo {volume} {46}},\ \bibinfo {pages} {561--584} (\bibinfo
  {year} {1995})}\BibitemShut {NoStop}%
\bibitem [{\citenamefont {Akaike}(1974)}]{akaike1974new}%
  \BibitemOpen
  \bibfield  {author} {\bibinfo {author} {\bibfnamefont {Hirotugu}\
  \bibnamefont {Akaike}},\ }\bibfield  {title} {\enquote {\bibinfo {title} {A
  new look at the statistical model identification},}\ }\href {\doibase
  doi:10.1109/TAC.1974.1100705} {\bibfield  {journal} {\bibinfo  {journal}
  {IEEE Trans. Autom. Control}\ ,\ \bibinfo {pages} {716--723}} (\bibinfo
  {year} {1974})}\BibitemShut {NoStop}%
\bibitem [{\citenamefont {Wimperis}(1994)}]{wimperis1994broadband}%
  \BibitemOpen
  \bibfield  {author} {\bibinfo {author} {\bibfnamefont {Stephen}\ \bibnamefont
  {Wimperis}},\ }\bibfield  {title} {\enquote {\bibinfo {title} {Broadband,
  narrowband, and passband composite pulses for use in advanced nmr
  experiments},}\ }\href
  {https://www.sciencedirect.com/science/article/pii/S1064185884711594}
  {\bibfield  {journal} {\bibinfo  {journal} {J. Magn. Reson. Series A}\
  }\textbf {\bibinfo {volume} {109}},\ \bibinfo {pages} {221--231} (\bibinfo
  {year} {1994})}\BibitemShut {NoStop}%
\bibitem [{\citenamefont {Merrill}\ and\ \citenamefont
  {Brown}(2012)}]{merrill2012progress}%
  \BibitemOpen
  \bibfield  {author} {\bibinfo {author} {\bibfnamefont {J}~\bibnamefont
  {Merrill}}\ and\ \bibinfo {author} {\bibfnamefont {Kenneth~R}\ \bibnamefont
  {Brown}},\ }\bibfield  {title} {\enquote {\bibinfo {title} {Progress in
  compensating pulse sequences for quantum computation},}\ }\href
  {https://arxiv.org/abs/1203.6392} {\bibfield  {journal} {\bibinfo  {journal}
  {Preprint at http://arxiv.org/abs/1203.6392}\ } (\bibinfo {year} {2012})}\BibitemShut
  {NoStop}%
\bibitem [{\citenamefont {Khodjasteh}\ and\ \citenamefont
  {Viola}(2009)}]{khodjasteh2009dynamical}%
  \BibitemOpen
  \bibfield  {author} {\bibinfo {author} {\bibfnamefont {Kaveh}\ \bibnamefont
  {Khodjasteh}}\ and\ \bibinfo {author} {\bibfnamefont {Lorenza}\ \bibnamefont
  {Viola}},\ }\bibfield  {title} {\enquote {\bibinfo {title} {Dynamical quantum
  error correction of unitary operations with bounded controls},}\ }\href
  {https://journals.aps.org/pra/abstract/10.1103/PhysRevA.80.032314} {\bibfield
   {journal} {\bibinfo  {journal} {Phys. Rev. A}\ }\textbf {\bibinfo {volume}
  {80}},\ \bibinfo {pages} {032314} (\bibinfo {year} {2009})}\BibitemShut
  {NoStop}%
\bibitem [{\citenamefont {Aharonov}\ \emph {et~al.}(1998)\citenamefont
  {Aharonov}, \citenamefont {Kitaev},\ and\ \citenamefont
  {Nisan}}]{aharonov1998quantum}%
  \BibitemOpen
  \bibfield  {author} {\bibinfo {author} {\bibfnamefont {Dorit}\ \bibnamefont
  {Aharonov}}, \bibinfo {author} {\bibfnamefont {Alexei}\ \bibnamefont
  {Kitaev}}, \ and\ \bibinfo {author} {\bibfnamefont {Noam}\ \bibnamefont
  {Nisan}},\ }\bibfield  {title} {\enquote {\bibinfo {title} {Quantum circuits
  with mixed states},}\ }in\ \href@noop {} {\emph {\bibinfo {booktitle}
  {Proceedings of the thirtieth annual ACM symposium on Theory of computing}}}\
  (\bibinfo {organization} {ACM},\ \bibinfo {year} {1998})\ pp.\ \bibinfo
  {pages} {20--30}\BibitemShut {NoStop}%
\bibitem [{\citenamefont {Nielsen}\ \emph
  {et~al.}(2020{\natexlab{a}})\citenamefont {Nielsen}, \citenamefont
  {Blume-Kohout}, \citenamefont {Saldyt}, \citenamefont {Gross}, \citenamefont
  {Scholten}, \citenamefont {Rudinger}, \citenamefont {Proctor},\ and\
  \citenamefont {Gamble}}]{pygstiversion0.9.9.1}%
  \BibitemOpen
  \bibfield  {author} {\bibinfo {author} {\bibfnamefont {Erik}\ \bibnamefont
  {Nielsen}}, \bibinfo {author} {\bibfnamefont {Robin}\ \bibnamefont
  {Blume-Kohout}}, \bibinfo {author} {\bibfnamefont {Lucas}\ \bibnamefont
  {Saldyt}}, \bibinfo {author} {\bibfnamefont {Jonathan}\ \bibnamefont
  {Gross}}, \bibinfo {author} {\bibfnamefont {Travis}\ \bibnamefont
  {Scholten}}, \bibinfo {author} {\bibfnamefont {Kenneth}\ \bibnamefont
  {Rudinger}}, \bibinfo {author} {\bibfnamefont {Timothy}\ \bibnamefont
  {Proctor}}, \ and\ \bibinfo {author} {\bibfnamefont {John~King}\ \bibnamefont
  {Gamble}},\ } {\enquote
  {\bibinfo {title} {Py{GST}i pre-release of Version 0.9.10: 7c6ddd1},}\ } (\bibinfo {year}
  {2020}{\natexlab{a}}),\ \bibinfo {note} {URL: https://github.com/pyGSTio/pyGSTi/tree/7c6ddd1de209b795e a39bfb69d010b687e812d07}\BibitemShut
  {NoStop}%
\bibitem [{\citenamefont {Nielsen}\ \emph
  {et~al.}(2020{\natexlab{b}})\citenamefont {Nielsen}, \citenamefont
  {Rudinger}, \citenamefont {Proctor}, \citenamefont {Russo}, \citenamefont
  {Young},\ and\ \citenamefont {Blume-Kohout}}]{nielsen2020probing}%
  \BibitemOpen
  \bibfield  {author} {\bibinfo {author} {\bibfnamefont {Erik}\ \bibnamefont
  {Nielsen}}, \bibinfo {author} {\bibfnamefont {Kenneth}\ \bibnamefont
  {Rudinger}}, \bibinfo {author} {\bibfnamefont {Timothy}\ \bibnamefont
  {Proctor}}, \bibinfo {author} {\bibfnamefont {Antonio}\ \bibnamefont
  {Russo}}, \bibinfo {author} {\bibfnamefont {Kevin}\ \bibnamefont {Young}}, \
  and\ \bibinfo {author} {\bibfnamefont {Robin}\ \bibnamefont {Blume-Kohout}},\
  }\bibfield  {title} {\enquote {\bibinfo {title} {Probing quantum processor
  performance with py{GST}i},}\ }\href
  {https://iopscience.iop.org/article/10.1088/2058-9565/ab8aa4} {\bibfield
  {journal} {\bibinfo  {journal} {Quantum Sci. Technol.}\ }\textbf {\bibinfo
  {volume} {5}},\ \bibinfo {pages} {044002} (\bibinfo {year}
  {2020}{\natexlab{b}})}\BibitemShut {NoStop}%
\bibitem [{\citenamefont {Fisk}\ \emph {et~al.}(1997)\citenamefont {Fisk},
  \citenamefont {Sellars}, \citenamefont {Lawn},\ and\ \citenamefont
  {Coles}}]{fisk1997accurate}%
  \BibitemOpen
  \bibfield  {author} {\bibinfo {author} {\bibfnamefont {Peter~TH}\
  \bibnamefont {Fisk}}, \bibinfo {author} {\bibfnamefont {MJ}~\bibnamefont
  {Sellars}}, \bibinfo {author} {\bibfnamefont {Malcolm~A}\ \bibnamefont
  {Lawn}}, \ and\ \bibinfo {author} {\bibfnamefont {G}~\bibnamefont {Coles}},\
  }\bibfield  {title} {\enquote {\bibinfo {title} {Accurate measurement of the
  12.6 {GHz} ``clock" transition in trapped $^{71}${Yb}$^{+}$ ions},}\ }\href
  {https://ieeexplore.ieee.org/abstract/document/585119} {\bibfield  {journal}
  {\bibinfo  {journal} {IEEE Trans. Ultrasonics, Ferroelectrics, and Frequency
  Control}\ }\textbf {\bibinfo {volume} {44}},\ \bibinfo {pages} {344--354}
  (\bibinfo {year} {1997})}\BibitemShut {NoStop}%
\bibitem [{\citenamefont {Olmschenk}\ \emph {et~al.}(2007)\citenamefont
  {Olmschenk}, \citenamefont {Younge}, \citenamefont {Moehring}, \citenamefont
  {Matsukevich}, \citenamefont {Maunz},\ and\ \citenamefont
  {Monroe}}]{olmschenk2007manipulation}%
  \BibitemOpen
  \bibfield  {author} {\bibinfo {author} {\bibfnamefont {Steve}\ \bibnamefont
  {Olmschenk}}, \bibinfo {author} {\bibfnamefont {Kelly~C}\ \bibnamefont
  {Younge}}, \bibinfo {author} {\bibfnamefont {David~L}\ \bibnamefont
  {Moehring}}, \bibinfo {author} {\bibfnamefont {Dzmitry~N}\ \bibnamefont
  {Matsukevich}}, \bibinfo {author} {\bibfnamefont {Peter}\ \bibnamefont
  {Maunz}}, \ and\ \bibinfo {author} {\bibfnamefont {Christopher}\ \bibnamefont
  {Monroe}},\ }\bibfield  {title} {\enquote {\bibinfo {title} {Manipulation and
  detection of a trapped {Yb}$^+$ hyperfine qubit},}\ }\href
  {https://journals.aps.org/pra/abstract/10.1103/PhysRevA.76.052314} {\bibfield
   {journal} {\bibinfo  {journal} {Phys. Rev. A}\ }\textbf {\bibinfo {volume}
  {76}},\ \bibinfo {pages} {052314} (\bibinfo {year} {2007})}\BibitemShut
  {NoStop}%
\bibitem [{\citenamefont {Ahmed}\ \emph {et~al.}(1974)\citenamefont {Ahmed},
  \citenamefont {Natarajan},\ and\ \citenamefont {Rao}}]{ahmed1974discrete}%
  \BibitemOpen
  \bibfield  {author} {\bibinfo {author} {\bibfnamefont {Nasir}\ \bibnamefont
  {Ahmed}}, \bibinfo {author} {\bibfnamefont {T.}~\bibnamefont {Natarajan}}, \
  and\ \bibinfo {author} {\bibfnamefont {Kamisetty~R}\ \bibnamefont {Rao}},\
  }\bibfield  {title} {\enquote {\bibinfo {title} {Discrete cosine
  transform},}\ }\href {https://ieeexplore.ieee.org/document/1672377/}
  {\bibfield  {journal} {\bibinfo  {journal} {IEEE Trans. Comput.}\ }\textbf
  {\bibinfo {volume} {100}},\ \bibinfo {pages} {90--93} (\bibinfo {year}
  {1974})}\BibitemShut {NoStop}%
 \bibitem [{\citenamefont {Rao}\ and\ \citenamefont
  {Yip}(2014)}]{rao2014discrete}%
  \BibitemOpen
  \bibfield  {author} {\bibinfo {author} {\bibfnamefont {K~Ramamohan}\
  \bibnamefont {Rao}}\ and\ \bibinfo {author} {\bibfnamefont {Ping}\
  \bibnamefont {Yip}},\ }\href@noop {} {\emph {\bibinfo {title} {Discrete
  cosine transform: algorithms, advantages, applications}}}\ (\bibinfo
  {publisher} {Academic press},\ \bibinfo {year} {2014})\BibitemShut {NoStop}%
  \bibitem [{\citenamefont {Feldman}(1987)}]{feldman1987fast}%
  \BibitemOpen
  \bibfield  {author} {\bibinfo {author} {\bibfnamefont {Frank~A}\ \bibnamefont
  {Feldman}},\ }\bibfield  {title} {\enquote {\bibinfo {title} {Fast spectral
  tests for measuring nonrandomness and the {DES}},}\ }in\ \href@noop {} {\emph
  {\bibinfo {booktitle} {Conference on the Theory and Application of
  Cryptographic Techniques}}}\ (\bibinfo {organization} {Springer},\ \bibinfo
  {year} {1987})\ pp.\ \bibinfo {pages} {243--254}\BibitemShut {NoStop}%
\bibitem [{\citenamefont {Zechmeister}\ and\ \citenamefont
  {K{\"u}rster}(2009)}]{zechmeister2009generalised}%
  \BibitemOpen
  \bibfield  {author} {\bibinfo {author} {\bibfnamefont {M}~\bibnamefont
  {Zechmeister}}\ and\ \bibinfo {author} {\bibfnamefont {M}~\bibnamefont
  {K{\"u}rster}},\ }\bibfield  {title} {\enquote {\bibinfo {title} {The
  generalised {L}omb-{S}cargle periodogram -- a new formalism for the
  floating-mean and keplerian periodograms},}\ }\href
  {https://www.aanda.org/articles/aa/abs/2009/11/aa11296-08/aa11296-08.html}
  {\bibfield  {journal} {\bibinfo  {journal} {Astron. Astrophys.}\ }\textbf
  {\bibinfo {volume} {496}},\ \bibinfo {pages} {577--584} (\bibinfo {year}
  {2009})}\BibitemShut {NoStop}%
\bibitem [{\citenamefont {VanderPlas}(2018)}]{vanderplas2018understanding}%
  \BibitemOpen
  \bibfield  {author} {\bibinfo {author} {\bibfnamefont {Jacob~T}\ \bibnamefont
  {VanderPlas}},\ }\bibfield  {title} {\enquote {\bibinfo {title}
  {Understanding the {L}omb--{S}cargle periodogram},}\ }\href
  {https://iopscience.iop.org/article/10.3847/1538-4365/aab766/meta} {\bibfield
   {journal} {\bibinfo  {journal} {Astro. J. Supplement Series}\ }\textbf
  {\bibinfo {volume} {236}},\ \bibinfo {pages} {16} (\bibinfo {year}
  {2018})}\BibitemShut {NoStop}%
  \bibitem [{\citenamefont {Benjamini}\ and\ \citenamefont
  {Hochberg}(1995)}]{benjamini1995controlling}%
  \BibitemOpen
  \bibfield  {author} {\bibinfo {author} {\bibfnamefont {Yoav}\ \bibnamefont
  {Benjamini}}\ and\ \bibinfo {author} {\bibfnamefont {Yosef}\ \bibnamefont
  {Hochberg}},\ }\bibfield  {title} {\enquote {\bibinfo {title} {Controlling
  the false discovery rate: a practical and powerful approach to multiple
  testing},}\ }\href {https://www.jstor.org/stable/2346101} {\bibfield
  {journal} {\bibinfo  {journal} {J. Roy. Statist. Soc. Ser. B}\ ,\ \bibinfo
  {pages} {289--300}} (\bibinfo {year} {1995})}\BibitemShut {NoStop}%
\bibitem [{\citenamefont {Billingsley}(1995)}]{billingsley2008probability}%
  \BibitemOpen
  \bibfield  {author} {\bibinfo {author} {\bibfnamefont {Patrick}\ \bibnamefont
  {Billingsley}},\ }\href@noop {} {\emph {\bibinfo {title} {Probability and
  measure}}},\ \bibinfo {edition} {3rd}\ ed.\ (\bibinfo  {publisher} {John
  Wiley \& Sons},\ \bibinfo {year} {1995})\BibitemShut {NoStop}%
\bibitem [{\citenamefont {Watrous}(2005)}]{watrous2005notes}%
  \BibitemOpen
  \bibfield  {author} {\bibinfo {author} {\bibfnamefont {John}\ \bibnamefont
  {Watrous}},\ }\bibfield  {title} {\enquote {\bibinfo {title} {Notes on
  super-operator norms induced by schatten norms},}\ }\href@noop {} {\bibfield
  {journal} {\bibinfo  {journal} {Quantum Information \& Computation}\ }\textbf
  {\bibinfo {volume} {5}},\ \bibinfo {pages} {58--68} (\bibinfo {year}
  {2005})}\BibitemShut {NoStop}%
\bibitem [{\citenamefont {Nielsen}(2002)}]{nielsen2002simple}%
  \BibitemOpen
  \bibfield  {author} {\bibinfo {author} {\bibfnamefont {Michael~A}\
  \bibnamefont {Nielsen}},\ }\bibfield  {title} {\enquote {\bibinfo {title} {A
  simple formula for the average gate fidelity of a quantum dynamical
  operation},}\ }\href
  {https://www.sciencedirect.com/science/article/pii/S0375960102012720}
  {\bibfield  {journal} {\bibinfo  {journal} {Phys. Lett. A}\ }\textbf
  {\bibinfo {volume} {303}},\ \bibinfo {pages} {249--252} (\bibinfo {year}
  {2002})}\BibitemShut {NoStop}%
\end{thebibliography}
\end{document}